\documentclass[]{pasj01}
\usepackage{url}
\Received{12-May-2017}
\Accepted{05-Sep-2017}
\Published{$\langle$publication date$\rangle$}
\SetRunningHead{}{}
\begin{document}

\title{Cosmic Magnetism in Centimeter and Meter Wavelength Radio Astronomy}
\author{
Takuya \textsc{Akahori}\altaffilmark{1}\thanks{
corresponding author: akahori@sci.kagoshima-u.ac.jp; 
$^{1}$Graduate School of Science and Engineering, Kagoshima University, 1-21-35 Korimoto, Kagoshima, Kagoshima 890-0065, Japan; 
$^{2}$Institute of Astronomy, The University of Tokyo, 2-21-1 Osawa, Mitaka, Tokyo 181-0015, Japan;
$^{3}$Graduate School of Science, Osaka University, Osaka 560-0043, Japan;
$^{4}$Kobayashi-Maskawa Institute, Nagoya University, Aichi 464-8602, Japan;
$^{5}$Department of Physics, UNIST, Ulsan 44919, Korea;
$^{6}$Mizusawa VLBI Observatory, National Astronomical Observatory of Japan, 2-12 Hoshigaoka, Mizusawa, Oshu, Iwate 023-0861, Japan;
$^{7}$Faculy of Education, Nagasaki University, Nagasaki 852-8521, Japan;
$^{8}$Faculty of Sciences, Chiba University, Chiba-shi 263-8522, Japan;
$^{9}$Faculty of Sciences, Kyushu University, Fukuoka 819-0395, Japan;
$^{10}$Department of Physics, Kumamoto University, Kumamoto 860-8555, Japan;
$^{11}$Tohoku Bunkyo College, Yamagata 990-2316, Japan;
$^{12}$National Astronomical Observatory of Japan, 2-21-1 Osawa, Mitaka, Tokyo 181-8588, Japan;
$^{13}$Department of Physics, Yamagata University, Yamagata 990-8560, Japan;
$^{14}$Institute for University Education and Student Support, Ibaraki University.
}, 
Hiroyuki \textsc{Nakanishi}\altaffilmark{1},
Yoshiaki \textsc{Sofue}\altaffilmark{2},\\
Yutaka \textsc{Fujita}\altaffilmark{3},
Kiyotomo \textsc{Ichiki}\altaffilmark{4},
Shinsuke \textsc{Ideguchi}\altaffilmark{5},
Osamu \textsc{Kameya}\altaffilmark{6},
Takahiro \textsc{Kudoh}\altaffilmark{7},
Yuki \textsc{Kudoh}\altaffilmark{8},
Mami \textsc{Machida}\altaffilmark{9},
Yoshimitsu \textsc{Miyashita}\altaffilmark{10},
Hiroshi \textsc{Ohno}\altaffilmark{11},
Takeaki \textsc{Ozawa}\altaffilmark{12},
Keitaro \textsc{Takahashi}\altaffilmark{10},\\
Motokazu \textsc{Takizawa}\altaffilmark{13}, and 
Dai. G. \textsc{Yamazaki}\altaffilmark{14,12}
}


\maketitle

%%%%%%%%%%%%%%%%%%%%%%%%%%%%%%%%%%%%%%%%%%%%
%%%%%%%%%%%%%%%%%%%%%%%%%%%%%%%%%%%%%%%%%%%%
%%%%%%%%%%%%%%%%%%%%%%%%%%%%%%%%%%%%%%%%%%%%
\begin{abstract}

Magnetic field is ubiquitous in the Universe and it plays essential roles in various astrophysical phenomena, yet its real origin and evolution are poorly known. This article reviews current understanding of magnetic fields in the interstellar medium, the Milky Way Galaxy, external galaxies, active galactic nuclei, clusters of galaxies, and the cosmic web. Particularly, the review concentrates on the achievements that have been provided by centimeter and meter wavelength radio observations. The article also introduces various methods to analyze linear polarization data, including synchrotron radiation, Faraday rotation, depolarization, and Faraday tomography.

\end{abstract}

%%%%%%%%%%%%%%%%%%%%%%%%%%%%%%%%%%%%%%%%%%%%
%%%%%%%%%%%%%%%%%%%%%%%%%%%%%%%%%%%%%%%%%%%%
%%%%%%%%%%%%%%%%%%%%%%%%%%%%%%%%%%%%%%%%%%%%
\section{Introduction}
\label{mag.s1}

%%%%%%%%%%%%%%%%%%%%%%%%%%%%%%%%%%%%%%%%%%%%
%%%%%%%%%%%%%%%%%%%%%%%%%%%%%%%%%%%%%%%%%%%%
\subsection{Magnetized Universe}
\label{mag.s1.ss1}

Magnetism plays substantial and often essential roles in astronomical objects. Most of known celestial objects, the Earth, planets, the Sun, stars, interstellar space and clouds, the Milky Way Galaxy, galaxies, accretion disks and active galactic nuclei (AGN), and clusters of galaxies, are known to be magnetized. An exception might be the Universe where the cosmological isotropy principle has denied the cosmological-scale uniform field, that defines the North and South of the Universe.

\begin{figure}
\begin{center}
\FigureFile(70mm,70mm){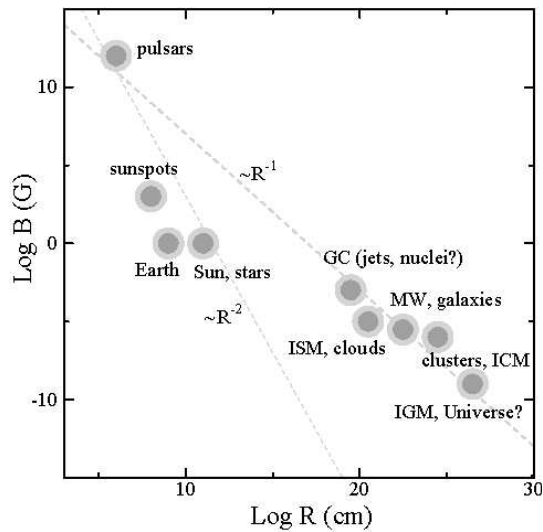}
\end{center}
\caption{
Cosmic magnetic fields from the strongest to weakest, accordingly from compact to large-scale objects in the Universe. The horizontal and vertical axe represent the object size, $R$, and the magnetic-field strength, $B$, respectively. Dashed lines indicates power-laws with indices of $-1$ and $-2$.
}\label{f01}
\end{figure}

The magnetic-field strength, $B$, is roughly related to the object size, $R$. Figure~\ref{f01} depicts the global distribution of magnetic fields in the log $B$ -- log $R$ plot. An inverse relation, $B \sim (R/10~{\rm kpc})^{-1}\ \mu{\rm {\rm G}}$, is seen in the plot. It may also be noticed that the stars and pulsars roughly obey a squared-inverse relation, $B \sim 10^{12}(R/10 \ {\rm km})^{-2}~{\rm G}$, suggestive of frozen-in amplification during stellar collapses. 

The strongest magnetic field observed so far in the Universe reaches $\sim 10^{13}$ G for magnetars among neutron stars. It is several orders of magnitude stronger than that achieved in the laboratories. Magnetic fields in the interstellar medium (ISM) are on the order of several $\mu$G, and those in the intra-cluster medium (ICM) are often observed with the strength of about $\mu$G. The largest-scale, hence the weakest, non-ordered magnetic fields may permeate the intergalactic medium (IGM) in the large-scale structure of the Universe, whilst the study of them is a challenging subject for cosmology as well as to polarization technology in radio and far-infrared astronomy. 

Magnetic fields induce fundamental astrophysical processes such as particle acceleration, non-thermal radiation, polarization, and impact on activities of astronomical objects through field tension, reconnection, instability, and turbulence. Such rich, diverse nature of magnetic phenomena is explained by common theories of magnetism, though various magnetic effects often make their appearance complicated. 

Magnetic fields often help researches of other science subjects. For instance, understanding Galactic magnetic field (GMF) assists studying interstellar physics such as the formation of molecular clouds and stars. That for spiral galaxies assists to investigate the origin of spiral arms. Magnetic fields around accretion disks assist high-energy cosmic jets. Inter-galactic magnetic field (IGMF) in the ICM and IGM are one of the keys to understand the acceleration mechanism of the high energy cosmic rays (CRs). The deepest magnetic fields may preserve information of the early Universe, so that they will assist studies of the epoch of reionization, the cosmic microwave background polarization, and ultimately the inflation and the Big Bang of the Universe.

%%%%%%%%%%%%%%%%%%%%%%%%%%%%%%%%%%%%%%%%%%%%
%%%%%%%%%%%%%%%%%%%%%%%%%%%%%%%%%%%%%%%%%%%%
\subsection{History of Cosmic Magnetism Research}
\label{mag.s1.ss2}

The dawn of contemporary magnetic view of the Universe occurred when Karl G. Jansky by chance detected Milky Way's radio emission in 1931. At the time, magnetic fields had been known only in the Earth and the Sun, and considered in high-energy astrophysics. The emission was proved to be originating from synchrotron radiation by the interaction of CR electrons and magnetic fields \citep{1965ARA&A...3..297G, 1969ARA&A...7..375G}, revealing that the Milky Way is a huge magnetized disk embedded in a stellar gravitational potential.

Measurements of radio emission from the sky have been extensively employed in the 1960's to estimate GMF strengths. They observed the emission in meter wavelengths, and hence mostly synchrotron radiation. Radio intensities in the North Galactic Pole were used to estimate the mean radio emissivity in the galactic disk, and were used to calculate field strength by assuming the equipartition of energy densities between CRs and magnetic fields. The field strength in the solar neighborhood was known to be on the order of a few $\mu$G.

Linearly polarized radio waves further advanced the study of cosmic magnetism in the 1970's. They provided more fundamental information of magnetic fields such as three dimensional configuration of magnetic lines of force. Galactic polarized emission indicated local field configurations, while Faraday depolarization made the analysis too complicated to reach a definite solution. Linear polarization observations of pulsars and extragalactic radio sources made the study of cosmic magnetism easier, more effective, and quantitative, and are today the major tool to study the GMF by using their Faraday rotation measures (RM) (e.g., \cite{1966ARA&A...4..245G}). 

In the 1980's, linear polarization mapping of nebulae, galaxies, and radio lobes made it possible to derive their internal field configurations using Faraday RM analyses. The following, various topologies of GMF configuration in the Milky Way and external galaxies were recognized:
\begin{itemize}
\item R .... Ring, or toroidal field,
\item BSS ... BiSymmetric spiral field,
\item ASS ... AxiSymmetric spiral field,
\item V ... Vertical, poloidal, or dipole field,
\end{itemize}
\citep{1986ARA&A..24..459S}. We follow these abbreviations in this paper. In addition, we also use
\begin{itemize}
\item MSS ... Multi-Symmetric Spiral field, including QSS (Quadri-Symmetrical spiral),
\end{itemize}
which was proposed recently \citep{2008A&A...480...45S}. One may also classify irregular field configuration into
\begin{itemize}
\item T ... Turbulent, flocculent, striated, ordered, or random field.
\end{itemize}
In the 1990's, with extensive development combined with sophisticated RM analyses of extragalactic radio sources, study of cluster magnetic fields became one of the most promising subjects as well. 

In the 2000's, high-sensitivity direct polarization mappings of Abell clusters of galaxies have revealed dynamical properties of the ICM. It was found polarization and magnetization stronger than that expected from some magneto-hydrostatic conditions. High-sensitivity observations also have provided the RM grid, which is a RM map consisting of pixels of RMs for compact background polarized sources. The RM grids have revealed magnetic-field structures of various objects including supernova remnants, the Milky Way, and external galaxies. 

In the last decade, various new methods of RM analyses and techniques to measure radio polarization have been proposed and applied to a large number of radio sources. These include depolarization analysis and Faraday tomography, both deal with high frequency-resolution, multi-wavelengths measurements of radio polarization.

%%%%%%%%%%%%%%%%%%%%%%%%%%%%%%%%%%%%%%%%%%%%
%%%%%%%%%%%%%%%%%%%%%%%%%%%%%%%%%%%%%%%%%%%%
\subsection{Layout of This Paper}
\label{mag.s1.ss3}

The purpose of this article is to review the current understanding and achievements in the studies of astrophysical and astronomical magnetism. The paper concentrates on low frequency radio astronomy at centimeter to meter wavelengths, because the wavelengths are suited to observe synchrotron radiation and Faraday rotation for many of astronomical objects. Although it is obvious that high frequency radio, optical, and infrared polarization observations are also the basic tools to investigate interstellar and galactic magnetic fields, these topics are beyond the scope of this paper. The paper also includes a thorough review of the various methods to analyze linear polarization data from radio observations, where are often employed sophisticated measurement technologies.

Sections in this review are as follows. 
\begin{itemize}
\item \S\ref{mag.s2} proposes a universal view on the theory of magnetism.
\item \S\ref{mag.s3} describes the methods to measure and analyze linearly polarized radio emission, particularly highlighting the new methods such as depolarization and Faraday tomography. 
\item \S\ref{mag.s4} summarizes magnetic fields in the ISM.
\item \S\ref{mag.s5} reviews magnetic fields in the Milky Way Galaxy.
\item \S\ref{mag.s6} highlights magnetic fields in external galaxies.
\item \S\ref{mag.s7} focuses on cosmic jets like AGN jets and activities related to magnetized nuclear disks. 
\item \S\ref{mag.s8} describes new magnetic views of clusters of galaxies. 
\item \S\ref{mag.s9} discusses cosmological implication of RM and linear polarization observations and analyses.
\item \S\ref{mag.s10} summarizes the paper. 
\end{itemize}

%%%%%%%%%%%%%%%%%%%%%%%%%%%%%%%%%%%%%%%%%%%%
%%%%%%%%%%%%%%%%%%%%%%%%%%%%%%%%%%%%%%%%%%%%
%%%%%%%%%%%%%%%%%%%%%%%%%%%%%%%%%%%%%%%%%%%%
\section{Basic Theory of Cosmic Magnetism}
\label{mag.s2}

%%%%%%%%%%%%%%%%%%%%%%%%%%%%%%%%%%%%%%%%%%%%
%%%%%%%%%%%%%%%%%%%%%%%%%%%%%%%%%%%%%%%%%%%%
\subsection{Overview of the Origin and Evolution}
\label{mag.s2.ss1}

From the cosmic-scale point of view, the origin of cosmic magnetic fields can be roughly classified into two major ideas. One is the primordial origin; magnetic field was formed in the early Universe and was permeating the interstellar and intergalactic spaces. The primordial fields were trapped to forming celestial bodies, and amplified during the contraction. The other is that the seed field was created inside a celestial body, e.g. a star, by a local electric current, and was amplified by dynamo and other amplification mechanisms. The amplified field was then expelled into circum-body, interstellar, and intergalactic spaces by winds, explosions, outflows, and so on. Here, the escaping flux may make larger-scale magnetic fields, though the mechanism, particularly its efficiency to create cluster-scale field in the Hubble time, is not well examined. Also, there remains a question whether the mechanism can create regular configuration of magnetic fields in galaxies such as the BSS topology. Although primordial models suffer from a lack of smaller-scale irregular magnetic fields seen in various objects, the above difficulties may be saved if there is a primordial field.

As to the amplification and regulation of cosmic magnetic fields, two major mechanisms have been considered. One is the primordial origin, where a field trapped into an object is wound up by the differential rotation of the object. The other is the dynamo mechanism by turbulence, convection, circulation, and/or differential rotation in a celestial object. Both are coupled to each other in most cases. Note that even without differential rotation, magnetic fields can be locally amplified by turbulence in the gas (ISM, IGM) and embedding objects (clouds, galaxies) due to their rotation and collisions. This mechanism is efficient in the local interstellar space where strong fields of scales on the order of cloud sizes can be created, and keeps the global field configuration.

%%%%%%%%%%%%%%%%%%%%%%%%%%%%%%%%%%%%%%%%%%%%
%%%%%%%%%%%%%%%%%%%%%%%%%%%%%%%%%%%%%%%%%%%%
\subsection{Magneto-hydrodynamics and Simulation}
\label{mag.s2.ss2}

%%%%%%%%%%%%%%%%%%%%%%%%%%%%%%%%%%%%%%%%%%%%
\subsubsection{MHD Equation}
\label{mag.s2.ss2.sss1}

In the circumstances considered in this paper, magnetic fields are practically frozen into partially or fully ionized gas such as the ISM, ICM, and the gas of AGN jets, because of the large magnetic Reynolds number and small resistivity. Such magnetized gas can be treated in the approximation of magneto-hydrodynamics (MHD). The MHD approximation is applicable even to interstellar ``neutral" gas such as molecular and HI clouds, because small fraction of particles ionized by CRs moves with the neutral gas by collisional resistivity.

The basic MHD equations are written as
\def\mb#1{\mbox{\boldmath $#1$}}
\begin{equation}\label{eq.mhd1}
\frac{\partial \rho}{\partial t} + \nabla \cdot (\rho \mb{v}) = 0, 
\end{equation}
\begin{equation}\label{eq.mhd2}
\rho \frac{\partial \mb{v}}{\partial t} + \rho (\mb{v} \cdot \nabla) \mb{v} + \nabla P = \frac{1}{c}\mb{J} \times \mb{B} + \rho \mb{a}, 
\end{equation}
\begin{equation}\label{eq.mhd3}
\frac{\partial U}{\partial t} + \nabla \cdot \left[ \left( \frac{\rho \mb{v}^2}{2}+\frac{\gamma}{\gamma-1}P \right) \mb{v} + \frac{c}{4 \pi} \mb{E} \times \mb{B} \right] = \rho \mb{v} \cdot \mb{a} + Q
\end{equation}
\begin{equation}\label{eq.mhd4}
\frac{\partial \mb{B}}{\partial t} = \ \nabla \times (\mb{v} \times \mb{B}) -c \nabla \times (\eta \mb{J}), 
\end{equation}
where $U$ is the internal energy,
\begin{equation}
U=\frac{\rho v^2}{2}+ \frac{B^2}{8 \pi}+\frac{P}{\gamma-1}, 
\end{equation}
$\mb{J}$ is the current density,
\begin{equation}\label{eq.mhd5}
\mb{J} = \frac{c}{4 \pi} \nabla \times \mb{B}, 
\end{equation}
and $\rho$, $\mb{v}$, $P$, $c$, $\mb{B}$, $\mb{a}$, $\gamma$, $\mb{E}$ and $\eta$ are the density, velocity, gas pressure, light speed, magnetic field, external force, specific heat ratio, electric field and resistivity, respectively.

The characteristic condition of the MHD approximation is that a typical timescale of plasma dynamics is longer than a period of plasma oscillation. In other words, the MHD approximation is applicable if the size of the system is much larger than the mean free path of the way perpendicular to magnetic field. The plasma effective mean free path is determined by the Larmor radius, which is in most cases smaller than the system size even in galactic halos and galaxy clusters possessing very weak magnetic fields and large mean free paths. Therefore, the MHD approximation is a reasonable assumption for the ISM and ICM. The MHD condition is an analogous to a fluid approximation for the fluid dynamics such as that the scale length is much larger than the mean free path and the typical timescale of the system is much longer than the collision timescale.

%%%%%%%%%%%%%%%%%%%%%%%%%%%%%%%%%%%%%%%%%%%%
\subsubsection{History and Issues of MHD Simulation}
\label{mag.s2.ss2.sss2}

The first application of MHD simulation to astrophysical magnetic-phenomena is about magnetic reconnection during solar flares (e.g., \cite{1977JPlPh..17..337U}). MHD simulations were then applied to the Parker instability \citep{1971ApJ...163..255P} to understand protostar jets and molecular-cloud formation (e.g., \cite{1984PASJ...36..105U, 1988PASJ...40..171M}). In 1990's, importance of the magnetic instability was pointed out in differentially rotating systems \citep{1991ApJ...376..214B}, and several authors have studied it in accretion disks (e.g., \cite{1995ApJ...440..742H}). 

In 2000's, relativistic MHD codes were developed \citep{2002Sci...295.1688K}, and general relativistic MHD codes (GRMHD) have been written (e.g., \cite{2003ApJ...589..444G}). In simulations of supernovae, further advanced GRMHD codes are developed, which include various physical processes such as the Einstein equation, self-gravity of the system, neutrino transports, and so on (e.g., \cite{2005PhRvD..72d4014S}). Radiative transfer also has been incorporated in MHD codes in the optically-thick regime (e.g., \cite{2004ApJ...605L..45T}). 

As for larger scales, MHD simulations of the cosmological structure formation were performed both with a grid-base code (uniform-grid or adoptive mesh refinement) and smoothed particle hydrodynamics (e.g., \cite{2008Sci...320..909R, 2008A&A...482L..13D, 2008SSRv..134..311D, 2009MNRAS.398.1678D, 2009ApJ...698L..14X, 2009MNRAS.392.1008D, 2010A&A...522..115S, 2010MNRAS.408..684S, 2015MNRAS.453.3999M}). The galactic magnetic fields have been studied under the cosmological structure formation (e.g., \cite{2013MNRAS.435.3575B, 2014ApJ...783L..20P}). 

Nowadays, several open MHD codes for astrophysics are available, e.g., Zeus \citep{1992ApJS...80..791S}, Athena \citep{2008ApJS..178..137S}, CANS+ \citep{2017ma}, and so on. A MHD simulation has become a commonly-used method in astrophysics, and been one of the most powerful tools to study magnetic fields in the Universe. On the other hand, modern simulations get complicated by incorporating many astrophysical processes such as star formation, supernova, galaxy formation, AGN feedback, and so on, as well as microscopic physics such as CR acceleration and plasma conduction/dissipation. These factors can also affect the magnetic fields evolution and structures. Moreover, they increase numerical uncertainty. It is thus necessary for modern MHD simulations to verify the numerical result by, for example, performing the same simulations with different spatial resolutions. 

Finally, a large computational cost is an unavoidable problem when one attempts to carry out a massive simulation. For example, a three-dimensional, radiative MHD simulation of an accretion disk was recently achieved \citep{2016ApJ...826...23T}. The simulation considered 4.5 million numerical grid points, and took 0.5 million time steps or more than one month CPU hours with 512 cores of K-computer in RIKEN, one of the world's top 10 supercomputers in 2017. Even with such a cost, it simulated disk evolution for only 0.3 second in real time. Therefore, not only to develop high-precision and robust codes but also to accelerate the calculation are critical in modern MHD simulation.

%%%%%%%%%%%%%%%%%%%%%%%%%%%%%%%%%%%%%%%%%%%%
%%%%%%%%%%%%%%%%%%%%%%%%%%%%%%%%%%%%%%%%%%%%
\subsection{Dynamo and Magnetic Field}
\label{mag.s2.ss3}

Astrophysical fluid is mostly ionized or partially ionized, including neutral gas such  as HI and molecular clouds, and hence electrically conducting. Its flow tends to be magnetized spontaneously by self-excited electric current. If the flow is magnetized and the field tension is not too strong to prevent the flow motion itself, the flow drags magnetic fields, stretch them, and increase the magnetic flux density -- called dynamo action. This process converts part of kinetic energy into magnetic energy.

Dynamo produces magnetic energy on a scale (i) smaller than the energy-carrying eddy scale (small-scale dynamo), and (ii) larger than the energy-carrying eddy scale (large-scale dynamo). The small-scale dynamo basically produces isotropic structure, but no helicity, while the large-scale dynamo can form anisotropy and helicity, which are further significant in stratified media such as a galactic gaseous disk. 

In early 1900's, it was already recognized that regular (laminar) flow becomes irregular (turbulent) flow in incompressible viscous fluids (e.g., liquid water) with a large Reynolds number,
\begin{equation}
Re=\frac{UL}{\nu},
\end{equation}
where $U$ and $L$ are the characteristic velocity and length, and $\nu$ is the kinematic viscosity coefficient. \citet{1941DoSSR..30..301K} proposed that the statistical nature of high-$Re$ flow motions is locally isotropic, similar, and universal in the inertial range ($L_{\rm d} \ll L \ll L_{\rm f}$), where the suffices f and d mean the ones at the forcing (energy injecting) and energy dissipating scales, respectively. With this hypothesis, dimensional analysis gives the flow kinetic energy spectrum,
\begin{equation}
E(k)=C_{\rm K} \varepsilon^{2/3}k^{-5/3},
\end{equation}
where $\varepsilon$ is the energy dissipation rate, $k=2\pi/l$ is the wavenumber for the physical scale $l$, and $C_{\rm K}$ is the normalization constant. Assuming that the dissipation-scale wavenumber, $k_{\rm d}$, is much larger than the forcing-scale wavenumber, $k_{\rm f}$, i.e. $k_{\rm d} \gg k_{\rm f}$, the Reynolds number can be written as
\begin{equation}
Re\approx \frac{3\sqrt{3}}{2}C_{\rm K}^{3/2}\left( \frac{k_{\rm d}}{k_{\rm f}} \right)^{4/3},
\end{equation}
(see e.g., \cite{2011RPPh...74d6901B}). The above nature can be summarized as follows: a flow with large $Re$ easily results in turbulence and such a flow has a broad ($k_{\rm d}/k_{\rm f} \propto Re^{3/4}$) inertial range.

In addition to the Reynolds number $Re=u_{\rm rms}/(\nu k_{\rm f})$, the magnetic Reynolds number $Re_{\rm M}=u_{\rm rms}/(\eta k_{\rm f})$ is another important parameter on the transition from laminar to turbulent flows. Here $u_{\rm rms}$ is the characteristic rms velocity of turbulence. Laboratory experiments show that the laminar/turbulent transition occurs around $Re_{\rm crit}\sim 2000$--4000, where $Re_{\rm crit}$ is called the critical Reynolds number. Meanwhile, the critical magnetic Reynolds number, $Re_{\rm M,crit}$, depends on the magnetic Prandtl number $Pm = \nu/\eta = Re_{\rm M}/Re$. Numerical simulations of small-scale dynamo suggest that $Re_{\rm M,crit}$ is $\sim 50$--500 and rather constant for $Pm >1$ (the resistive scale lies in the viscous scale) \citep{2005ApJ...625L.115S}. As $Pm$ decreases below unity (the resistive scale lies in the kinetic inertial range), the range of laminar flows (i.e. $Re_{\rm M,crit}$) sharply increases. 

In small-scale dynamo, eddy cascading proceeds from larger to smaller eddies, and terminates at the smallest spatial scale at which kinetic energy is dissipated into thermal energy due to the viscosity and/or resistivity. Here, as the same analogy of the kinetic inertial range $k_\nu/k_{\rm f}\approx Re^{3/4}$, the magnetic inertial range depends on $k_\eta/k_{\rm f}\approx Re_M^{3/4}$. Magnetic fields at smallest scales first grows exponentially by stretching of field lines therein. Then, larger-scale magnetic fields are amplified like inverse cascade. The growth proceeds gradually and is called the linear growth. Finally, dynamo amplification is saturated when magnetic energy and kinetic energy are comparable to each other. The timescale and efficiency of dynamo amplification depend on various physical parameters such as not only $Re$, $Re_{\rm M}$, $Pm$ but also the plasma $\beta$ (the ratio between thermal pressure and magnetic tension) and the rms Mach number (the ratio between flow velocity and sound velocity).

As for large-scale dynamo, onset of large-scale dynamo occurs at scales with very large magnetic Reynolds number and is essentially independent on $Pm$. A well-known large-scale dynamo is the Parker dynamo mechanism in a galactic disk. In the mechanism, perturbed GMF inflates into the halo and form an $\Omega$ shape under the gravity vertical to the disk. By the buoyancy and expansion, as well as by the angular momentum conservation, it begins to rotate in the opposite direction of the galactic rotation, where the differential rotation of the disk and local epicyclic motion are the driving force, which stretches the field lines, and increase the net field strength. In this mechanism, the local field strength increases, but averaged field strength and configuration does not change. The convective instability plays a key role in evolution of large-scale dynamo (e.g., \cite{2008A&A...491..353K}). Also the magneto rotational instability (MRI) acts similarly to the Parker instability for amplifying the local field strengths (e.g., \cite{2013ApJ...764...81M}), while it does not change the global configuration.

%%%%%%%%%%%%%%%%%%%%%%%%%%%%%%%%%%%%%%%%%%%%
%%%%%%%%%%%%%%%%%%%%%%%%%%%%%%%%%%%%%%%%%%%%
\subsection{Turbulence and Magnetic Field}
\label{mag.s2.ss4}

%%%%%%%%%%%%%%%%%%%%%%%%%%%%%%%%%%%%%%%%%%%%
\subsubsection{MHD Turbulence}
\label{mag.s2.ss4.sss1}

The classical theory of hydrodynamic turbulence is expanded into the MHD. The induction equation (\ref{eq.mhd4}) introduces the magnetic diffusivity (resistivity), $\eta$, and the magnetic Prandlt number, $Pm=\nu/\eta$, which satisfy $\eta \ll \nu$ and $Pm \gg 1$ in the ideal MHD approximation. In hydrodynamic turbulence, kinetic energy is transferred due to eddy cascade. In MHD, on the hand, such energy transfer is reduced by the ratio of a parallel Alfv\'en transit time ($1/k_{||} v_{\rm A}$) for the Alfv\'en velocity,
\begin{equation}
v_{\rm A} = \frac{B_{||}}{\sqrt{4\pi \rho}}
\end{equation}
($\rho$ is the plasma mass density), to a perpendicular eddy shearing rate ($l_\perp/v_\perp$). This Alfv\'en effect is important in a magnetized system (e.g., \cite{2010rdla.book.....D}). For weakly magnetized isotropic cascade, the energy spectrum is rescaled as,
\begin{equation}
E(k_\perp)=C_{\rm IK} (\varepsilon v_{\rm A})^{1/2}k^{-3/2},
\end{equation}
where $C_{\rm IK}$ is the normalization constant \citep{1964SvA.....7..566I, 1965PhFl....8.1385K}. As for a strongly magnetized system in which there is a large-scale anisotropy, the relation is rescaled as,
\begin{equation}
E(k_\perp)=C_{\rm GS} (\varepsilon k_{||}v_{\rm A})^{1/2}k_\perp^{-2},
\end{equation}
where $C_{\rm GS}$ is the normalization constant \citep{1995ApJ...438..763G,1997ApJ...485..680G}. Transition between strongly and weakly magnetized systems appears when the parallel Alfv\'en wave transit time is equal to the perpendicular eddy turnover time (the critical balance). The energy spectrum under the critical balance can be written as
\begin{equation}
E(k_\perp)=C_{\rm GS} \varepsilon^{2/3}k_\perp^{-5/3}.
\end{equation}
Interestingly, the slope is identical to the Kolmogorov slope (e.g., \cite{2010rdla.book.....D}).

%%%%%%%%%%%%%%%%%%%%%%%%%%%%%%%%%%%%%%%%%%%%
\subsubsection{Structure Function and Intermittency}
\label{mag.s2.ss4.sss2}

The structure function (SF) is useful to quantify the amplitude of spatial structures at scale $r=|$\mbox{\boldmath $r$}$|$. For a physical quantity $A$(\mbox{\boldmath $x$}) at \mbox{\boldmath $x$}, the $n$-th order SF, $S_n(r)$, is defined as the $n$-th order statistical moment of the difference,
\begin{equation}
S_n(r) = \langle \delta A(r)^n \rangle,
\end{equation}
\begin{equation}
\delta A(r) = | A(\mbox{\boldmath $x$} + \mbox{\boldmath $r$}) - A(\mbox{\boldmath $x$}) |,
\end{equation}
where $\langle \rangle$ means the ensemble average. We often quantify the SF exponent, $\xi_n$, assuming $S_n(r) \propto r^{\xi_n}$.

For example, the velocity SF is given by
\begin{equation}
S_n(r) = \langle \delta v(r)^n \rangle = C_n \varepsilon^{n/3}r^{n/3},
\end{equation}
where $C_n$ is the normalization constant. The latter equation, $\xi_n = n/3$, appears in Kolmogorov turbulence as follows  (e.g., \cite{2010rdla.book.....D}). In a sequential cascading of eddies, $m$-th sequence takes place during the eddy turnover time, $t_m=l_m/v_m$, and its energy is transferred to the energies of eddies in the next sequence with $\varepsilon_m \sim E_m/t_m \sim (v_m)^3/l_m$. For high-$Re$ incompressible fluid, the momentum equation becomes 1st order differential equation and it can be parameterized such as $x\rightarrow \delta \hat{x}$, $t\rightarrow \delta^{1-\alpha/3} \hat{t}$, and $v\rightarrow \delta^{\alpha/3} \hat{v}$, where $\alpha$ is an arbitral scaling exponent. Thus, the velocity SF, $\langle \delta v(r)^n \rangle$, gives $\xi_n=n\alpha/3$. Since energy transfer from large to small scales is spatially isotropic and $\varepsilon$ is scale invariant, $ \varepsilon_m/\varepsilon_0 \sim (v_m^3/l_m)/(v_0^3/l_0)=\delta_m^\alpha/\delta_m=\delta_m^{\alpha-1}$ and $\alpha=1$. Therefore, $\xi_n=n/3$. 

In generic turbulence, energy-transfer is not spatially isotropic, and intermittent energy-transfer exists in space and time, i.e. $\xi_n \neq n/3$. Based on experiments and simulations, the phenomenological, She-Leveque relation is proposed:
\begin{equation}
\xi_n = \frac{n}{9} + C\left[ 1-\left(1-\frac{2/3}{C}\right)^{n/3} \right]
\end{equation}
\citep{1994PRL....72..336S, 2011RPPh...74d6901B}. Here, $C$ is interpreted as the co-dimension of the dissipative structures; $C\sim 2$ with 1D tube-like dissipative structures for weakly compressible or incompressible turbulence, and $C\sim 1$ with 2D sheet-like dissipative structures for compressible or highly supersonic turbulence. 

In MHD, the relation between the SF exponent and turbulence properties is still under discussion. For instance, the Iroshnikov-Kraichnan scaling gives the exponent of the longitudinal velocity SF as $\xi_n=n/4$. MHD turbulence simulations with the modest $Pm \sim 1$ indicated the energy spectrum close to the Iroshnikov-Kraichnan relation, but it showed $\xi_n=n/3$ \citep{2003ApJ...597L.141H}.

%%%%%%%%%%%%%%%%%%%%%%%%%%%%%%%%%%%%%%%%%%%%
%%%%%%%%%%%%%%%%%%%%%%%%%%%%%%%%%%%%%%%%%%%%
%%%%%%%%%%%%%%%%%%%%%%%%%%%%%%%%%%%%%%%%%%%%
\section{Method and Analysis of Measurements of Magnetic Fields}
\label{mag.s3}

%%%%%%%%%%%%%%%%%%%%%%%%%%%%%%%%%%%%%%%%%%%%
%%%%%%%%%%%%%%%%%%%%%%%%%%%%%%%%%%%%%%%%%%%%
\subsection{Measurement of Stokes Parameters}
\label{mag.s3.ss1}

If an electro-magnetic wave travels in the $z$-direction in the Cartesian three-dimensional coordinates, electric and magnetic fields oscillate in the $xy$-plane. Defining functions of the electric field as 
\begin{equation}
E_x(t)=E_{x0}(t)e^{i \{2\pi \nu t +\delta_x(t)\}},
\end{equation}
\begin{equation}
E_y(t)=E_{y0}(t)e^{i \{2\pi \nu t +\delta_y(t)\}}, 
\end{equation}
where the phase difference is $\delta_x(t)-\delta_y(t)=n\pi$ ($n$: integer) for a linearly polarized wave and $\pm \pi/2+2n\pi$ for a circularly polarized wave. 

Radio receivers are generally designed to detect either linear or circular polarization. Orthogonal dipoles detect horizontal and vertical components of linearly-polarized radio waves. The Stokes parameters are given by time-averaged auto-correlation and cross-correlation of $E_{x}(t)$ and $E_{y}(t)$ as expressed below using the following matrix:
\begin{equation}
\left( 
\begin{array}{c}
I\\Q\\U\\V
\end{array}
\right)=
\left( 
\begin{array}{cccc}
1 & 0 & 0 &  1\\
1 & 0 & 0 & -1\\
0 & 1 & 1 &  0\\
0 &-i & i &  0\\
\end{array}
\right)
\left( 
\begin{array}{c}
\langle E_xE_x^*\rangle\\
\langle E_xE_y^*\rangle\\
\langle E_yE_x^*\rangle\\
\langle E_yE_y^*\rangle\\
\end{array}
\right).
\end{equation} 

Here, $\langle E_i E_j^*\rangle$ represents the auto/cross correlation. 
Stokes parameter $I$ corresponds to the total intensity of radiation. Fractions of $\sqrt{Q^2+U^2}/I$ and $V/I$ indicate degrees of linear and circular polarizations, respectively. The polarization angle $\chi$ is given by $\chi=\arctan(U/Q)/2$. 

Similarly, right-handed and left-handed helical antennae detect right and left circularly-polarized radio waves. Defining the electric field of right and left circularly-polarized components as $E_r(t)$ and $E_l(t)$, Stokes parameters are given as follows:
\begin{equation}
\left( 
\begin{array}{c}
I\\Q\\U\\V
\end{array}
\right)=
\left( 
\begin{array}{cccc}
1 & 0 & 0 &  1\\
0 & 1 & 1 &  0\\
0 & i &-i &  0\\
-1 & 0 & 0 &  1\\
\end{array}
\right)
\left( 
\begin{array}{c}
\langle E_rE_r^*\rangle\\
\langle E_rE_l^*\rangle\\
\langle E_lE_r^*\rangle\\
\langle E_lE_l^*\rangle\\
\end{array}
\right).
\end{equation} 

These expressions show that either a diagonal linearly polarized feed or a diagonal helical feed can measure all Stokes parameters. Linearly polarized feed is widely used to realize broad-band observation \citep{2010A&A...509A..23D} and it is sensitive to the circular polarimetry, which is necessary to detect Zeeman effect \citep{2014JAI.....350010M}. If an antenna mount is alt-azimuth, circular polarization is advantageous because it is not necessary to rotate the feed even though the parallactic angle of the object changes during an observation \citep{1969MNRAS.142...11C}.

Above description on the derivation of polarimetry is an ideal case where no leakage exists between two polarized components. However, there is non-negligible leakage between two polarized components in an actual observation, which has to be eliminated. Measured signal voltages $v'_x$ and $v'_y$ are written with intrinsic values $v_x$ and $v_y$ as 
\begin{equation}
v'_x = v_x + D_x v_y, \\
v'_y = v_y + D_y v_x, 
\end{equation}
where subscriptions $x$ and $y$ denote horizontal (h) and vertical (v) polarizations, or right-hand (r) and left-hand (l) polarizations, respectively. The second terms including $D_x$ and $D_y$ are called $D$-term indicating leakage. These $D$-terms are calibrated by observing calibrators whose polarization is known well.

%%%%%%%%%%%%%%%%%%%%%%%%%%%%%%%%%%%%%%%%%%%%
%%%%%%%%%%%%%%%%%%%%%%%%%%%%%%%%%%%%%%%%%%%%
\subsection{Synchrotron Radiation and Faraday Rotation}
\label{mag.s3.ss2}

Synchrotron radiation and Faraday rotation are conventional tools of radio astronomy to study cosmic magnetic fields (for a text book, see \cite{1979rpa..book.....R}). Synchrotron radiation is emitted from relativistic charged particles gyrating around magnetic fields. Assuming isotropic distribution of relativistic electrons and their energy spectrum of the form,
\begin{equation}
N(\gamma)d\gamma = \mathcal{N}(r) \gamma^{-p} d\gamma,
\end{equation}
where $\gamma$ is the Lorentz factor, $\mathcal{N}(r)$ is the proportional constant at the position $r$, and $p$ is the spectral index, the synchrotron emissivity can be written as
\begin{equation}
\epsilon (r) \propto \mathcal{N}(r) B_\perp(r)^{(1+p)/2} \nu^{(1-p)/2},
\end{equation}
where $B_\perp$ is the strength of magnetic fields perpendicular to the line-of-sight (LOS) and $\nu$ is the frequency. The synchrotron intensity at frequency $\nu$ is often fitted with a power-law form:
\begin{equation}
I_\nu \propto \nu^{-\alpha}.
\end{equation}
The spectral index is thus related to the electron energy spectral index as $p=2\alpha+1$.

Assuming a reasonable distribution of $\mathcal{N}(r)$ and the size of emission region, we can estimate the value of $B_\perp$ and its orientation on the sky from the synchrotron intensity. The equipartition between total energy densities of CRs and that of the magnetic field ($\epsilon_{\rm B}=\epsilon_{\rm CR}$) is often used, and the magnetic-field strength is estimated as follows (e.g., \cite{2016arXiv161106647A}):
\begin{equation}
B_{\rm eq}~[{\rm G}]={4\pi (2\alpha+1)(K_0+1)I_\nu E_{\rm p}^{1-2\alpha} (\nu/2c_1)^\alpha \over (2\alpha -1)c_2 c_4}, 
\end{equation}
where
\begin{eqnarray}
c_1 &=& \frac{3e}{4\pi m_{\rm e}^3c^5}=6.26428\times 10^{18}~[{\rm erg}^{-2}~{\rm s}^{-1}~{\rm G}^{-1}], \\
c_2 &=& \frac{c_3}{4} {p+7/3 \over p +1} \Gamma \left[ \frac{3p-1}{12} \right] \Gamma \left[ \frac{3p+7}{12} \right], \\
c_3 &=& {\sqrt[]{3}e^3\over 4\pi m_{\rm e}c^2} = 1.86558\times 10^{-23} ~[{\rm erg}~{\rm G}^{-1}~{\rm sr}^{-1}], \\
c_4 &=& [\cos{i}]^{(p+1)/2}, 
\end{eqnarray}
$K_0$ the ratio of the number density of protons and electrons, $E_{\rm p}$ the proton rest energy (938.28 MeV), $l$ the LOS length, $i$ the inclination angle of the magnetic field with respect to the sky plane ($i=0$ in the case of face-on view), $e$ the elementary charge, $m_{\rm e}$ the electron mass, and $c$ is the velocity of light. 

When a linearly-polarized electromagnetic wave passes through magneto-ionic media, its polarization angle rotates as
\begin{equation}\label{eq:polangle}
\chi(\lambda^2) = \chi_0 + RM \lambda^2,
\end{equation}
where $\chi_0$ is the initial polarization angle and $\lambda$ is the wavelength. This phenomenon is called Faraday rotation and the coefficient $RM$ is the rotation measure, 
\begin{equation}
RM~{\rm (rad~m^{-2})} \approx 811.9 \int 
\left(\frac{n_{\rm e}}{\rm cm^{-3}}\right)
\left(\frac{B_{||}}{\rm \mu G}\right)
\left(\frac{dr}{\rm kpc}\right),
\end{equation}
where $n_{\rm e}$ is the density of free electrons, $B_{||}$ is the strength of magnetic fields parallel to the LOS, and $RM$ is defined to be positive if the magnetic field points toward the observer. Therefore, once we observe a polarized source and obtain the RM, we can estimate the integral of $n_{\rm e} B_{||}$. RM can be obtained if polarization angles are measured at more than one wavelength. 

In summary, we can guess the magnetic field parallel to the LOS from the RM. Considering inverse Faraday rotation with the RM, the intercept of $\chi$-$\lambda^2$ plots gives the intrinsic polarization angle. We can guess the magnetic field perpendicular to the LOS from the synchrotron intensity and the intrinsic polarization angle. Synchrotron radiation is hence an important observable to construct the three-dimensional magnetic-field model. However, only an average over the emission region is obtained and the spatial distribution along the LOS is difficult to evaluate.

%%%%%%%%%%%%%%%%%%%%%%%%%%%%%%%%%%%%%%%%%%%%
%%%%%%%%%%%%%%%%%%%%%%%%%%%%%%%%%%%%%%%%%%%%
\subsection{Depolarization}
\label{mag.s3.ss3}

Depolarization is a phenomenon in which we observe a weaker polarization degree than that at the origin. If polarizations with different polarization angles are observed simultaneously, depolarization takes place. Such a situation appears in several ways and is categorized as follows (see e.g., \cite{1998MNRAS.299..189S} for more details):
\begin{itemize}
\item {\bf Wavelength independent depolarization:} Consider an observation of a magnetized medium emitting polarization, depolarization can take place if the magnetic fields perpendicular to the LOS are not aligned with each other in the medium. This depolarization does not depend on the wavelength of polarization.
\item {\bf Differential Faraday rotation depolarization:} In the above situation, even if the magnetic fields perpendicular to the LOS are aligned with each other, depolarization can take place when non-zero RM, i.e. the net magnetic fields parallel to the LOS, is present along the LOS in the medium. In such a case, polarizations emitted at different depths experience different Faraday rotation, so that they can cancel out with each other.
\item {\bf Beam depolarization:} If RM is not uniformly distributed in the medium and/or in front of the medium, depolarization can take place because polarizations inside an observing beam experience different Faraday rotation. This depends on the observing beam size.
\item {\bf Bandwidth depolarization:} The degree of Faraday rotation depends on the wavelength. Therefore, depolarization can take place when one integrates broadband polarizations.
\end{itemize}

\citet{1966MNRAS.133...67B} formulated depolarization in several simple situations. For example, in the case of a matter with uniform electron density and uniform magnetic fields, the polarization degree $\Pi$ of the matter for a specific LOS can be written as 
\begin{equation}
\Pi(\lambda^2) = \Pi_{0}(\lambda^2)
\frac{\sin \left( RM \lambda^2 \right) }{RM \lambda^2}
\exp{2i \left( \chi_0 + \frac{1}{2}RM \lambda^2 \right) },
\end{equation}
where $\Pi_{0}$ is the intrinsic polarization degree. This case is differential Faraday rotation depolarization.

Another example is beam depolarization. We suppose that RM distribution inside a beam area follows the Gaussian with the standard deviation of $\sigma_{\rm RM}$. In the case where this RM source itself is an emitter of polarization, $\Pi$ can be written as 
\begin{equation}
\Pi(\lambda^2) = \Pi_{0}(\lambda^2)
\frac{1-\exp \left(-2\sigma_{\rm RM}^2 \lambda^4\right)}{2\sigma_{\rm RM}^2 \lambda^4},
\end{equation}
and is called internal Faraday dispersion depolarization. Otherwise, in the case where this RM source is a foreground of a polarized source, $\Pi$ can be written as 
\begin{equation}\label{eq:EFD}
\Pi(\lambda^2) = \Pi_{0}(\lambda^2)\exp \left(-2\sigma_{\rm RM}^2 \lambda^4\right),
\end{equation}
and is called external Faraday dispersion depolarization. 

\begin{figure}[tbp]
\begin{center}
\FigureFile(70mm,70mm){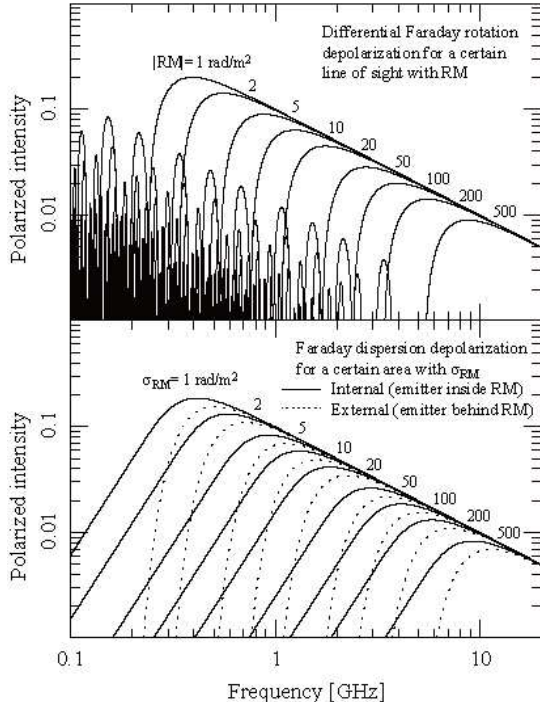}
\end{center}
\caption{
Depolarization of differential Faraday rotation (top) and dispersion (bottom). See \citet{2011MNRAS.418.2336A} for the original figures and details.
}
\label{f02}
\end{figure}

Figure~\ref{f02} shows the Burn law \citep{2011MNRAS.418.2336A}, which indicates that the degree of depolarization depends on the frequency. Wide frequency coverage is hence essential to capture the feature of depolarization. The fact that depolarization is weaker at higher frequencies implies that the observed polarization degree can be larger for higher redshift sources, because depolarization happens at their rest-frame (higher) frequencies.

Depolarization is thought to have the capability to inform three-dimensional magnetic-field structures both along the LOS and within an observing beam. For example, depolarization depends on the amount of magnetic helicity, which has been claimed to be a tool for determining magnetic helicity in galaxies (e.g., \cite{2011A&A...530A..89O, 2014ApJ...786...91B}). However, it is hard to recognize depolarization without wideband polarimetric data, which are not easy to obtain. Therefore, application of depolarization diagnostics to real observations has been limited. Future radio telescopes should facilitate wideband polarimetric observations and break through the situation in order to advance the study of cosmic magnetism.

%%%%%%%%%%%%%%%%%%%%%%%%%%%%%%%%%%%%%%%%%%%%
%%%%%%%%%%%%%%%%%%%%%%%%%%%%%%%%%%%%%%%%%%%%
\subsection{Faraday Tomography}
\label{mag.s3.ss4}

Faraday Tomography is a decomposition technique proposed by \citet{1966MNRAS.133...67B}. An observed polarized intensity is an integration of the synchrotron emissivity $\epsilon (r)$ along the LOS, and it can be decomposed as 
\begin{equation}
P(\lambda^2)=\int^\infty_{0} \varepsilon(r)e^{2i\chi(r,\lambda^2)}dr
=\int^\infty_{-\infty}F(\phi)e^{2i\phi\lambda^2}d\phi,
\end{equation}
where
\begin{equation}
\chi(r,\lambda^2) = \chi_0(r) + \phi(r)\lambda^2,
\end{equation}
is the polarization angle at an observer, $\chi_0(r)$ is the intrinsic polarization angle at $r$, and $\phi(r)$ is the Faraday depth in rad~m$^{-2}$. $F(\phi) \equiv \varepsilon(\phi)e^{2i\chi_0(\phi)} dr/d\phi$ is the Faraday dispersion function (FDF) or the Faraday spectrum, which represents the sum of the emissivity through the regions with a specific value of $\phi$. Note that we changed the integration variable from $r$ to $\phi(r)$, resulting in the form of Fourier transformation with conjugate variables $\phi$ and $\lambda^2$.

In theory, $F(\phi)$ can be precisely derived from $P(\lambda^2)$ by the inverse Fourier transformation. This is, however, not the case because the observable $P(\lambda^2)$ is available only for a limited range of $\lambda^2$. Using a window function $W(\lambda^2)$ for observable wavelengths, the reconstructed FDF, $\tilde F(\phi)$, can be written as
\begin{equation}
\tilde F(\phi)=\frac{1}{\pi}\int_{-\infty}^\infty W(\lambda^2)P(\lambda^2)e^{-2i\phi\lambda^2}d\lambda^2.
\end{equation}
Using the convolution theorem, this equation becomes
\begin{eqnarray}
&& \tilde F(\phi) = K^{-1} R(\phi) \ast F(\phi), \\
&& R(\phi) = K \int^\infty_{-\infty} W(\lambda^2) e^{-2i\phi\lambda^2} d\lambda^2, \\
&& K = \left[ \int^\infty_{-\infty} W(\lambda^2) d\lambda^2 \right]^{-1},
\end{eqnarray}
where $R(\phi)$ is called the rotation measure spread function (RMSF). 

\begin{figure}[tbp]
\begin{center}
\FigureFile(80mm,80mm){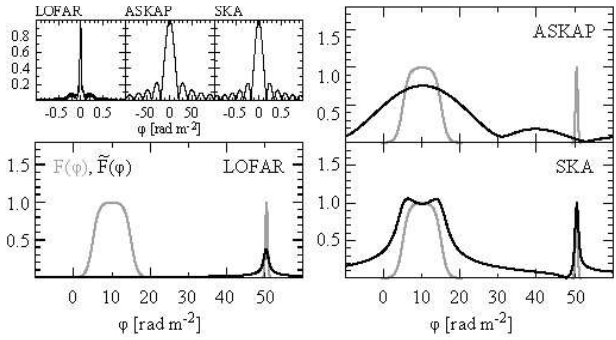}
\end{center}
\caption{
(Top left) The rotation measure spread function. (The others) The model (gray) and reconstructed (black) Faraday spectra \citep{2014PASJ...66...65A}. LOFAR and ASKAP frequency coverages are sensitive to only thin (compact source) and thick (diffuse source) Faraday structure, respectively, while SKA's seamless, broad bandwidth can provide sensitivities for RM structures of $O(1-1000)$~rad~m$^{-2}$. 
}
\label{f03}
\end{figure}

If the window function is unity for all the value of $\lambda^2$, $R(\phi)/K$ is reduced to the delta function $\delta(\phi)$ and the reconstruction is perfect. But, in reality, $P(\lambda^2)$ is unphysical for negative values of $\lambda^2$, and coverage of positive $\lambda^2$ is imperfect in observation. Thus, the RMSF has a finite width depending on wavelength coverage, as shown in figure~\ref{f03}.

The full width at half maximum (FWHM) of the RMSF corresponds to the resolution in the Faraday depth space. Considering the case where $W(\lambda^2)=1$ for $\lambda^2_{\rm min} \leq \lambda^2 \leq \lambda^2_{\rm max}$ and otherwise $W(\lambda^2) = 0$, the FWHM is given by,
\begin{equation}
{\rm FWHM~(rad~m^{-2})} = \frac{2\sqrt3}{\Delta \lambda^2({\rm m^2})}.
\end{equation}
This indicates that the resolution is determined by the $\lambda^2$-space coverage, $\Delta \lambda^2 = \lambda^2_{\rm max} - \lambda^2_{\rm min}$. In particular, going up to a longer wavelength expands the $\lambda^2$ coverage effectively, though it tends to suffer from depolarization more seriously. Differential Faraday rotation depolarization can be significant when the polarization angle rotates by $\pi$ within a source extended in Faraday depth space. Hence the maximum observable width in the Faraday depth space is,
\begin{equation}
L_{\rm \phi, max}~{\rm~(rad~m^{-2})} \approx \frac{\pi}{\lambda^2_{\rm min}({\rm m^2})}
\end{equation}
In order to make Faraday tomography feasible, the above two frequency conditions are at least needed to be improved.

The simple inverse transformation mentioned above cannot perfectly reconstruct the true FDF due to the presence of the side lobe in the RMSF. An effective method to reduce the side lobe, called RM CLEAN, was proposed by \citet{2009A&A...503..409H}. RM CLEAN was shown to work, if multiple polarized sources are separated by more than the FWHM of RMSF each other in Faraday depth space \citep{2014PASJ...66...61K, 2015AJ...149...60S, 2016PASJ...68...44M}. 

Another technique to estimate the true FDF is QU-fitting. We compare a model of polarized intensity with observed one, where we can avoid to perform incomplete inverse transformation. We explore the best-fit model parameters using, for example, the Markov Chain Monte Carlo method. This exploration can be extended to various models, and we can argue the best model among them using, for example, Akaike's Information Criterion (AIC) or Bayesian Information Criterion (BIC):
\begin{eqnarray}
{\rm AIC} &=& -2\log L + 2k, \\
{\rm BIC} &=& -2\log L + k \log n,
\end{eqnarray}
where $L$ is the biggest likelihood, $k$ is the number of parameters, and $n$ is the number of data. These criteria evaluate the adaptability of a model to the data and the simplicity of the model in well balance. The above QU-fitting approach showed the results better than RM CLEAN in a recent benchmark challenge (see \cite{2015AJ...149...60S} for details).

Faraday tomography is a powerful tool to study the Faraday structure along a LOS, and is superseding classical RM study. Indeed, multiple polarized sources and their Faraday depths were successfully resolved, even though they were not spatially resolved in radio/polarization images \citep{2012MNRAS.421.3300O, 2015PASJ...67..110O}. The FDF intrinsically contain rich information about distributions of magnetic fields, thermal electrons, and CR electrons along the LOS in the sense that the FDF indicates the synchrotron polarization as a function of Faraday depth. Thus, if we can understand how to extract such information from the FDF, the Faraday tomography technique maximizes its potential.

The interpretation of the FDF is, however, not straightforward, mainly because there is no one-to-one correspondence between the Faraday depth and the physical distance. Hence the structure in physical space is not directly obtained from the FDF. Particularly, the presence of turbulent magnetic fields makes the interpretation difficult. There are some attempts to consider the effects of turbulence on the FDF of galaxies using simple models (e.g., \cite{2011A&A...535A..85B,2011MNRAS.414.2540F,2012A&A...543A.113B}). It is found that the effects of turbulence appears as many small components in FDF, called ``Faraday forest''. 

\begin{figure}[tbp]
\begin{center}
\FigureFile(75mm,75mm){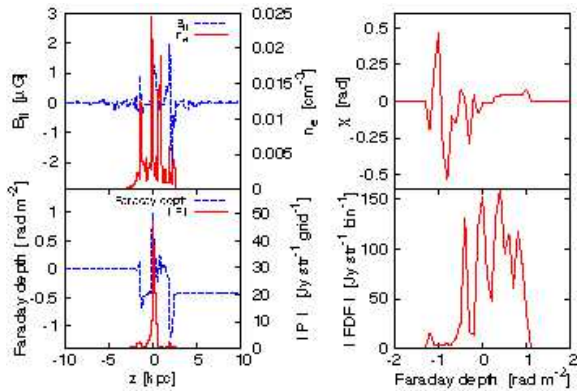}
\end{center}
\caption{
The Faraday spectrum of a simulated face-on spiral galaxy \citep{2014ApJ...792...51I}. (Top left) distributions of the LOS component of magnetic fields and the thermal electrons, (bottom left) distributions of the polarized intensity and Faraday depth, (top right) the polarization angle, (bottom right) the absolute Faraday spectrum.
}
\label{04}
\end{figure}

\citet{2014ApJ...792...51I} studied the realistic form of the galactic FDF using the Galactic model produced by \citet{2013ApJ...767..150A}. They found very complicated FDFs and Faraday forests, which cannot be approximated by Gaussian or other simple analytic functions (figure~\ref{04}). The complexity is mainly due to the stochasticity of turbulence in magnetic fields and gas density. \citet{Ideguchi16} studied the FDF of face-on galaxies using simpler model than former works. The turbulent magnetic field was expressed as a random field with single coherence length. They showed that the stochasticity can be significantly reduced if we consider a beam size about ten times larger than the coherence length squared, and that the global properties of galaxies such as coherent magnetic fields and characteristic scale of turbulence could be extracted from the Faraday spectrum. These studies are also important for model functions of the QU-fitting and base functions of the compressed sensing.

%%%%%%%%%%%%%%%%%%%%%%%%%%%%%%%%%%%%%%%%%%%%
%%%%%%%%%%%%%%%%%%%%%%%%%%%%%%%%%%%%%%%%%%%%
%%%%%%%%%%%%%%%%%%%%%%%%%%%%%%%%%%%%%%%%%%%%
\section{Interstellar Medium}
\label{mag.s4}

Magnetic fields play an important role in ISM's kinematics and energetics. While magnetic fields assist contraction of molecular clouds through transporting the angular momentum outward, they prevent the contraction against the self-gravity. Magnetic fields affect the evolution of HII region and are essential for radio emission from supernova remnants (SNRs) and pulsar wind nebulae (PWN), both known as sources of Galactic CRs. It is considered that magnetic fields relate directly to filamentary and loop structures in the ISM. In this section, we review magnetic fields in various discrete objects in the ISM.

%%%%%%%%%%%%%%%%%%%%%%%%%%%%%%%%%%%%%%%%%%%%
%%%%%%%%%%%%%%%%%%%%%%%%%%%%%%%%%%%%%%%%%%%%
\subsection{Thermal/Pressure Equilibrium}
\label{mag.s4.ss1}

The ISM is composed of gases, dusts, CRs, magnetic fields, and radiation fields. The steady-state gaseous ISM is in pressure equilibrium among various species with multiple temperatures, e.g., warm and cold neutral materials \citep{1969ApJ...155L.149F}. Ionized gas is in pressure equilibrium with the UV radiation field from OB stars, while low-temperature gas containing dust is balanced in pressure with the starlight radiation field. The kinetic energy of the turbulent ISM is converted to magnetic energy through the dynamo action, reaching a pressure equilibrium between the gas and the magnetic field. The magnetic pressure is further in equilibrium with the pressure of interstellar CRs, which are accelerated and supplied by shock-compressed SNR shells and pulsar magnetospheres.

The condition for the stationary ISM is, therefore, ascribed to the equilibrium among the energy densities, which are equivalent to the pressures, of gases in various phases, magnetic fields, CRs, and the starlight UV radiation field:
\begin{equation}
u_{\rm mag}\sim u_{\rm CR}\sim u_{\rm gas}\sim u_{\rm HII}\sim u_{\rm HI} \sim u_{\rm MC} \sim u_{\rm UV}
\end{equation}
where 
$u_{\rm mag},\ u_{\rm CR},\ u_{\rm gas},\ u_{\rm HII},\ u_{\rm HI}, u_{\rm MC}$ and $u_{\rm UV}$ are the energy densities of magnetic field, CR, gas (which are either in ionized gas, HI, or molecular gas clouds), and UV radiation. In typical interstellar conditions, except for giant molecular clouds and dense molecular cores, the gravitational energy is neglected as a reasonable approximation.

If one of the equilibrium conditions is broken, the ISM becomes unstable, resulting in local expansion, contraction, and/or ejection. The region containing such unstable ISM is regarded as an active region. Generally, astrophysical activity is defined as the state that the local condition is significantly displaced from the thermal/dynamical equilibrium, as often recognized in expanding HII regions, SNRs, jets and/or various types of instabilities including the Parker magnetic inflation.

%%%%%%%%%%%%%%%%%%%%%%%%%%%%%%%%%%%%%%%%%%%%
%%%%%%%%%%%%%%%%%%%%%%%%%%%%%%%%%%%%%%%%%%%%
\subsection{Local magnetic field}
\label{mag.s4.ss2}

The local interstellar magnetic field is simply estimated from the synchrotron radio emissivity toward the Galactic poles, assuming that the magnetic and CR energy densities corresponding to the observed frequency are in equilibrium. Taking a LOS depth of $\sim 200$ pc in the poles, where the brightness temperature after subtracting the 2.7 K due to the CMB is measured to be $\sim 10$ K at 1 GHz, the field strength is estimated to be $\sim 3\ \mu$G. This leads to $u_{\rm mag}\sim 10^{-12}$ ergs $\sim 1$ eV cm$^{-3}$ \citep{sofue2017.springer.6}.

A slightly stronger magnetic field, several $\mu$G, has been found in the local ISM at $\sim 100-200$ pc from the Sun along the HI arch in the Aquila Rift, where a correlation analysis between the HI column density and Faraday RM of extragalactic radio sources were obtained to yield the LOS magnetic strength through HI filaments \citep{2017MNRAS.464..783S}.

More direct measurements of the local magnetic field have been obtained by linear polarization observations of 0.4 to 1.4 GHz all-sky continuum surveys (\cite{reich2007.springer.63} for a review). They obtained the local magnetic field of a few $\mu$G from an analysis of the polarization intensities, degrees, and depolarization as well as the Faraday screen effects at different frequencies.

The current estimations of the local magnetic field yielded a strength of a few $\mu$G, corresponding to an energy density of $\sim 1$ eV cm$^{-3}$. These values are comparable to the energy densities of the ISM of $u_{\rm gas}\sim 1$ eV cm$^{-3}$ with $n_{\rm gas} \sim 1$ cm$^{-3}$ and turbulent velocity $\sim 5$ km s$^{-1}$ observed in the solar vicinity, indicating that the local magnetic field and gas are in pressure equilibrium.

%%%%%%%%%%%%%%%%%%%%%%%%%%%%%%%%%%%%%%%%%%%%
%%%%%%%%%%%%%%%%%%%%%%%%%%%%%%%%%%%%%%%%%%%%
\subsection{Molecular Clouds}
\label{mag.s4.ss3}

Stars form in molecular clouds in which large-scale magnetic fields are observed. The direction of the magnetic fields in the clouds has been studied with linear polarization of thermal millimeter/sub-millimeter dust emission and extinction of light from background stars. The polarized thermal emission probes the high-density region of the molecular cloud (e.g., \cite{1998ApJ...502L..75R}, \cite{2006Sci...313..812G}), and the optical polarized light from background stars is sensitive to the magnetic fields in the ISM (e.g., \cite{1990ApJ...359..363G}, \cite{2011ApJ...741...21C}). Recent observations show the strong correlation between the directions of magnetic fields of the interstellar gas and those for the cores in the molecular clouds \citep{2009ApJ...704..891L}. Since the weak magnetic fields are easily distorted by turbulence in the clouds, this result strongly suggests that the magnetic energies are larger enough than the random kinetic energies in molecular clouds.

\begin{figure}[tbp]
\begin{center}
\FigureFile(80mm,80mm){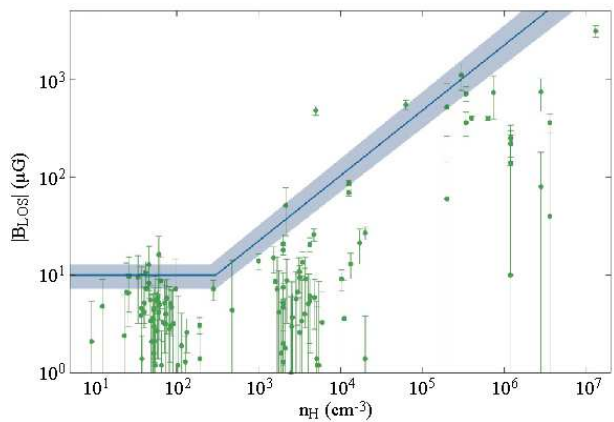}
\end{center}
\caption{
Correlation between the hydrogen density and the LOS magnetic field strength (by the Zeeman effect) in molecular clouds (taken from \cite{2012ARA&A..50...29C}).
}
\label{f05}
\end{figure}

The magnetic-field strength in molecular clouds is often estimated with the Chandrasekhar-Fermi method \citep{1953ApJ...118..113C} assuming that the velocity fluctuation of the clouds is related to the Alfv\'en waves that were assumed by the fluctuation of the magnetic-field lines observed from polarized light. This method estimates the strength of the magnetic fields along the plane of sky. The method tends to overestimate the field strength; some molecular clouds show that their magnetic energies are larger than the gravitational energies \citep{2008A&A...486L..13A, 2011ApJ...741...21C, 2015ApJ...807....5F}. 

The Zeeman effect is used to measure the strength of magnetic fields along the LOS (\cite{2012ARA&A..50...29C}; figure~\ref{f05}). According to the observations of the OH and CN Zeeman effect, the magnetic energies in the molecular clouds or cores are estimated to be slightly smaller than the gravitational energies of the clouds or cores on average (e.g., \cite{1999ApJ...520..706C}, \cite{2008ApJ...680..457T}). However, \citet{Nakamura2017} found a molecular cloud core whose magnetic energy is significantly larger than the gravitational energy by using the CSS Zeeman effect.

The magnetic-field strength in molecular clouds is also estimated from the difference of line widths of neutral and ionized molecular lines \citep{2008ApJ...677.1151L}, assuming that the ions have a steeper turbulent energy spectrum than that of the neutrals at ambipolar diffusion scale.  The method of the line width can measure the strength of the magnetic-field component on the plane of the sky. The magnetic-field strengths obtained by the line width are consistent with those estimated from the Zeeman effect (e.g., \cite{2010ApJ...720..603H}).

%%%%%%%%%%%%%%%%%%%%%%%%%%%%%%%%%%%%%%%%%%%%
%%%%%%%%%%%%%%%%%%%%%%%%%%%%%%%%%%%%%%%%%%%%
\subsection{Star Formation and Magnetic Field}
\label{mag.s4.ss4}

In molecular clouds, the energy equipartition among turbulence, magnetic field, and self-gravity is roughly satisfied. Once the self-gravity dominates over the others, stars are formed. If the magnetic force is strong enough to prevent the gravitational collapse in the clouds, stars are not formed. In this case, one of the important physical processes is the ambipolar diffusion (e.g., \cite{1978PASJ...30..671N, 1994ApJ...432..720B, 1999ASIC..540..305M}). Since the magnetic field is frozen only to the ionized gas, neutral gas can pass through the magnetic field and contract by gravity. When the contraction makes gravity strong enough to collapse, stars begin to be formed (e.g., \cite{2004ApJ...607L..39B, 2007MNRAS.380..499K}).

Since the ambipolar-diffusion time is normally larger than the free-fall time, stars are formed slowly in the molecular cloud if magnetic force is dominated, even though the ambipolar diffusion is enhanced by small-scale turbulent or large-scale flows in the molecular clouds (e.g., \cite{2002ApJ...570..210F, 2004ApJ...609L..83L, 2011ApJ...728..123K}). The suppression of the star formation by the magnetic field may explain the low star-formation rates and efficiencies in the clouds, although there are no direct evidence that the magnetic force dominates the gravity in the molecular clouds. 

Once the collapse happens for the star formation, the magnetic field plays an important rule for removing angular momentum of the contracting gas (e.g. \cite{1956MNRAS.116..503M, 1980ApJ...237..877M, 1989MNRAS.241..495N}) and forming outflows (e.g. \cite{1985PASJ...37..515U, 1986ApJ...301..571P, 1998ApJ...508..186K, 2002ApJ...575..306T, 2008ApJ...676.1088M}). The outflows from young stars can be the origin of turbulence in molecular clouds \citep{2007ApJ...662..395N}.

%%%%%%%%%%%%%%%%%%%%%%%%%%%%%%%%%%%%%%%%%%%%
%%%%%%%%%%%%%%%%%%%%%%%%%%%%%%%%%%%%%%%%%%%%
\subsection{HII Region}
\label{mag.s4.ss5}

\citet{1980ApJL...235L..105H} first estimated the magnetic-field strength in HII regions from the Very Large Array (VLA) measurement of the Faraday rotation of extragalactic radio sources. Typical value of magnetic-field strength in HII regions is between several $\mu$G to 12 $\mu$G (e.g., \cite{2007A&A...463..993S, 2010A&A...515A..64G}). \citet{1981ApJL...247L..77H} measured magnetic-field strengths in the HII regions, S117, S119, and S264 to be 1 $\mu$G to 50 $\mu$G. 

\citet{1999ApJ...514..221G} achieved polarimetric imaging around the W3/W4/W5/HB3 in the Perseus Arm in order to study the ISM in the Milky Way. The images were obtained at 1420~MHz with an angular resolution of $1'$ over more than 40~deg$^2$ with the Dominion Radio Astronomy Observatory Synthesis Telescope. They identified (i) mottled polarization arising from random fluctuations in a magneto-ionic screen in the vicinity of the HII regions themselves, and (ii) depolarization arising from very high RMs and RM gradients. 

\citet{2011ApJ...736..83H} studied the LOS magnetic field in five large-diameter Galactic HII regions. Using the Faraday rotation with background polarized radio sources, they estimated the field strengths in the regions, 2~$\mu$G to 6~$\mu$G, which is similar to the values in the diffuse ISM. Using the same method, \citet{2012MNRAS.420..279R} estimated the LOS-averaged magnetic field of 36~$\mu$G in an HII region NGC6334A in the Milky Way. This value is consistent with the former trial estimation of this source 40~$\mu$G \citep{1989MNRAS.242..209K}.

Interaction process between HII regions and adjacent molecular clouds in strong magnetic fields have been studied by many researchers (e.g. \cite{1986PhDT.........3Y}). Using IRAM 30m telescope, \citet{1991A&A...242..376S} observed the interacting region between HII region G0.18-0.04 and an associated molecular cloud. They showed that the magnetic field seems to work as a braking force on the cloud in the interaction. \citet{2007ApJ...671..518K} developed a three-dimensional MHD code for simulating the expansion of an HII region into a magnetized gas. They showed that the magnetic fields distort the HII region and reduce the strength of the shock. \citet{2012ApJ...745..158G} extended the simulation to a blister-type HII region driven by stars on the edge of magnetized gas clouds. They found that magnetized blister HII regions can inject the energy into clouds. \citet{2007ApJ...658.1119P} found that the magnetic field in M17 is strong enough to halt the expansion of the HII region. However, observational results of the magnetic fields in blister HII regions are still very limited.

%%%%%%%%%%%%%%%%%%%%%%%%%%%%%%%%%%%%%%%%%%%%
%%%%%%%%%%%%%%%%%%%%%%%%%%%%%%%%%%%%%%%%%%%%
\subsection{Supernova Remnant}
\label{mag.s4.ss6}

The ejecta of a supernova interacts with a high-density circum-stellar medium and create shocks observed as a SNR. The shock waves can convert a SNR's kinetic energy of $\sim 10^{51}$~erg into not only the thermal energy of ionized plasma around the SNR but also various non-thermal energies and radiation. SNR shocks are thought to be the most plausible site of Galactic CR acceleration \citep{2013A&ARv..21...70B}, and also excite Galactic turbulence. Magnetic fields play crucial roles in all these phenomena.

SNRs are found in the Milky Way and Local Group galaxies with angular sizes from minutes to degrees and ages from $10^2$ to $10^5$ yr. \citet{2014BASI...42...47G} listed 294 SNRs\footnote{\url{http://www.mrao.cam.ac.uk/surveys/snrs}} in the Milky Way, based on bright and large characteristic shell-like structures in radio survey images. The radio spectral index $\alpha$ (section~\ref{mag.s3.ss2}) is in the range of $\alpha \sim 0.3$--0.8 in the Green catalog. The mean is close to $\alpha \sim 0.5$ or the CR electron energy spectral index $p = 2\alpha+1 \sim 2$, which can be broadly explained as a test-particle, strong-shock case of the diffusive shock acceleration (DSA), $p = (r+2)/(r-1) \sim 2$, where $r$ is the shock compression ratio (see e.g. \cite{1987PhR...154....1B} for a review). There is also evidence of electron acceleration based on the detection of synchrotron X-rays from shell of young SNRs \citep{1995Natur.378..255K, 2003ApJ...589..827B}.

In general, $O(100)$~$\mu$G magnetic fields have been measured from radio total and polarized intensities. In some bright SNRs, $\sim $~mG has been observed, indicating amplified magnetic fields (see e.g. \cite{2012SSRv..166..231R, 2015aska.confE..46G, 2015aska.confE.096G}). Observations of non-thermal X-ray emissions in young SNRs provide evidence of magnetic-field amplification in the shock regions (e.g., \cite{2004MNRAS.353..550B, 2007Natur.449..576U, 2012ApJ...744...71I, 2014ApJ...790...85R}). Radio observation can also provide information of shock microphysics. \citet{2016MNRAS.462L..31B} recently found that the product of the energy fractions of non-thermal electrons ($\varepsilon_{\rm e}$) and magnetic fields ($\varepsilon_{\rm B}$) is around $\varepsilon_{\rm e}\varepsilon_{\rm B} \sim 0.001$ of the total shocked fluid energy for radio SNRs in the Magellanic Clouds.

Structure of magnetic fields in SNRs have been studied in the literature (e.g., \cite{2012SSRv..166..231R, 2015aska.confE..46G, 2015aska.confE.096G}). \citet{2012SSRv..166..231R} suggested that radial and tangential magnetic fields are generally predominant in young and old SNRs, respectively. Moreover, toroidal magnetic fields exist in some SNRs and they are interpreted as the fields associated with the shock-swept, past stellar wind of the progenitor \citep{2002ApJ...565.1022U}. It is, on the other hand, not clear how global Galactic magnetic-fields and circum-stellar magnetic-fields affect SNR's magnetic-fields (e.g., \cite{2015ApJ...804...22P, 2015ApJ...811...40S, 2016A&A...587A.148W}) and how turbulence alters properties of them \citep{2016MNRAS.459..178B}. These questions will be addressed with more spatially-resolved samples. Moreover, understanding foreground/background structures are essential to resolve structure of the targets. Broadband polarimetry will be a key strategy to distinguish them.

Finally, supernovae themselves are also interesting targets for radio observation. The lack of radio emission from Type Ia supernovae has been argued; no-radio detection in a near ($\sim 6.4$~Mpc) extragalactic Type Ia SN 2011fe ruled out a symbiotic progenitor system and a system with a high accretion rate onto a white dwarf \citep{2012ApJ...761..173C}. Another important topic is dust obscuration; \citet{2011ApJ...738..154H} found that the measured core-collapse supernova (CCSN) rate is a factor of 2 smaller than that predicted from the massive-star formation rate. While optical observations suffer from dust obscuration, radio observations may have advantage to find a missing CCSN hidden in the inner region of the host galaxy and solve the problem.

%%%%%%%%%%%%%%%%%%%%%%%%%%%%%%%%%%%%%%%%%%%%
%%%%%%%%%%%%%%%%%%%%%%%%%%%%%%%%%%%%%%%%%%%%
\subsection{Pulsar Wind Nebulae}
\label{mag.s4.ss7}

A pulsar generates a relativistic, magnetized outflow called a pulsar wind. It seems a transition from a cold, magnetic-energy-dominated flow into a hot, particle-energy-dominated flow around a strong termination shock. Relativistic particles in the downstream of the shock emit non-thermal radiations, which are identified as pulsar wind nebulae (PWN, see e.g. \cite{2012SSRv..166..231R, 2015aska.confE..46G}). Magnetic reconnection is thought to play a key role in accelerating particles in PWN. There are models of magnetic reconnection before/at/after the termination shock. 

Radio PWN are often seen around young ($10^4$--$10^5$ yr) pulsars. Pulsars can be kicked by a supernova explosion and hence often have high velocities of up to and beyond 1000~${\rm km~s^{-1}}$, but some young pulsars can be still inside the SNR created by the progenitor explosion. In very young ($<10^4$ yr) SNR, there are two streams inside PWN. The first zone is between the pulsar and the wind termination shock, where the wind energy is radiation-dominated. The second zone is beyond the termination shock up to the outer bow shock, where the wind energy is particle-dominated. Radio synchrotron emission is bright at the second zone in general. PWN have radio spectral indices in the range $\alpha \sim 0.0$--0.3, which is shallower than those of SNRs and is too flat to be explained by simple models of diffusive shock acceleration (e.g., \cite{2010ApJ...715.1248T}). 

Observations of inverse Compton emission in gamma-rays suggest a wide range of nebular magnetic fields, from $\sim 5$~$\mu{\rm G}$ to $> 1$~mG \citep{2009ASSL..357..451D, 2012SSRv..166..231R}). Orientation of magnetic fields in PWN can be studied with polarization; some PWN show a broadly toroidal magnetic field \citep{2006A&A...457.1081K} and some others show complex or tangled appearance \citep{1976A&A....53...89W}. \citet{2006A&A...457.1081K} suggested that these variations result from differences in viewing angle with respect to pulsar's spin axis. 

The number of detected PWNs is small compared to SNRs. A lack of samples remains some outstanding problems of PWN open. The PWN magnetic field configuration will be best studied through depolarization and Faraday tomography with wideband data, and by dense RMs of background sources.

A binary pulsar provides an opportunity to observe an inter-binary plasma very close to the neutron star magnetosphere. For instance, \citet{2011ApJ...742...97B} observed a strongly-magnetized wind compared to that seen further from the neutron star, and constrained possible models for magnetic reconnection in the pulsar wind. A binary pulsar also gives us an opportunity to observe eclipses and such eclipses have been observed in $\sim 50$ binaries pulsars (see \cite{2015aska.confE..46G}, references therein). We can measure changes in the dispersion measure during the eclipse of a binary pulsar \citep{2015aska.confE..46G}, and makes it possible to study the density and magnetic field structure of the intervening plasma.

%%%%%%%%%%%%%%%%%%%%%%%%%%%%%%%%%%%%%%%%%%%%
%%%%%%%%%%%%%%%%%%%%%%%%%%%%%%%%%%%%%%%%%%%%
\subsection{Loop Structure}
\label{mag.s4.ss8}

Galactic HI survey data exhibit a lot of filamentary structures, some of which were called ``worms'' crawling out of the Galactic plane \citep{1984ApJS...55..585H}. Some of them vertically extending from the Galactic plane are interpreted as a part of expanding shells. Recent works show that such filamentary structures can be produced in the shock-compressed diffuse ISM and the orientation of HI filaments is controlled by the directions of shock propagation and magnetic field \citep{2016ApJ...833...10I}. Meanwhile, some part of filamentary structures are thought to be magnetically floating loops generated by the Parker instability, which is thought to work in the Galactic dynamo \citep{1971ApJ...163..255P}. Candidates of such magnetic floating loops have been found in nearby galaxies such as M31 \citep{1989A&A...222...58B}, NGC253 \citep{1994AJ....108.2102S}, and IC342 \citep{2015A&A...578A..93B}. 

\citet{2006Sci...314..106F} identified two Galactic molecular loops named Loops 1 and 2 from wide-field imaging observations of $^{12}$CO~($J=1$--0) line with NANTEN 4-m telescope in Chile. The molecular loops are located within 1~kpc from the Galactic center and its length and width were measured several hundreds~pc and 30~pc, respectively. The total mass and kinetic energy of the molecular loops were estimated to be $1.7\times 10^5$~M$_\odot$ and $0.9\times 10^{51}$~erg, respectively. This energy is comparable to the energy supplied by a single supernova explosion, though it may not be able to convert all the energy to the loop. Therefore, they concluded that the magnetic floatation is more plausible mechanism to explain the physical characteristics of the molecular loop. \citet{2009PASJ...61.1039F} extended this study and found another more massive molecular loop, called Loop 3, whose total mass and kinetic energy were estimated to be $3.0\times 10^6$ M$_\odot$ and $1.7\times 10^{52}$ erg, respectively. This observational evidence supports that molecular loops are formed by magnetic field. 

\citet{1996ApJ...458L..25K} demonstrated that Parker instability can be easily triggered by a single supernova explosion forming $\Omega$-shaped structure based on two-dimensional MHD simulations. \citet{2009PASJ...61..957T} conducted two-dimensional MHD simulations and found that loop-like structure can be formed through the Parker instability. They showed that the molecular loops emerging from the low-temperature layer are similar to the dark filaments observed in the solar surface. Moreover, \citet{2009PASJ...61..411M} conducted three-dimensional MHD simulations of the gas disk in the Galactic center and they found that buoyantly rising magnetic loops are formed at the typical height of 200~pc from the Galactic plane. The typical length and width were 0.1--2~kpc and 50--300~pc, respectively, which are consistent with physical parameters of the molecular loop found in the Galactic center. 

Loop 3 is considered as a loop in the earlier evolutionary phase than Loops 1 and 2. \citet{2010PASJ...62..675T} carried out sensitive CO observations for the foot points of Loops 1 and 2, where gas is expected to be accumulated by the falling motion along the loops, and found that sharp intensity gradient is characterized by U shape, which suggests existence of a shock caused by the falling gas. The gas density and temperature are found higher at these foot points because of the shock heating. The $b-v$ diagram shows that foot point found as a U-shape, an L shape or a mirrored-L shape, which can be explained by a simple kinematic model incorporating the expansion of loop and the Galactic rotation \citep{2011PASJ...63..171K}. 

Loop structures are also found in the HI line in the outer Galactic disk \citep{Nakanishi17} by studying Galactic All Sky Survey (GASS) data \citep{2009ApJS..181..398M} and Southern Galactic Plane Survey (SGPS) data \citep{2005ApJS..158..178M}. They showed that integrated intensity map and longitude-velocity diagram can be explained with a toy model of HI cloud helically moving along a surface of tube with 60~pc radius and 2.1~kpc length. \citet{2017MNRAS.464..783S} studied the RM of radio sources behind the arch of Aquila-Rift to show that the magnetic field with strength of  10~$\mu$G is aligned along the Aquila-Rift arch and they suggested that the it is formed by the Parker instability.

%%%%%%%%%%%%%%%%%%%%%%%%%%%%%%%%%%%%%%%%%%%%
%%%%%%%%%%%%%%%%%%%%%%%%%%%%%%%%%%%%%%%%%%%%
%%%%%%%%%%%%%%%%%%%%%%%%%%%%%%%%%%%%%%%%%%%%
\section{The Milky Way}
\label{mag.s5}

The galactic magnetic field (GMF) in the Milky Way is introduced in this section. We first briefly summarize properties of the diffuse ionized medium (DIG), because the GMF is tightly related to the DIG. We then review the GMFs toward the galactic plane, toward the halo or high galactic latitudes, and toward the galactic center in this order. Regular magnetic fields and random magnetic fields are introduced separately, since the origins and properties of them are different. Hereafter, for convenience, we define the galactic plane (the Galactocentric cylindrical coordinate $|z|\lesssim 1$--2$~{\rm kpc}$), the galactic halo ($|z| > 1$--2$~{\rm kpc}$), high galactic latitudes (the galactic latitude $|b|>50^\circ$), and the galactic center ($|r|< 1$~kpc).

%%%%%%%%%%%%%%%%%%%%%%%%%%%%%%%%%%%%%%%%%%%%
%%%%%%%%%%%%%%%%%%%%%%%%%%%%%%%%%%%%%%%%%%%%
\subsection{Properties of Diffuse Ionized Gas}
\label{mag.s5.ss1}

\begin{figure*}[tbp]
\begin{center}
\FigureFile(160mm,160mm){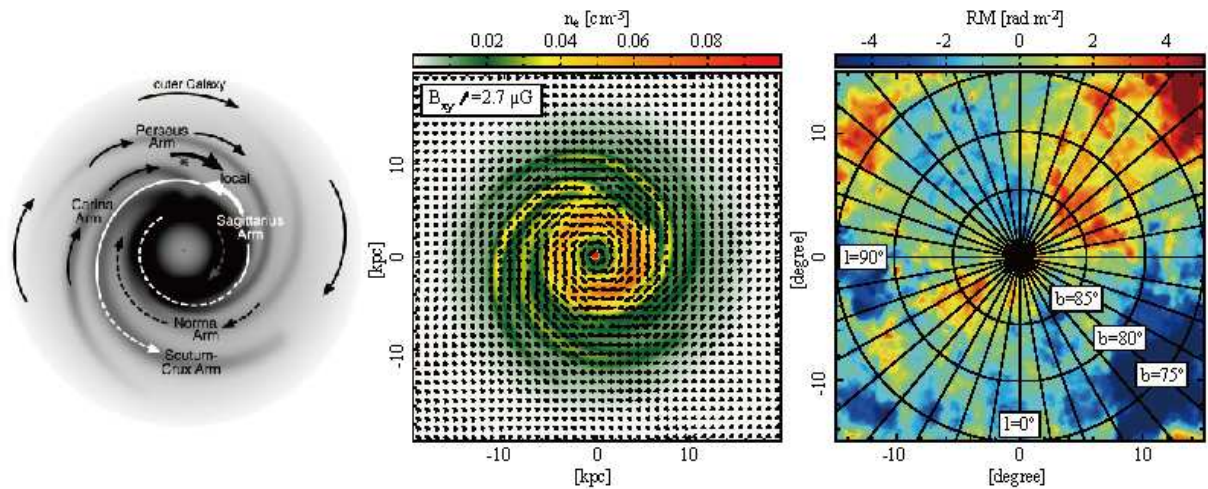}
\end{center}
\caption{
(Left) Sketch of the magnetic field in the disk of the Milky Way based on RM observations \citep{2011ApJ...728...97V}. (Middle) The color shows the modified NE2001 thermal electron density, and arrows depict the ASS+RING disk magnetic fields \citep{2008A&A...477..573S}. (Right) RM toward high galactic latitudes based on a model of regular and turbulent components (electron density and magnetic fields) of the Milky Way \citep{2013ApJ...767..150A}.
}
\label{f06}
\end{figure*}

Observations of pulsars allow us to estimate the DIG's electron density through pulsar DMs. \citet{2002astro.ph..7156C} and \citet{2003astro.ph..1598C} developed the NE2001 model, which is currently the standard model of three dimensional distribution of the electron density in the Milky Way (see also Schnitzeler 2013 for recent work). \citet{2008PASA...25..184G} proposed a modification of the parameters for the so-called thick disk component, the scale height (modified from $0.97~{\rm kpc}$ to $1.8~{\rm kpc}$) and the mid-plane electron density (modified from $0.034~{\rm cm^{-3}}$ to $0.014~{\rm cm^{-3}}$), which better reproduce both DM and emission measure (EM) toward high Galactic latitudes. Figure~\ref{f06} shows the modified NE2001 model.

Due to the presence of turbulence, there also exists local electron density fluctuation. The electron volume filling factor, $\sim 0.05-0.5$, which quantifies the clumpiness, has been estimated from DMs and emissions/absorptions (e.g., \cite{2002ApJ...575..217P, 2006AN....327...82B, 2008ApJ...686..363H, 2008PASA...25..184G, 2008A&A...477..573S}). 

The plasma $\beta_0$, the ratio of the gas pressure to the magnetic pressure due to regular magnetic field, is one of the key parameters which characterize magnetic turbulence. The gas pressure can be derived from observations of electron temperature, $T_e$. According to H$_\alpha$ observations, the distribution of $T_e$ can be approximated by $T_e(R,z)=5780+287R-526|z|+1770z^2$, where $T_e$ is in units of K and the Galactocentric cylindrical coordinates $R$ and $z$ are in kpc \citep{2008A&A...477..573S}. With observations of magnetic field strength, the plasma $\beta_0$ is expected in the range of $\sim 0.01-100$ and becomes smaller away from the Galactic plane \citep{2013ApJ...767..150A}. 

The rms Mach number $M_{\rm rms}\equiv u_{\rm rms}/c_s$, the ratio of the rms speed of random flow motions $u_{\rm rms}$ to the sound speed $c_s$, is another key parameter which characterizes magnetic turbulence. A weak constraint on the plausible range of $M_{\rm rms}$ is only provided by previous ${\rm H}\alpha$ observations toward high galactic latitudes. \citet{2008ApJ...686..363H} studied the distribution of EM, and found that $M_{\rm rms} \sim 1.4-2.4$ for $|b|>10^\circ$ and $M_{\rm rms}$ is smaller at higher Galactic latitudes. Studies of polarization gradients also broadly constrain $M_{\rm rms} \sim 0.5-2$ \citep{2011Nature...478..214G, 2012ApJ...749..145B}. There is so far no observational evidence which motivates to introduce a large-scale gradient of $u_{\rm rms}$ toward high Galactic latitudes. A simple approximation would be an uniform rms speed with $u_{\rm rms}\sim 15-50~{\rm km~s^{-1}}$ (or $M_{\rm rms}\sim 0.5-2$).

%%%%%%%%%%%%%%%%%%%%%%%%%%%%%%%%%%%%%%%%%%%%
%%%%%%%%%%%%%%%%%%%%%%%%%%%%%%%%%%%%%%%%%%%%
\subsection{Magnetic Fields in Galactic Plane}
\label{mag.s5.ss2}

%%%%%%%%%%%%%%%%%%%%%%%%%%%%%%%%%%%%%%%%%%%%
\subsubsection{Regular Magnetic Field}
\label{mag.s5.ss2.sss1}

The large-scale field in the Galactic disk has been studied by using RMs of pulsars and extragalactic radio sources (\cite{2001SSRev...99..243B}, \cite{2002ChJAA...2..293H}, \cite{2015ASSL...407..483H} for review), and using total and polarized synchrotron intensities of the Galactic synchrotron emission as well as RMs (\cite{2004A&A...427..169H, 2010MNRAS...401..1013J, 2012ApJ...757..14J, 2016JCAP...5..56B}). Also, the GMF has been constrained through observation of near-infrared star light polarization (e.g., \cite{2010ApJ...722L..23N}, \cite{2011ApJ...740..21P}, \cite{2012ApJ...749..71P}).

The total magnetic-field strength including regular and random components has been constrained from the synchrotron radio intensity by assuming equipartition between energy density of CRs and that of magnetic fields. According to \citet{2001SSRev...99..243B}, the total magnetic-field strength is estimated to be $\sim 6~\mu$G. \citet{2010MNRAS...401..1013J} simulated total and polarized synchrotron intensities using a GMF model which consists of coherent, isotropic random, and anisotropic random (``ordered" or ``striated") components. They found that the peak strength of the coherent component is $\sim 1-3~\mu$G, and found that the relative energy density ratio of these components is roughly 1:5:3, respectively.

The GMF orientation has been studied using RMs of pulsars and extragalactic radio sources. Studies of nearby pulsars show that the local field is directed toward the Galactic longitude $l \sim 70^\circ - 90^\circ$ and the field strength is $\sim 2-3~\mu$G (e.g., \cite{1974ApJ...188..637M, 1980MNRAS...191..863T}). A large-scale GMF model in the Galactic disk, where the R or BSS configuration is assumed, is determined so that the observed distribution of RMs is reproduced. The studies suggest at least one field reversal interior to the solar circle (e.g., \cite{1980ApJ...242..74S, 1983ApJ...265..722S, 1989ApJ...343..760R}). By using large data sets of RMs compiled by \citet{2009ApJ...702.1230T} (37,543 sources, figure~\ref{f09}) and \citet{2011PASA...28..171K} (2257 sources), it has been possible to develop the GMF model which consists of a disk field and a halo field. \citet{2011ApJ...738..192P} found that the spiral-field models fit the observed RM distribution better than the ring-field model, and also found that the disk field is symmetric with respect to the Galactic plane and in contrast the halo field is antisymmetric.

Pulsar DMs and RMs allow us to evaluate the GMF orientation at different positions in the Galactic disk, because the distance to a pulsar is estimated from DM (e.g., \cite{2005ApJ...619..297V}, \cite{2008ApJ...728..303V}). \citet{2006ApJ...642..868H} estimated the mean LOS component of the GMF within regions near the tangential points of the spiral arms and the inter-arms, using the mean slope of the DM--RM plot, $\langle B_\parallel \rangle = 1.232 \langle\Delta {\rm RM} / \Delta {\rm DM}\rangle$, for 554 RMs of pulsars, where $\Delta {\rm RM}$ and $\Delta {\rm DM}$ are differences in RMs and DMs, respectively. The obtained distribution of $\langle B_\parallel \rangle$ suggests that the large-scale magnetic fields in the spiral arms are counterclockwise, but in the inter-arm regions the fields are clockwise. As a result, they suggest that the large-scale GMF has a BSS configuration. The GMF structure derived from RMs of pulsars which include new data and that of extragalactic sources is presented in \citet{2015AASKA14...041..41H}.

Several attempts have been made to determine the large-scale field in the arm and inter-arm regions (e.g., \cite{2007ApJ...663..258B, 2010A&A...513..65N, 2011ApJ...728...97V}). \citet{2011ApJ...728...97V} developed an empirical model of the GMF in the disk with using a multi-sector model. In this model, the Galactic disk is divided into three sectors: A ($100^\circ < l < 260^\circ$), B ($260^\circ < l < 360^\circ$), and C ($ 0^\circ < l < 100^\circ$). They considered some models in each sector. The best-fit parameters of each model are determined independently so as to reproduce the observed RMs of both pulsars and extragalactic sources. They do not use any boundary matching condition between the sectors to see whether there are any common features among the sectors. In the Galactic outer region (sector A), the ring field model without field reversals reproduce the RM variation along the Galactic longitude better than the spiral model. In the sector B and C, the large-scale field in the inner disk follows the spiral arms, i.e. the spiral magnetic field with a pitch angle of $11.5^\circ$ except for the innermost and the outermost regions, in which azimuthal field is assumed. They conclude that the large-scale GMF is predominantly clockwise while a single reversed region exists in the inner Galaxy (left panel of figure~\ref{f06}). 

One field reversal interior to the solar circle is a common feature between observational studies on the GMF. However, the existence of more field reversals is still controversial. One of difficulties in determining the GMF orientation is an effect of SNRs and HII regions. Since these sources significantly affect RMs and DMs (\cite{1992ApJ...386..143C, 2003A&A...398..993M, 2010A&A...513..65N}), their influence has to be reduced as much as possible. Theoretical explanation of field reversals is also under debate as described in \S\ref{mag.s6}. Furthermore, it seems difficult to represent the large-scale GMF by the beautiful spirals or rings that have been used. \citet{2008A&A...486..819M} and \citet{2008MNRAS...386..1881N} examined whether the widely used models of the large-scale magnetic field (R, ASS, and BSS) are consistent with the observed data or not, but could not successfully reproduce the observations. These results suggest that the GMF has a more complex configuration and/or more complex random field components. 

%%%%%%%%%%%%%%%%%%%%%%%%%%%%%%%%%%%%%%%%%%%%
\subsubsection{Turbulent Magnetic Field}
\label{mag.s5.ss2.sss2}

The random field strength has been estimated from residual RMs (e.g., \cite{1969ApJ...157..1137J, 1980MNRAS...191..863T, 1989ApJ...343..760R, 2001ApJ...563..L31B}). A best fit model of the large-scale magnetic field is determined so that the variance of residuals $\langle ({\rm RM}_{\rm obs}-{\rm RM}_{\rm model})^2 \rangle$ is minimized, where RM$_{\rm obs}$ is RM from observation and RM$_{\rm model}$ is that derived from an assumed field model, and the brackets denote the averaging over all observed sources. If the residual is assumed to be caused by the random field component, the residual is expressed in terms of random walks (a single-size cell model) as $|{\rm RM}_{\rm obs}-{\rm RM}_{\rm model}| \propto (n_e \delta B L)\sqrt{d/L}$, where $n_e$ is the electron density, $\delta B$ is the random filed strength, $L$ is the cell size which corresponds to the outer scale of turbulence (the energy forcing scale), and $d$ is the distance to a radio source. \citet{1989ApJ...343..760R} used RMs of 163 pulsars. They exclude pulsars with $|B_\parallel| > 2~\mu$G in the region $0^\circ < l < 60^\circ$ affected by the North Polar Spur. They found that a concentric-ring model of the GMF reproduce the observed RM distribution better than a BSS model. As a result, they obtained the random field strength $\delta B \sim 5~\mu$ G and the cell size $L \sim 55$~pc from the variance and the covariance of best-fit residuals with the single-cell-size model.

As noted above, the residuals of RMs increase with pulsar distance $|{\rm RM}_{\rm obs}-{\rm RM}_{\rm model}| \propto \sqrt{d}$. However, the obtained residuals do not show the expected correlation (\cite{1989ApJ...343..760R, 2003A&A...398..993M}). Since this lack of the correlation suggests that the large-scale field is not represented by the beautiful spirals or rings, \citet{1993MNRAS...262..953O} evaluated the random field without using any large-scale field models. They used pairs of pulsars which are seen in almost the same directions on the sky. The electron density weighted magnetic field $B_\parallel$ in the region between the pair is obtained in terms of differences between their RMs ($\Delta {\rm RM}$) and DMs ($\Delta {\rm DM}$), $B_\parallel = \Delta {\rm RM}/(0.81 \Delta {\rm DM})~\mu$G. The obtained correlation between $|B_\parallel|$ and $\Delta {\rm DM}$ was interpreted by means of a Monte Carlo simulation with the single-cell-size model considering effects of random magnetic fields, electron density fluctuation and finite angular separations of the pulsar pairs. They found the random field strength to be $\delta B \sim 4-6~\mu$G with the cell size $L \sim 10$--$100$~pc.

\citet{2004ApJ...610..820H} also analyzed the random magnetic field using pulsar pairs. They sampled 1200 pulsar pairs from 490 pulsars and evaluated the power spectrum of the turbulent fields from the correlation between $|B_\parallel|$ and distance between paired pulsars. The obtained spectrum is $E_{\rm B}(k) \propto k^{-0.37}$ over the scale range 0.5--15~kpc. This suggests that the power spectrum of the interstellar turbulent fields becomes flatter than the Kolmogorov spectrum at the scales larger than the outer scale ($\sim 10 - 100$~pc).

Fluctuations of interstellar magnetic fields and electron density on scales smaller than $\sim 100$~pc have been investigated through the second order structure function (\S \ref{mag.s2.ss4.sss2}) of RMs, $S_{\rm 2,RM}(\delta \theta)$ defined as $S_{\rm 2,RM}(\delta \theta) = \langle [ {\rm RM}(\theta) - {\rm RM}(\theta+\delta \theta)]^2\rangle_\theta$, where $\theta$ is the position of a source in angular coordinates and $\delta \theta$ is the separation between sources. $S_{\rm 2,RM}$ is related to the power spectrum of random fields and density fluctuations (\cite{1984ApJ...284..126S, 1996ApJ...458..194M}); $S_{\rm 2,RM}(\delta \theta) \propto \delta \theta^{5/3}$ on scales smaller than the outer scale when the power spectrum of the ISM turbulence follows the Kolmogorov scaling (\S \ref{mag.s2.ss3}).

\citet{2009A&A...507..1087S} found that the slope of $S_{\rm 2,RM}$ varies depending both on the outer scale and the integral length of RMs. They found $S_{\rm 2,RM}(\delta \theta) \propto \delta \theta^{5/3}$ when the outer scale is comparable to the integral length. On the other hand, $S_{\rm 2,RM}$ has a shallower slope $S_{\rm 2,RM}(\delta \theta) \propto \delta \theta^{2/3}$ when the integral length exceeds the outer scale, because the fluctuations are smeared out. We can estimate the outer scale from the scale at which $S_{\rm 2,RM}$ becomes flat. \citet{2008ApJ...680..362H} evaluated $S_{\rm 2,RM}$ from RMs of extragalactic sources within an area of $253^\circ < l < 357^\circ$ and $|b| < 1.5^\circ$. They found that $S_{\rm 2,RM}$ in the Carina and Crux spiral arms have a flat slope, and estimated that the outer scale is smaller than $\sim 10$~pc. This suggests that stellar sources are the main energy source of the turbulence in the spiral arms. Meanwhile, $S_{\rm 2,RM}$ in the inter-arm regions have a shallower slope than the Kolmogorov, suggesting an additional energy sources on larger scales $\sim 100$~pc.

Also, the spatial gradient of linearly polarized synchrotron emission is used to constrain the sonic Mach number of the ISM turbulence (\cite{2011Nature...478..214G, 2012ApJ...749..145B}). \citet{2014A&A...566..5I} found that the observational data from S-band Polarization All Sky-Survey is consistent with transonic turbulence through comparison with MHD simulations.

%%%%%%%%%%%%%%%%%%%%%%%%%%%%%%%%%%%%%%%%%%%%
%%%%%%%%%%%%%%%%%%%%%%%%%%%%%%%%%%%%%%%%%%%%
\subsection{Magnetic Fields toward High Galactic Latitudes}
\label{mag.s5.ss3}

%%%%%%%%%%%%%%%%%%%%%%%%%%%%%%%%%%%%%%%%%%%%
\subsubsection{Regular Magnetic Field}
\label{mag.s5.ss3.sss2}

Regular, coherent magnetic field toward high galactic latitudes can be combinations of disk spiral (e.g., \cite{2003A&A...410....1P, 2008A&A...477..573S, 2011ApJ...728...97V}, middle panel of figure~\ref{f06}), halo toroidal (e.g., \cite{2010ApJ...724.1456T, 2010RAA....10.1287S, 2011ApJ...738..192P, 2012ApJ...755...21M}, and halo poloidal (or X-) fields (e.g., \cite{2010JCAP...08..036G, 2012ApJ...757..14J}). By definition, the spiral field dominates regular magnetic field near the Galactic plane, while the toroidal field dominates above. Transition from disk to halo fields arises at $\sim 1.25$~kpc from the mid-plane.

RMs toward the Galactic poles have been investigated in the literature. Using RM data from the NRAO VLA Sky Survey (NVSS), \citet{2009ApJ...702.1230T} estimated non-zero vertical strengths of the GMF, about $-0.14\pm 0.02~\mu{\rm G}$ and $+0.3\pm 0.03~\mu{\rm G}$ toward the North galactic pole (NGP) and South galactic pole (SGP), respectively.  \citet{2010ApJ...714.1170M} used RM data from the Westerbork Radio Synthesis Telescope (WSRT) and the Australia Telescope Compact Array (ATCA), and found that the median value of RMs toward the SGP is $+6.3\pm 0.5$~rad~m$^{-2}$ (corresponding to V field strength of $+0.31\pm 0.02~\mu{\rm G}$), while that toward the NGP is $0.0\pm 0.5$~rad~m$^{-2}$ ($+0.00\pm 0.02~\mu{\rm G}$). The origin of this field anomaly is not resolved (see \cite{2013ApJ...767..150A}, some discussion therein).

%%%%%%%%%%%%%%%%%%%%%%%%%%%%%%%%%%%%%%%%%%%%
\subsubsection{Turbulent Magnetic Field}
\label{mag.s5.ss3.sss3}

Highly disturbed distributions of RM and polarization angle clearly indicate turbulent structures of the GMF (\cite{2008A&A...477..573S, 2009ApJ...702.1230T, 2009A&A...495..697W, 2011Nature...478..214G}). Using the WSRT and ATCA data, \citet{2010ApJ...714.1170M} claimed the standard deviations of RMs, $\simeq 9.2$~rad~m$^{-2}$ and $8.8$~rad~m$^{-2}$ toward the NGP and SGP, respectively. They put an upper limit of $\sim 1~\mu{\rm G}$ on the strength of random magnetic field at high Galactic latitudes. Based on the latitude dependence of RM, \citet{2010MNRAS.409L..99S} examined the Galactic and extragalactic contributions to RM in the NVSS data. He estimated that the Galactic contribution (including both the disk and halo components) is $\bar{\sigma}_{\rm RM,MW}\sim 6.8 \pm 0.1 (8.4 \pm 0.1)$~rad~m$^{-2}$ and the extragalactic contribution (including those intrinsic to the polarized background radio sources and due to the IGM) is $\bar{\sigma}_{\rm RM,EG}\sim 6.5 \pm 0.1 (5.9 \pm 0.2)$~rad~m$^{-2}$ for the northern (southern) hemisphere. \citet{2011ApJ...726....4S} examined the NVSS data in detail, and found that $S_{\rm 2,RM}$ at $\delta \theta \gtrsim 1^\circ$ has a value $\sim 100-200$~rad~m$^{-2}$ toward the NGP and $\sim 300-400$~rad~m$^{-2}$ toward the SGP. 

The random, turbulent component has been modeled analytically using power-law spectra with random phases in Fourier space, i.e. uniform turbulence. For instance, \citet{2009A&A...507..1087S} used the publicly-available HAMMURABI code \citep{2009A&A...495..697W}, and adopted a Kolmogorov-like power spectrum with average amplitude $3~\mu{\rm G}$ in a box of 10 pc size. They found that $S_{\rm 2,RM}$ has a magnitude of up to a few $\times 100$~rad~m$^{-2}$ at angular scales of $>10'$ at Galactic latitudes $|b|\sim 70^\circ$.

Constant rms amplitudes for electron density fluctuations and turbulent magnetic field, however, would not be justified, because the amplitudes should depend on $\beta_0$ and $M_{\rm rms}$ and they distribute broadly toward high altitudes. In addition, in turbulent flows, phases are not really random. Toward high Galactic latitudes, the random components are the dominant contribution to the RM. \citet{2013ApJ...767..150A} developed a sophisticated model of the Milky Way (right panel of figure~\ref{f06}). While they modeled the regular component based on a number of observations, they used the data of three-dimensional MHD turbulence simulations and first considered latitude dependences of $\beta_0$ and $M_{\rm rms}$ to model the random component.

\citet{2013ApJ...767..150A} found that the observed medians of RMs toward the north and south Galactic poles are difficult to explain with any of many alternate GMF models. The standard deviation of observed RMs is clearly larger than that of simulated RMs. $S_{\rm 2,RM}$ of the observed RMs is substantially larger than that of the simulated RMs, especially at small angular scales. They suggested that reproducing the observed medians may require additional components or/and structures of the GMF that are not present in their models. They also pointed out the RM due to the IGMF may account for a substantial fraction of the observed RM.

%%%%%%%%%%%%%%%%%%%%%%%%%%%%%%%%%%%%%%%%%%%%
%%%%%%%%%%%%%%%%%%%%%%%%%%%%%%%%%%%%%%%%%%%%
\subsection{Magnetic Fields toward the Galactic Center}
\label{mag.s5.ss4}

%%%%%%%%%%%%%%%%%%%%%%%%%%%%%%%%%%%%%%%%%%%%
\subsubsection{Magnetic Fields around the Galactic Center}
\label{mag.s5.ss4.sss1}

Magnetic fields around the Galactic Center are observed to be as strong as up to 2~mG, an order of magnitude stronger than those in the local Galactic disk \citep{1996A&ARv...7..289M}. This region has many strong radio continuum sources, Sgr A, B, C, D, and E along the Galactic plane in the order of their flux densities \citep{1978A&AS...35...23A, 1996A&ARv...7..289M, 1996ARA&A..34..645M}.

Sgr A consists of the non-thermal Radio Arc, the Bridge of thermal filaments, Sgr A East, Sgr A West, and Sgr A$^*$ (e.g., \cite{1996A&ARv...7..289M}). \citet{1984Natur.310..557Y} carried out VLA observations of Sgr A at 6~cm and 20~cm. Their results indicate that the Radio Arc consists of a system of narrow filamentary structures having lengths larger than 30~pc but typical widths of as narrow as 1~pc. The filaments consist of two groups: one is the ``Radio Arc" emitting non-thermal polarized emission, which is a bunch of numerous vertical filaments perpendicular to the Galactic plane and crosses the plane. The other is ``arched" filaments emitting thermal radiation, which arises in the halo of Sgr A, diverges curving eastward, and joins the vertical filaments of the Radio Arc.

Linear polarization with the polarization degree of $\sim 20$~\% is seen along the vertical filaments in the Radio Arc, and they show a high $RM\sim 10^3$ rad~m$^{-2}$ \citep{1984PASJ...36..633I, 1986AJ.....92..818T, 1987PASJ...39...95S}. The observed RM sharply increases along the Arc toward the Galactic plane, but suddenly drops to zero near the plane due to strong beam and bandwidth depolarization. The observed RM is then followed by negative steep increase toward the opposite side of the plane. Such a reversal of the RM value across the Galactic plane means that the vertical magnetic field is pinched at the disk due to twisting rotation with respect to the off plane field. On the other hand, the thermal arched filaments show no linear polarization \citep{1988ApJ...329..729Y}.

\citet{2006Natur.440..308M} obtained a near infrared image near the Galactic Center by using Spitzer Space telescope at 24~$\mu$m, and found an intertwined double helix nebula with a length of 25~pc. Each of the two continuous helically-wound strands rounds about 1.25 full turns. They interpreted that this feature is explained as a torsional Alfv\'en wave propagating vertically away from the Galactic plane.

The Radio Arc is a part of a larger-scale off-plane radio lobe over the Galactic Center \citep{1984Natur.310..568S}. The height of the Galactic Center Lobe is 1.2~deg from the Galactic plane or about 200~pc in size. The formation mechanism is still controversial, whether it is produced by a vertical field twisted by the rotation of an accreting gas disk, due to inflation of giant loop of magnetic tube filled with ionized gas, or a result of some explosive events at the Center.  The western ridge of the lobe apparently emerges from Sgr C and has a filament and polarization indicating existence of magnetic fields \citep{1988ApJ...329..729Y, 1989IAUS..136..213S}. 

\citet{1991MNRAS.249..262A} conducted 90~cm VLA observation for 2~deg $\times$ 2~deg field and found many non-thermal filaments and found that all of them are perpendicular to the Galactic plane. These structures trace a vertical (V; poloidal), or a dipole magnetic field \citep{1996ARA&A..34..645M}. \citet{2000AJ....119..207L} enlarged the 90~cm VLA observation area for 4~deg $\times$ 5~deg and catalogued over a hundred sources including SNRs, filaments, threads, the Snake, the Mouse that show structures of a variety of type of magnetic structures and activity near the Galactic Center.

An MHD model on the magnetic field structure near the Galactic Center was proposed by \citet{1985Natur.317..699U} and extended by \citet{1987PASJ...39..559S}. Assuming that magnetic fields are frozen into the interstellar gas, a differentially rotating and infalling gaseous disk can shear an initially poloidal field into toroidal, and produce a twisted/helical field.

Far-infrared and submillimeter polarization observations suggested that magnetic fields are parallel to the Galactic plane, indicating toroidal magnetic fields (e.g., \cite{1992ApJ...399L..63M}. \citet{2009ApJ...690.1648N} and \citet{2010ApJ...722L..23N} conducted wide-field near-infrared polarimetry of point sources in J, H, Ks bands, and found that the magnetic fields near the Galactic plane are almost parallel, but the fields are nearly perpendicular to the Galactic plane at the high Galactic latitude. It suggests transition from a toroidal to a poloidal magnetic field.

%%%%%%%%%%%%%%%%%%%%%%%%%%%%%%%%%%%%%%%%%%%%
\subsubsection{At Sgr A* and the Black Hole}
\label{mag.s5.ss4.sss2}

Earth's nearest candidate supermassive black hole, whose mass is estimated to be $\sim 4\times 10^6 M_\odot$, lies at the exact center of Sgr A* \citep{2008ApJ...689.1044G, 2009ApJ...692..1075G}. If strong magnetic fields exist around the block hole, they can influence the dynamics of accretion, and can transport angular momentum from the infalling gas to relativistic jets.

\citet{1995ApJ...445L..113P} detected strong magnetic fields of $-3.0\pm 0.5$~mG at $1'$ north of Sgr A* by measurement of Zeeman splitting of the HI absorption line. \citet{2005ApJ...618L..L29B} conducted polarimetric observations between 215~GHz and 230~GHz using BIMA array and detected a variable linear polarization from Sgr A*, evidencing a hot turbulent accretion flow. \citet{2006ApJ...646L.111M} observed RM of $-4.4 \pm 0.3 \times 10^5$~rad~m$^{-2}$ at 82 GHz and 86 GHz using BIMA array and detected polarization variability on a timescale of days of Sgr A*. They argued that such high Faraday rotation occurs external to the polarized source at all wavelengths, and it may imply an accretion rate $\sim (0.2$--$4)\times 10^8$~M$_\odot$~yr$^{-1}$. \citet{2007ApJ...654L..57M} detected an unambiguous Faraday rotation of $-5\times 10^5$~rad~m$^{-2}$ in Sgr A* at 227~GHz and 343~GHz using SMA.

\citet{2015Sci...350..1242J} conducted VLBI observations at 1.3 millimeter and spatially resolved the linearly polarized emission from Sgr A*. They found evidence of partially ordered magnetic fields near the event horizon. This kind of high-frequency VLBI approach is important because the circum black hole region should have dense electron density and so electron scattering is large. There will be another way of research in the Galactic Center using the supposed thousands of pulsars in the central region, some of which may transports Faraday rotation information in the vicinity of the black hole.

In fact, Swift discovered a new soft gamma-ray repeater,SGR J1745-2900 near Sgr A* \citep{2013ApJ...770L..24K}, and the source was identified as an X-ray as well as a radio pulsar, PSR J1745-2900, which has a pulsation period of 3.76 seconds (e.g., \cite{2013ApJ...770L..23M, 2013ATel.5040....1E} ). \citet{2013Natur.501..391E} reported multi-frequency radio measurements of the newly discovered pulsar and showed the pulsar's unusually large RM of $(-6.696 \pm 0.005) \times 10^4$~rad~m$^{-2}$, which indicates a dynamically relevant magnetic field near the black hole.

Recently, it was found by IR precision astrometry observations that a small gas cloud, G2, is approaching Sgr A* \citep{2009ApJ...692..1075G}. It seemed that there would be any emission enhancement if G2 give some perturbation to the accretion flow around the supermassive black hole. However, no significant microwave enhancement of Sgr A* was observed \citep{2015ApJ...798L...6T}. There are some possible explanations of the result. In the case that G2 is a gas cloud, one possible explanation of the no significant microwave enhancement is that the magnetic fields in the accretion flow is too strong to make bow shock wave in the accretion flow. If the Alfv\'en velocity around Sgr A* is faster than the velocity of the G2 cloud, 6000 km s$^{-1}$, the result suggests the lower limit of the magnetic fields, 30 mG \citep{2015ApJ...798L...6T}. \citet{2017arXiv170207903K} carried out MHD simulations of an accretion disk for Sgr A* to study the effect of pericenter passage of the G2 cloud. They found that magnetic fields in the accretion disk are enhanced by the G2 perturbation in 5-10 years after the pericenter passage of G2. Since the enhancement boosts synchrotron emission from the disk and the outflow, the radio and the infrared luminosity of Sgr A* is expected to increase around A.D. 2020.

%%%%%%%%%%%%%%%%%%%%%%%%%%%%%%%%%%%%%%%%%%%%
%%%%%%%%%%%%%%%%%%%%%%%%%%%%%%%%%%%%%%%%%%%%
%%%%%%%%%%%%%%%%%%%%%%%%%%%%%%%%%%%%%%%%%%%%
\section{Galaxies}
\label{mag.s6}

Studying the GMF in external galaxies is important to understand how the GMF were formed and evolved in the Universe as well as to understand the correlation between the GMF and various properties of galaxies such as the morphology, degree of star-formation, supernova and stellar wind feedbacks, AGN activity, and so on. In this section, we review the GMF in normal spiral galaxies and other galaxies in this order, and then summarize cosmological evolution of the GMF. Similar to the Milky Way, the GMF in external galaxies can be described as a combination of large-scale regular fields and small-scale turbulent fields.

%%%%%%%%%%%%%%%%%%%%%%%%%%%%%%%%%%%%%%%%%%%%
%%%%%%%%%%%%%%%%%%%%%%%%%%%%%%%%%%%%%%%%%%%%
\subsection{Normal Spiral Galaxies}
\label{mag.s6.ss1}

Detection of synchrotron emission indicates the existence of magnetic fields as well as CR electrons in spiral galaxies. The magnetic-field strength can be estimated by assuming the equipartition between the total energy densities of CRs and that of the magnetic field (see section~\ref{mag.s3.ss2}). Applying such an equipartition assumption to the observational data for one of the well-studied nearby spiral galaxy IC342, typical strength of the total magnetic field was estimated to be 15 $\mu$G assuming $K_0=100$, $i=31^\circ$, $l=1/ \cos i$, and $\alpha=1.0$ \citep{2015A&A...578A..93B}. \citet{2012ApJ...756..141B} argue that condition of the equipartition is satisfied at scale of $>1$ kpc while it does not hold on small scale \citep{2014MNRAS.437.2201S}. 

Typical magnetic-field strength is estimated to be 1--10~$\mu$Jy by measuring the Faraday rotation. It is known that synchrotron and far-infrared luminosities for normal galaxies are well correlated with each other \citep{1991ApJ...376...95C} and that this correlation holds for high-redshift galaxies \citep{2009ApJ...706..482M}. Current possible scenarios of the origin of the radio-FIR correlation are (i) calorimeter model \citep{1989A&A...218...67V} and (ii) equipartition model \citep{1997A&A...320...54N}. The former model suggests that a power of the radio emission is independent of the magnetic energy density in assuming that the life time of CR electrons is inversely proportional to the magnetic energy density.

%%%%%%%%%%%%%%%%%%%%%%%%%%%%%%%%%%%%%%%%%%%%
\subsubsection{Regular Magnetic Field}
\label{mag.s6.ss1.sss1}

Polarization observation shows that magnetic fields are globally coherent along spiral arms with the same pitch angle \citep{2011MNRAS.412.2396F}. The global configuration of the coherent magnetic field is roughly classified into ASS and BSS fields. Typical examples of ASS configuration are M31 \citep{1981PASJ...33...47S} and IC342 \citep{1988A&A...192...66G}. Typical cases for BSS are M51 \citep{1978PASJ...30..315T} and M81 \citep{1980A&A....91..335S}. Note that clear magnetic spiral arm is sometimes found in the inter-arm region in such a case as NGC 6946 \citep{2007A&A...470..539B}. 

\begin{figure*}[tbp]
\begin{center}
\FigureFile(160mm,160mm){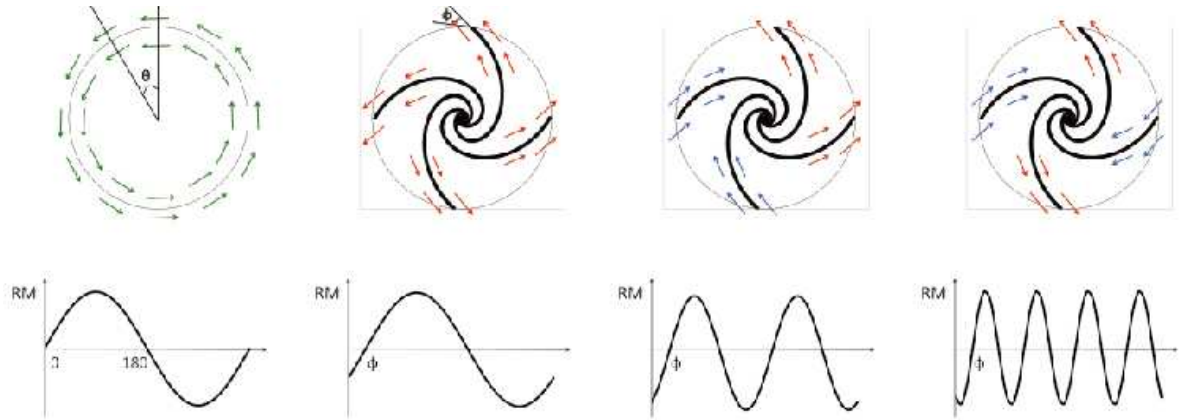}
\end{center}
\caption{
Left, middle-left, middle-right, and right panels show magnetic fields of Ring, ASS, BSS, and QSS, respectively. In the case of BSS and QSS, there are two and four reversals in the magnetic fields, respectively, but no reversal in the cases of Ring and ASS.
}\label{f07}
\end{figure*}

Most simple example of the axisymmetric field is the R field shown in the left-most panel of figure~\ref{f07}, though in most cases magnetic fields are aligned with spiral arms so that the R field is rarely observed. The other axisymmetric field is the ASS field shown in the middle-left panel of figure~\ref{f07}. In the case of ASS, the magnetic field is symmetric about the galactic center and there is no field reversal. The middle-right panel of figure~\ref{f07} shows the case of the BSS field, where single magnetic reversal is observed along a annulus. Recently, topology with more field reversals including the QSS (MSS) field, shown in the right-most panel of figure~\ref{f07}, is suggested for several galaxies \citep{2008A&A...480...45S}.

Configuration of magnetic fields are identified by studying how the RM varies with azimuthal angle about the galactic center. Azimuthal variations of RM in the cases of ASS, BSS, and QSS are shown in the bottom panels in figure~\ref{f07}. In the case of the R field, the RM sinusoidally changes with the azimuthal angle with a single reversal of positive and negative. Its maximum and minimum are symmetric about $RM=0$. The RM of the ASS field changes in the same manner with the R field while the phase is shifted by a pitch angle of spiral patterns. In the case of BSS, the RM also changes sinusoidally but the RM reaches maximum (minimum) twice. In the case of QSS, the RM changes sinusoidally but the RM reaches maximum (minimum) four times. The maximum and minimum are not necessarily symmetric about $RM=0$ in the cases of BSS and QSS. 

Observations of edge-on galaxies show that magnetic field in the halo region is found to be X-shaped \citep{2009A&A...506.1123H}. Possible explanations of such a vertical magnetic field are (i) galactic wind which blow the ISM up to the halo \citep{1993A&A...271...36B} and (ii) primordial magnetic field which runs vertically relative to the disk \citep{2010PASJ...62.1191S}.

Note that the large scale BSS and V configurations of the GMF cannot be created by dynamo, but are considered to be created during the formation of primeval galaxies. On this primordial-origin  hypothesis, the BSS and central vertical fields are interpreted as the fossil of an intergalactic field wound up by the primordial galaxy disk during its formation and contraction.

%%%%%%%%%%%%%%%%%%%%%%%%%%%%%%%%%%%%%%%%%%%%
\subsubsection{Random Magnetic Field}
\label{mag.s6.ss1.sss2}

Depolarization seen in a spiral galaxy NGC 6946 implies that there is random (turbulent) magnetic fields whose strength is 10~$\mu$G and coherence length is $\sim 50$~pc \citep{2007A&A...470..539B}. The regular magnetic field tends to be along with spiral structure and ordered in the regions where turbulence is small. The turbulent field can be also described as a combination of the components parallel and perpendicular to the regular field. A high resolution 5~GHz observation unveiled that the characteristic length of the parallel component is $\sim 100$~pc, while the perpendicular component is $\sim 50$~pc \citep{2013ApJ...766...49H}, indicating the existence of anisotropic turbulence.

External galaxies often intervene the LOS toward background polarized sources and alter their polarization properties by magnetic fields. MgII absorption observations indicate the existence of absorbers in front of about a half of SDSS galaxies including polarized sources such as radio galaxies and quasars \citep{2013ApJ...770..130Z}. MgII surveys highlight intervener's effects and allow us to investigate the strength, growth time, and coherence length of galactic dynamo in such galaxies as functions of redshift, beam covering fractions, and spatial sizes of the galaxies \citep{2008ApJ...676...70K, 2012ApJ...761..144B, 2013ApJ...772L..28B}. Investigation of such depolarizing intervening galaxies (DINGs) can allow unbiased survey of magnetic fields in galaxies.

\begin{figure}[htbp]
\begin{center}
\FigureFile(60mm,60mm){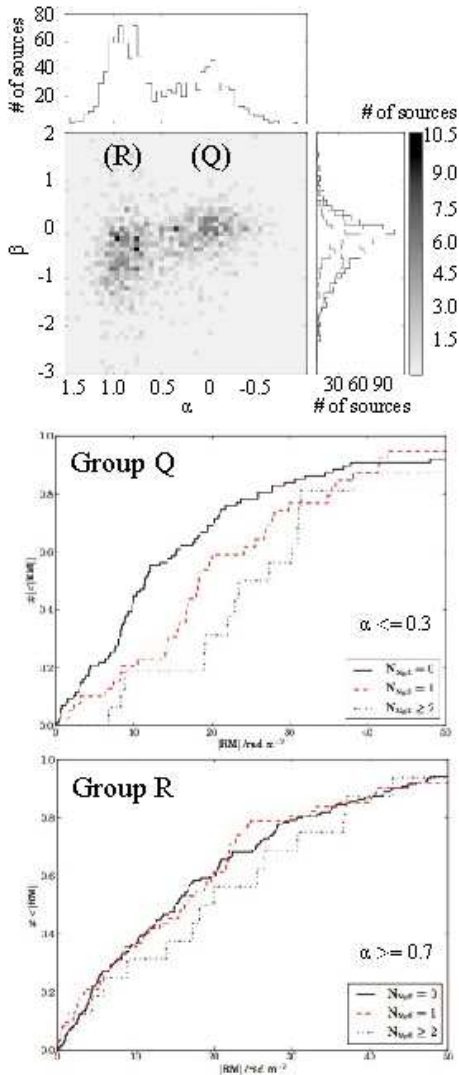}
\end{center}
\caption{
(Top panel) The relation between the total intensity spectrum index $\alpha$ and the polarization spectrum index $\beta$ for 951 sources \citep{2014ApJS..212...15F}, where the indices were derived from multiple observations between 400~MHz and 100~GHz. (Bottom two panels) The correlation between RM and the existence of foreground MgII absorption systems \citep{2014ApJ...795...63F}.
}
\label{f08}
\end{figure}

Figure~\ref{f08} shows the relation between the total intensity spectrum index $\alpha$ and the polarization spectrum index $\beta$ \citep{2014ApJS..212...15F}. There are two populations, an optically-thick AGN-core type ($\alpha\sim 0$) and an optically-thin radio-lobe type ($\alpha\sim 0.7$). The behavior of $\beta$ depends on $\alpha$, therefore, part of depolarization takes place at the source because $\alpha$ should depend only on the source nature. Meanwhile, \citet{2014ApJ...795...63F} suggested that there is a clear correlation between RM and the existence of foreground MgII absorption systems for the AGN-core type (figure~\ref{f08}). Note that DING's depolarization is caused by both regular and random magnetic fields. The latter may become more predominant at higher redshift, according to a relatively longer timescale to form the latter.

\citet{2013MNRAS.435.3575B} recently performed a sophisticated MHD simulation of an evolving (star-forming) galaxy including effects of star formation, supernova magnetic seeding, and galactic wind. They demonstrated that the entire diffuse ionized gas in the galactic halo is magnetized by redshift $z\approx 0$. The magnetic-field strength is at a level of a few $\mu {\rm G}$ in the halo center and ${\rm nG}$ at the virial radius. The mean halo intrinsic RM peaks between $z\approx 4$ and $z\approx 2$ and reaches absolute values around 1000~rad~m$^{-2}$. While the halo virializes towards $z\approx 0$, the intrinsic RM values decline to a mean value below 10~rad~m$^{-2}$.

\citet{Akahori17} tested depolarization caused by DINGs, using a model of the Milky Way. They found that both the global and local magnetic fields can contribute to depolarization, and its significance depends on the observing frequency, the observing beam size, and the pointing center. The Burn law is not satisfied, because RM distribution within the beam does not follow the Gaussian. DING's contribution to the observed RM decreases significantly as the DING's redshift increases.

%%%%%%%%%%%%%%%%%%%%%%%%%%%%%%%%%%%%%%%%%%%%
%%%%%%%%%%%%%%%%%%%%%%%%%%%%%%%%%%%%%%%%%%%%
\subsection{Other Galaxies}
\label{mag.s6.ss2}

%%%%%%%%%%%%%%%%%%%%%%%%%%%%%%%%%%%%%%%%%%%%
\subsubsection{Barred Galaxies}
\label{mag.s6.ss2.sss1}

\citet{2002A&A...391...83B} investigated 17 barred galaxies and found a positive correlation between the mean radio intensity of the galaxy and the length of its bar structure. They also found a tight correlation between radio and far-infrared intensities. Such a relation is similar to that seen in non-barred galaxies, in which the relation is thought to be originated from star formation activity. Magnetic fields around the bar is stretched by shearing gas flow where the field strength is relatively large and aligned along the bar (\cite{2005A&A...444..739B}). 

There have been a number of MHD simulations in barred galaxies. The bar-shape gravitational potential is assumed to produce the bar structure. \citet{2012ApJ...751..124K} carried out two-dimensional MHD simulations and showed that the magnetic-field strength enhances the mass inflow rate to the galactic center. In the three-dimensional simulations, \citet{2010A&A...522A..61K} showed the formation of magnetic arms with low density around the bar. \citet{2015A&A...575A..93K} and \citet{2015A&A...575A..93K} simulated galactic dynamo driven by CRs, and presented that a saturation level of magnetic-field strength is $3-10$~$\mu$G and the field structure form the quadrupole-like symmetry with respect to the galactic plane.

%%%%%%%%%%%%%%%%%%%%%%%%%%%%%%%%%%%%%%%%%%%%
\subsubsection{Elliptical Galaxies}
\label{mag.s6.ss2.sss2}

Magnetic fields of elliptical galaxies have been observed based on the Faraday rotations of background radio sources behind elliptical galaxies \citep{1984ApJS...54..291P, 1986A&A...156..234L, 1987ApJ...316...95O, 1987MNRAS.228..557L, 1990ApJ...360...41T, 1992ApJ...395..444C}. Elliptical galaxies have little cold ISM forming stars and therefore little synchrotron emissions from their halo regions. 

Differential Faraday rotation along a LOS of gaseous halo of a parent elliptical galaxy shows that random magnetic field is dominant \citep{1988A&A...194...79S}. Meanwhile, they did not find evidence of large-scale magnetic fileds in the elliptical galaxy. The random magnetic field is thought to be generated by turbulent motion which are driven by type I supernovae and the stellar motion relative to the ISM \citep{1996MNRAS.279..229M}.

%%%%%%%%%%%%%%%%%%%%%%%%%%%%%%%%%%%%%%%%%%%%
\subsubsection{Dwarf Galaxies}
\label{mag.s6.ss2.sss3}

Dwarf galaxies are low-mass systems whose rotational velocities are relatively smaller than those of spiral galaxies. Therefore, the small-scale dynamo is thought to be more significant than large-scale dynamo in the sense that the $\Omega$-effect is weak. Observations of IC10 and NGC6822, both small and low-mass galaxies, support this idea since large-scale magnetic fields are not identified \citep{2003A&A...405..513C}. 

Larger dwarf galaxies such as Large Magellanic Cloud (LMC) and NGC4449, however, seem to have regular magnetic field due to $\alpha$-$\Omega$ dynamo since they have large-scale spiral structure \citep{2000A&A...355..128C,2005Sci...307.1610G}. Whereas a regular magnetic field is found in the Small Magellanic Cloud (SMC), it is unlikely to be generated with the large-scale dynamo because of its asymmetric configuration \citep{2008ApJ...688.1029M}. Additionally, \citet{2012ApJ...759...25M} found that the HI filament in the southeast of the LMC is directed to the SMC and suggested that it formed $\sim 1$~Gyr ago due to the tidal interaction between the LMC and SMC. The tidal interaction is thought to be one of the possible mechanisms to form a coherent magnetic field as found in the dwarf galaxy NGC2976 \citep{2016A&A...589A..12D}.

%%%%%%%%%%%%%%%%%%%%%%%%%%%%%%%%%%%%%%%%%%%%
%%%%%%%%%%%%%%%%%%%%%%%%%%%%%%%%%%%%%%%%%%%%
\subsection{Cosmological Evolution}
\label{mag.s6.ss3}

The cosmological primordial fields and the turbulent seed fields are considered to be the seed for global fields of galaxies. The advantage of the primordial field model is that the existence of the global magnetic fields are running through the disk plane and that the global structure such as R, ASS, BSS and V fields are explained without contradiction (e.g., \cite{2010PASJ...62.1191S}). 

Magnetic fields generated at the inflation epoch are estimated $<10^{-9}$ G observed by the CMB polarization (see \S\ref{mag.s9}). During the reionization epoch, these magnetic fields are attenuated to be $10^{-10}$~G, and the scale of the coherent length is estimated about kpc (e.g., \cite{2016RPPh...79g6901S}, \cite{2015MNRAS.451.1692P}). Magnetic fields generated by the Biermann buttery effect \citep{1950ZNatA...5...65B} are up to $10^{-21}$~G and pc scale coherent length \citep{2007PhRvD..75j3501K}. In either case, it is necessary to amplify the weak seed fields to explain the present magnetic fields of galaxies.

\citet{2008Natur.454..302B} studied quasars in a redshift range between 0.6 and 2.0, and found that quasars with strong MgII line have large rotation measure. This observational fact implies that MgII absorption is attributed to halos of normal galaxies in the LOS to the quasar and such normal galaxies have large magnetic field (see also \cite{2014ApJ...795...63F}).

There are roughly three theoretical ways to study the cosmological evolution of the GMF. First, \citet{2009A&A...494...21A} studied the evolution of magnetic fields in galaxies based on a dynamo theory. They suggested that $\mu$G turbulent magnetic field can be quickly (a few $10^8$ yr) generated in halos of protogalaxies from a weak seed field by the small-scale dynamo. The turbulent field can become a seed to the large-scale dynamo, and in a Milky-Way-type galaxy, $\mu$G coherent magnetic fields with scales of kpc and tens of kpc are established until $z\sim 3$ and $z\sim 0.5$, respectively, by the large-scale dynamo. 

Second, MHD simulations of the galactic dynamo have been carried out (e.g., \cite{2006ApJ...641..862N}, \cite{2009ApJ...706L.155H}). Although the main mechanisms of amplification were not necessarily the same, initial weak magnetic fields can be successfully amplified up to a few $\mu$~G and the amplified fields are maintained during several billions of years. \citet{2013A&A...560A..93G} included the effect of the amplification of the supernovae by $\alpha-\Omega$ dynamo terms and turbulent dynamo was treated by MHD (called Hybrid dynamo). They demonstrated a mixture of modes with even and odd parities which creates a strong localized vertical field on one side of the Galactic disk. \citet{2013ApJ...764...81M} presented the all-sky map of RM obtained from their numerical simulation, and suggested that the magnetic fields generated by MRI-Parker dynamo explain the tendency of observed RM distribution.

Third, cosmological simulations of galaxy formations have been performed by AREPO group (e.g., \cite{2014ApJ...783L..20P}). They assumed weak random magnetic fields and the standard $\Lambda$CDM cosmology. They suggested that ordered magnetic fields of strength about 6 $\mu$G were formed until $z \sim 2$.

Finally, an environmental effect would be another important factor of cosmological evolution of the GMF. It is well known that galaxies in the Virgo cluster suffer the environmental effect such as ram pressure stripping \citep{1990AJ....100..604C, 2006ApJ...651..804N}. \citet{2012A&A...545A..69W} and \citet{2013A&A...553A.116V} investigated magnetic fields of Virgo cluster galaxies based on radio observations with 100-m Effelsberg telescope and Jansky Very Large Array. They studied asymmetric polarized emission indicating distorted magnetic-field structures. According to their investigations, galaxies such as NGC 4294, 4298, 4302, 4303, 4321, 4568 seem to have experienced tidal interactions, NGC 4532 and 4808 are likely to have huge accreting HI envelopes, NGC 4388 and 4505 seem to have experienced strong ram-pressure and shearing effect, and NGC4457 seems to have a recent minor merger. The magnetic field is a good indicator to trace even weak interactions, which are difficult to detect with other observations.

%%%%%%%%%%%%%%%%%%%%%%%%%%%%%%%%%%%%%%%%%%%%
%%%%%%%%%%%%%%%%%%%%%%%%%%%%%%%%%%%%%%%%%%%%
%%%%%%%%%%%%%%%%%%%%%%%%%%%%%%%%%%%%%%%%%%%%
\section{AGN and Jet}
\label{mag.s7}

Magnetic fields significantly affect physics of accretion disks, black holes, and jets. They also induce radio emission, which is used to classify AGN and jets. We review a role of magnetic field in AGN and astronomical jets in this section.

\begin{figure*}[tbp]
\begin{center}
\FigureFile(140mm,140mm){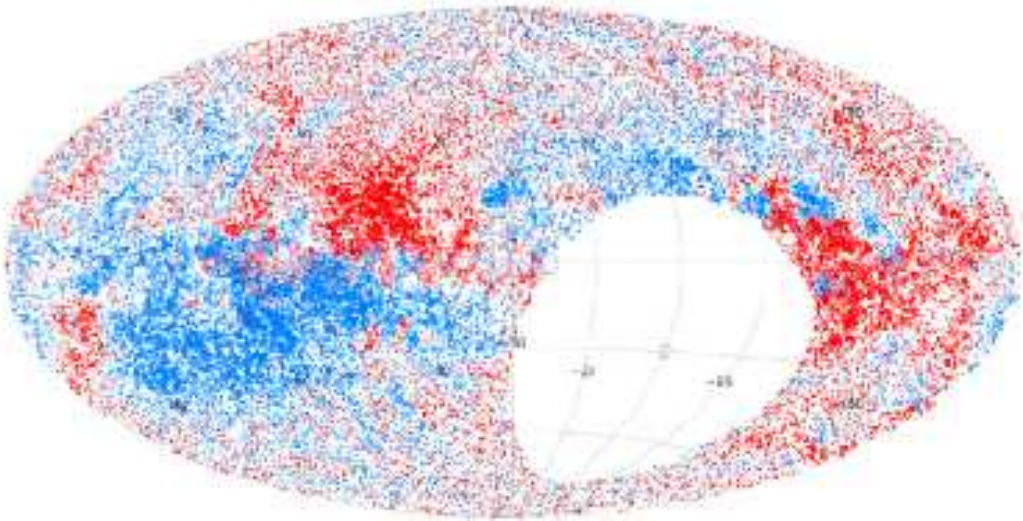}
\end{center}
\caption{
The longitude-latitude map of RMs obtained from the NRAO VLA Sky Survey (NVSS, \cite{2009ApJ...702.1230T}). The size and color of each circle shows the amplitude of RM and its sign (red: positive, blue: negative), respectively, of each extragalactic source (mostly radio galaxy or quasar, i.e. AGN).
}
\label{f09}
\end{figure*}

%%%%%%%%%%%%%%%%%%%%%%%%%%%%%%%%%%%%%%%%%%%%
%%%%%%%%%%%%%%%%%%%%%%%%%%%%%%%%%%%%%%%%%%%%
\subsection{Radio Classification of AGN and Jets}
\label{mag.s7.ss1}

AGN has been known as a radio source since Cygnus A had been detected at 160~MHz in 1944 \citep{1944ApJ...100..279R} and identified as a radio point source in 1948 \citep{1948AuSRA...1...58B}. In 1950--60's, radio source catalogs, known as 3C members, were complied by Cambridge \citep{1959MmRAS..68...37E, 1962MmRAS..68..163B}. The catalogs opened the door to the research field of AGN; generation mechanisms of the energy, objects responsible for the emission (supermassive black holes), and magnetic fields have been intensively studied. Radio sources 3C 273 and 3C 48 were discovered as enormously high-redshift sources in 1963 \citep{1963Natur.197.1037H, 1963Natur.197.1040S, 1963Natur.197.1040O, 1963Natur.197.1041G}, which means that they are very distant objects and are producing a huge amount of energy, and established a new classification ``quasars".

One of the classes of AGN is based on radio emission. AGN are classified into two groups, conventionally called radio-quiet and radio-loud. Radio-quiet AGN include low-ionization nuclear emission-line regions (LINERs), Seyfert galaxies, and radio-quiet quasars. Radio-loud AGN include radio galaxies, blazars, and radio-loud quasars. Details of the above classifications can be found in the literature \citep{2015ARA&A..53..365N}. We note that AGN's radio emission can be tightly related to magnetic fields. Therefore, the above classification based on radio emission can be influenced by magnetic fields.

AGN often exhibit collimated jets which emit synchrotron radiation over wide range in radio (\cite{2015A&A...579A..27S}, \cite{2016ApJ...817..131H}). \citet{1974MNRAS.167P..31F} measured the distance between two peaks of radio lobes, and suggested that galaxies possessing radio lobes can be classified into two classes based on $R_{\rm FR}$, the ratio of the distance to the extent of the contour of lowest brightness (FANAROFF-RILEY classification). Galaxies with $R_{\rm FR}<0.5$ and $R_{\rm FR}>0.5$ are classified into FR I and FR II, respectively. FR II galaxies have relatively larger radio luminosity than FR I galaxies. Jets from FR II galaxies have larger opening angles and have hot spots at their terminal. On the contrary, jets from FR I galaxies have small opening angles and no hot spot is found. Based on observations of synchrotron emission, typical magnetic-field strength inside radio lobes is 3--10~$\mu$G \citep{2007ApJ...668..203M}.

Based on the distribution of RMs and the intrinsic polarization angles, \citet{2002PASJ...54L..39A} pointed out helical magnetic fields operating along the jet in 3C273. A few tens of AGN possess such helical magnetic-field structures of jets \citep{2015MNRAS.450.2441G}. VLBI observations showed that a jet has cylindrical structure \citep{2014ApJ...785...53N}. Proper motions of jet nots were detected through VLBI monitoring observations and super luminous motions were found in some cases \citep{2014ApJ...781L...2A}. Recently, using ALMA, \citet{2015Sci...348..311M} found a strong polarization signal from the jet of a distant AGN, PKS 1830-211. They estimated magnetic fields of at least tens of Gauss on scales of $0.01~{\rm pc}$ from observed large RMs. \citet{2016ApJ...817..131H} reported a highly polarized ($\sim 20~\%$) feature at about 50 Schwarzschild radius from the core. Because depolarization is expected during the propagation along the LOS, they suggested that the intrinsic polarization degree must be higher than the observed one and hence a well-ordered magnetic field is present in this region. 

The current plausible scenarios of the formation of relativistic jets are (1) magnetic acceleration model \citep{1985PASJ...37..515U, 2002Sci...295.1688K} and (2) radiation driving model \citep{2004ApJ...601...78I, 2009ApJ...690L..81A, 2012ApJ...754..148T}. The origin of AGN activities and jets is considered to occur on the surface of an accretion disk surrounding a central super massive black hole. According to the standard theory of accretion disks, stable branch of accretion disks can be classified into the radiatively inefficient accretion flows (RIAF), the standard disk, and the slim disk \citep{1995ApJ...438L..37A}. Here, the RIAF describes an optically thin disk, and the latter two exhibit optically thick disks. The standard disks are thermal-pressure supported, while slim disks are supported by radiative pressure. A link between AGN activities and accretion disk state is that the low luminosity AGN, Seyfert, and narrow line Seyfert I possess a RIAF-like disk, a standard disk, and a slim disk, respectively. No drastic difference of magnetic-field strength and structure is reported among the three disks. But magnetic field plays an essential role in the accretion disk state in the sense that magnetic turbulence produced by MRI governs the efficiency of viscosity and the viscous energy (e.g., \cite{1991ApJ...376..214B}).

%%%%%%%%%%%%%%%%%%%%%%%%%%%%%%%%%%%%%%%%%%%%
%%%%%%%%%%%%%%%%%%%%%%%%%%%%%%%%%%%%%%%%%%%%
\subsection{Seyfert Galaxies and Magnetic Fields}
\label{mag.s7.ss2}

\begin{figure*}[tbp]
\begin{center}
\FigureFile(160mm,160mm){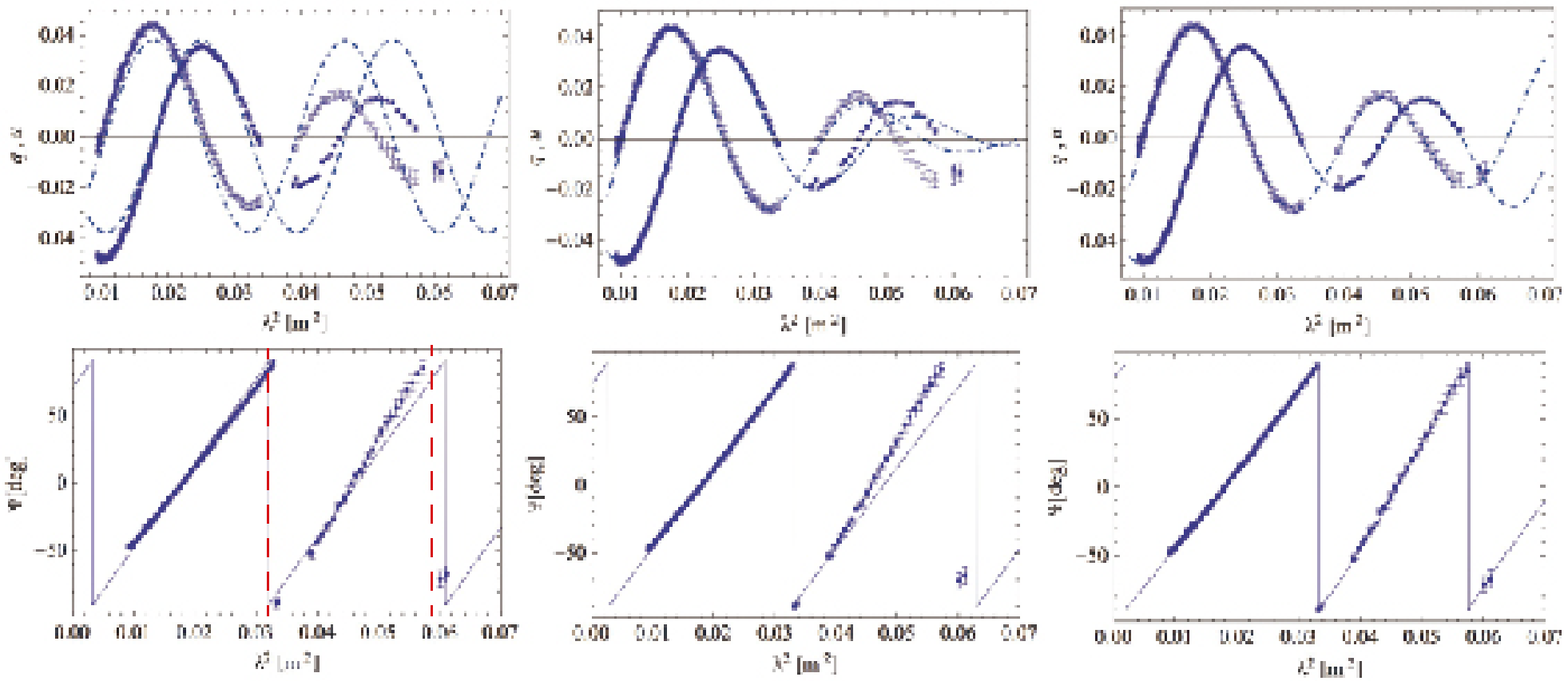}
\end{center}
\caption{
Polarization spectra of Quasar PKS B1610-771 \citep{2012MNRAS.421.3300O, 2015aska.confE.103G}. Top panels show normalized Stokes $q$ ($=Q/I$, open circles) and $u$ ($=U/I$, filled circles) and the bottom panels show the polarization angle $\Psi$ (filled circles) against the wavelength squared. Vertical red solid lines indicate the 350 MHz bandwidth centered at 1.4 GHz. Dot-dashed, dashed, and solid lines show the best-fits of $q$, $u$, and $\Psi$, respectively. Left, middle, and right panels show the results with a single RM component, a single depolarization screen, and two RM components, respectively.
}
\label{f10}
\end{figure*}

According to features of optical and ultraviolet absorption lines, some galaxies are classified into Seyfert Type I or II; spectra of Type I show both broad and narrow lines while spectra of Type II show only narrow lines. The broad lines have widths of up to $10^4$~km~s$^{-1}$ and the narrow lines show to much smaller velocities (a few $\times$ 100~km~s$^{-1}$). In addition, about 10--20~\% of quasars exhibit ``ultra fast outflow", blue-shifted absorption lines whose velocity is about 10,000~km~s$^{-1}$ with respect to the galaxy's rest frame and the line width is about 2,000~km~s$^{-1}$ \citep{1991ApJ...373...23W}. Similar blue-shifted absorption lines were also observed in X-rays \citep{2003MNRAS.345..705P, 2010A&A...521A..57T}. 

The ``unification scheme" of Seyfert galaxies is often cited to explain the difference between Type I and II \citep{1985ApJ...297..621A}. In this theory, high velocity clouds are moving around the supermassive black hole and they are responsible for the broad lines. A dusty molecular gas torus partially surrounds the ``broad line region". The difference between Type I and II is made by the direction of the AGN axis with respect to the LOS. In Type II the torus is inclined and blocks the observer's direct view to the broad line region. In type I the torus is almost face-on and the observer can see the broad line region. Radio-quiet quasars and QSOs are more luminous than Seyferts. Since the optical luminosity is so large that the host galaxy is not visible. Like Seyferts, quasars are classified as Type I and II. The spectra of Type I have both broad and narrow lines, while those of Type II have only narrow lines. In addition, so as to explain the ultra fast outflow, some theoretical models, e.g. the line-driven acceleration \citep{2016PASJ...68...16N} and the magnetic disk wind \citep{2016AN....337..454F}, have been proposed.

The polarization vector of type II Seyfert galaxy tends to be aligned along the rotation axis, and the polarization degree is significant, 1--10 \%. On the other hands, type I Seyfert galaxy shows low polarization degree (0.1--1 \%) and the polarization vector becomes perpendicular to the rotation axis. For example, \citet{2002MNRAS.335..773S} observed 36 Seyfert I galaxies with Willam Hershel and Anglo-Anstralian Telescopes to obtain the optical spectropolarimetry. They found that 20 out of the 36 Seyfert I galaxies exhibit linear polarization of the broad line region. \cite{2014MNRAS.441..551M} compiled 53 AGN from archival data, and suggested that the above tendency can originates from the inclination angle of the AGN.

The magnetic-field strength around the emission region is presumed to be about $O(10)$~G \citep{2015Sci...348..311M}. The azimuthal magnetic fields is considered to the dominant component \citep{2013AstBu..68...14S}. The method of the black hole mass estimation is proposed using the relation between the polarization degree and the inclination angle \citep{2015MNRAS.454.1157P}.

%%%%%%%%%%%%%%%%%%%%%%%%%%%%%%%%%%%%%%%%%%%%
%%%%%%%%%%%%%%%%%%%%%%%%%%%%%%%%%%%%%%%%%%%%
\subsection{Cosmological Evolution of AGN Magnetic Fields}
\label{mag.s7.ss4}

Figure~\ref{f09} shows the largest (37,543) RM catalog of extragalactic polarized sources to date \citep{2009ApJ...702.1230T}. Some of the sources are nearby ($z<0.2$) radio galaxies and others are quasars distributing in a wide range of redshift, $0 < z < 5$ \citep{1209.1438v3}. Using such a massive extragalactic RM catalog, a number of works have attempted to extract cosmological evolution of magnetic fields associated with AGN and quasars (e.g., \cite{2008ApJ...676...70K, 2014MNRAS.442.3329X}). Although these works still suffer from large errors of RM measurement, a current consensus would be that at least the standard deviation of the residual (observed $-$ Galactic foreground) RMs is constant in redshift at a $\sim 10$~rad~m$^{-2}$ level. Since contributions from DINGs and the IGM can be minor in nearby Universe, the standard deviation for nearby sources can be associated mostly with AGN, AGN host galaxies, or ambient media, and it decreases by $(1+z)^{-2}$ in a passive evolution scenario (\cite{2014ApJ...790..123A}). Note that careful collection of RM data is important because RM strongly depends on the frequency and the beam size due to depolarization (e.g. \cite{2012ApJ...761..144B, 2013ApJ...772L..28B, 2014ApJS..212...15F, 2014ApJ...795...63F}).

Wideband polarimetry is powerful for studying structures of AGN jets which are located too far to be resolved. Figure~\ref{f10} introduces the result for a distant quasar PKS B1610-771 \citep{2012MNRAS.421.3300O, 2015aska.confE.103G}. Vertical dashed lines indicates the 350 MHz bandwidth centered at 1.4 GHz. If the data only in this bandwidth is available, one may notice a simple linear relation between the polarization angle and $\lambda^2$, and obtain the best-fit of $RM = +135$~rad~m$^{-2}$ with a single RM component. However, full bandwidth data (1.1 -- 3.1 GHz) reveals that a single RM component does not fit the full data well (left panels) and remind us complex RM structures. An external Faraday dispersion depolarization model (equation~\ref{eq:EFD}) improves the fit at the short wavelengths but fails to adequately fit the long-wavelength data (middle panels). Accordingly, the full data can be nicely fitted with a two-component model with RMs of $+107.1 \pm 0.2$~rad~m$^{-2}$ and $+78.7 \pm 0.4$~rad~m$^{-2}$ (right panels). \citet{2012MNRAS.421.3300O} suggested that these components are polarized knots of unresolved jets.

%%%%%%%%%%%%%%%%%%%%%%%%%%%%%%%%%%%%%%%%%%%%
%%%%%%%%%%%%%%%%%%%%%%%%%%%%%%%%%%%%%%%%%%%%
%%%%%%%%%%%%%%%%%%%%%%%%%%%%%%%%%%%%%%%%%%%%
\section{Galaxy Clusters}
\label{mag.s8}

Information on cluster magnetic fields is obtained from observations of polarized radio sources inside or behind galaxy clusters. It is well-known that some galaxy clusters contain diffuse synchrotron sources which allow to estimate the IGMF in the ICM. In addition, since polarization from the polarized radio sources is affected by Faraday rotation due to thermal electrons within magnetic fields in the ICM, we can calculate RM and infer properties of the magnetic fields in the ICM. In this section, we review three types of diffuse synchrotron emissions from galaxy clusters, radio halo, radio mini-halo, and radio relic in this order, and summarize observations of RM toward galaxy clusters.

%%%%%%%%%%%%%%%%%%%%%%%%%%%%%%%%%%%%%%%%%%%%
%%%%%%%%%%%%%%%%%%%%%%%%%%%%%%%%%%%%%%%%%%%%
\subsection{Radio Halo}
\label{mag.s8.ss1}

Radio halos are diffuse synchrotron emission observed in the central region of galaxy clusters. According to the review by \citet{2012A&ARv..20...54F}, radio halos have low surface brightness ($\sim 0.1$ -- 1~$\mu$Jy/arcsec$^2$ at 1.4~GHz) and a steep radio spectrum ($\alpha > 1$). The linear size ranges from several hundred kpc to a few Mpc. A deep survey of X-ray selected galaxy clusters with the GMRT shows that only 30 $\%$ of clusters with $L_{\rm X} > 5 \times 10^{44}$ erg/s host radio halos \citep{2011JAA...32..519C}. Most radio halos are observed in clusters which are undergoing sub-cluster merging. Several properties from radio and X-ray observations can be interpreted in terms of a merging scenario as will be described below.

It is possible to constrain the magnetic-field strength in clusters with radio halos through observations of synchrotron radio and non-thermal X-ray radiation due to inverse Compton (IC) scattering of cosmic microwave background (CMB) photons (for reviews, \cite{2008SSR...134..71R, 2012RAA...12..973O}). If we detect non-thermal X-ray emission, we can estimate mean field strength for radio halo clusters using the relation between the synchrotron radio flux $F_{\rm radio}$ and the X-ray flux $F_{\rm ICX}$, $F_{\rm radio}/F_{\rm ICX} = U_{\rm mag} / U_{\rm CMB}$ where $U_{\rm mag}$ and $U_{\rm CMB}$ are the energy density of magnetic field and CMB photons, respectively. Suzaku has searched for non-thermal hard X-ray in several clusters hosting radio halos with the Hard X-ray Detector (HXD) which has low detector background and narrow field of view. 

The Coma radio halo, which has the flux density $F_\nu \sim 530$ mJy at $\nu=1.4$ GHz, is the well-known radio halo source. The magnetic-field strength is estimated to be a few~$\mu$G by using the RMs of the radio emission \citep{1990ApJ...355..29K}. \citet{2009ApJ...696..1700W} analyzed hard X-ray observations from the Coma cluster with Suzaku HXD-PIN and XMM-Newton data and failed to find statistically significant contribution from the non-thermal component. The upper limit they obtained on the flux of the non-thermal component is $F_X < 6 \times 10^{-12}$ erg s$^{-1}$ cm$^{-2}$ in the $20-80$ keV band, corresponding to the lower limit of the magnetic field of 0.15 $\mu$G. 

Abell 2319 possesses a giant radio halo ($F_\nu \sim 1$ Jy at 610 MHz, $\alpha \sim 0.92$) which is more powerful than that of the Coma cluster. In this cluster two subgroups are merging almost along the LOS with the velocity difference of $\sim 3000$ km s$^{-1}$. \citet{2009PASJ...61..1293S} found the upper limit of the IC component to be $F_X < 2.6 \times 10^{-11}$ erg s$^{-1}$ cm$^{-2}$ in the $10-40$ keV band, which means that $ B > 0.19~\mu$G. 

Abell 2163 is known as the hottest cluster and also hosts a powerful radio halo($F_\nu \sim 155$ mJy at 1.4 GHz, $\alpha \sim 1.18$). A complex structure of the hot gas temperature suggests that this cluster is undergoing sub-cluster merging. Weak lensing observation also supports the merging \citep{2011ApJ...741..116O}. \citet{2013A&A...562..60O} analyzed the hard X-ray spectrum and concluded that the observed spectrum is well explained by the multi temperature thermal emission model. The obtained upper limit is $F_X < 1.2 \times 10^{-11}$ erg s$^{-1}$ cm$^{-2}$ in the $12-60$ keV band, which means that $ B > 0.098~\mu$G.

The origin of CR electrons responsible for the radio halo emission is still under debate. Since the life time of the CR electrons, mainly due to the IC losses, is short ($\sim 10^7 - 10^8$ yr), CR electrons have to be injected or (re)accelerated in the clusters which possess radio halos (\cite{1999ApJ...520..529S, 2001MNRAS...320..365B}). Shock waves which propagate in the ICM during sub-cluster merging can accelerate CRs (e.g., \cite{2000ApJ...535..586T, 2001ApJ...562..233M, 2009MNRAS...395..1333V, 2016MNRAS...459..70V}). Numerical simulations show that MHD turbulence is developed during the sub-cluster merging (\cite{2008ApJ...687..951T, 2011A&A...529..17V, 2013ApJ...771..131B}). Driven MHD turbulence can also accelerate CRs (e.g., \cite{2002CosmicRayAstrophysicsS}). In addition, inelastic collisions between CR protons and thermal protons generate secondary CR electrons (e.g., \cite{1980ApJ...239L..93D, 1999APh...12..169B}). \citet{2014IJMP...23..1430007B} discuss these injection and acceleration mechanisms of CRs in detail.

Radio and X-ray observations for radio halo clusters indicate the correlation of $P_{1.4} \propto L_{\rm X}^s$, where $P_{1.4}$ is the monochromatic radio power at 1.4 GHz and $L_{\rm X}$ is the X-ray luminosity, and the index is $s \sim 2$. \citet{2006MNRAS...369..1577C} and \citet{2010A&A...517..10C} interpreted this slope by assuming a magnetic field dependence on the cluster mass. \citet{2007ApJ...670..L5B} found that upper limits of the radio power for clusters without radio halos lie about one order of magnitude below this correlation, and suggested the bimodal distribution in the $P_{1.4} - L_{\rm X}$ plane. Since radio halos are associated with merging clusters and clusters with the upper limits are relaxed clusters, \citet{2009A&A...507..661B} interpreted this bimodality in terms of the turbulent acceleration model during cluster merging as below. Cluster merging can supply energy both to the hot gas and CRs through shocks and/or turbulence. The CR electrons, which are accelerated by turbulence generated through cluster merging emit synchrotron radiation in radio halo clusters. In relaxed clusters after merging events, the CR electrons lose their energy and then radio halos disappear.

Several attempts to investigate MHD turbulence which explain the radio properties have been made based on the fact that the power spectrum of the MHD turbulence gives the acceleration efficiency and then the acceleration determines the synchrotron spectrum. MHD turbulence which consists of Alfv\'{e}n waves or magnetosonic waves has been assumed, because these waves accelerate CRs through resonant interaction (e.g., \cite{2001ApJ...557..560P, 2003ApJ...584..190F, 2004MNRAS...350..1174B, 2007MNRAS...378..245B, 2011MNRAS...410..127B}). \citet{2002ApJ...577..658O} constrained the power spectrum $P(k)$ so as to reproduce the radio spectrum of the Coma radio halo. They obtained $P(k) \propto k^{-2.8}$ which is steeper than the Kolmogorov power spectrum. Recently, \citet{2015ApJ...815..116F} proposed a model in which turbulence behind a shock reaccelerates CR electrons that had been weakly accelerated at the shock. They applied their model to the clusters 1RXS J0603.3+4214 (Toothbrush Cluster) and 1E0657-56 (Bullet Cluster) and found that the effective mean free path for efficient acceleration must be much smaller than the Coulomb mean free path.

The observed integrated spectra from several radio halos show spectral steepening at high frequency ($\nu \sim 1$ GHz) (\cite{2012A&ARv..20...54F}). Also, spatial distribution of the spectral index has been measured from several radio halos. We may investigate MHD turbulence through measurements of spectral steepening at high frequency and spatial variation of the radio spectrum combined with theoretical studies.

%%%%%%%%%%%%%%%%%%%%%%%%%%%%%%%%%%%%%%%%%%%%
%%%%%%%%%%%%%%%%%%%%%%%%%%%%%%%%%%%%%%%%%%%%
\subsection{Radio Mini-Halo}
\label{mag.s8.ss2}

Radio mini-halos are diffuse radio synchrotron emission that are found around the cores of galaxy clusters. They are often observed in non-merging clusters in contrast with cluster-scale giant radio halos \citep{2012A&ARv..20...54F,2014ApJ...781....9G}. This may indicate that the origin of the mini-halos is different from that of the giant radio halos. The mini-halos are generally dim, which has prevented us from understanding their origin.

Hadronic and leptonic models have been considered as the origin of mini-halos. In hadronic models, synchrotron emissions come from secondary electrons created through $pp$-interaction between CR protons and thermal protons in the ICM \citep{2004A&A...413...17P, 2007ApJ...663L..61F, 2010ApJ...722..737K, 2011ApJ...738..182F, 2011A&A...527A..99E}. In leptonic models, electrons are re-accelerated by turbulence in the ICM \citep{2002A&A...386..456G, 2013ApJ...762...78Z}.

For the hadronic models, the origin of protons are often thought to be the AGN in the core of the cluster. In fact, some studies indicate that jets launched by the AGN have a large fraction of CR protons \citep{2005ApJ...625...72S}. Moreover, CR protons may be efficiently accelerated in a RIAF disk of the AGN \citep{2015ApJ...806..159K}. Since most of the AGN at cluster centers at low redshifts are dim and are expected to have RIAFs \citep{2013MNRAS.432..530R}, a huge amount of CR protons may be flowing out of them in cluster cores. Those CR protons cannot move freely because they interact with magnetic fields in the ICM. Moreover, they are likely to excite Alfv\'en waves in the ICM through the streaming instability \citep{1975MNRAS.173..255S}. Since the CR protons are well scattered by the waves, the protons as a whole slowly move with the waves at the Alfv\'en velocity (CR streaming). The slow bulk motion of the CRs works as a $PdV$ work, for which the heating rate is proportional to the production of the Alfv\'en velocity and the gradient of CR pressure. Thus, CR streaming effectively heats the ICM and compensates radiative ICM cooling in the core \citep{1988ApJ...330..609B, 2008MNRAS.384..251G, 2013MNRAS.432.1434F, 2017MNRAS.467.1449J}. 

Since the above heating mechanism does not require turbulence, it is consistent with the results of Hitomi that showed only a low level of turbulence ($\sim 164\rm\: km\: s^{-1}$) in the core of the Perseus cluster \citep{2016Natur.535..117H}. At the same time, those CR protons create secondary electrons via $pp$-interaction, and they radiate synchrotron radio emissions. 

These process proceeds on a time scale of the Alfv\'en crossing time of the core:
\begin{eqnarray}
\label{eq:TA}
T_{\rm A} &=& L/v_{\rm A} \nonumber\\
&\sim& 4.5\times 10^8{\rm\: yr}\left(\frac{L}{100\rm\: kpc}\right)
\left(\frac{n}{10^{-2}\rm\: cm^{-2}}\right)^{1/2}
\left(\frac{B}{10\mu\rm\: G}\right)^{-1}\:,
\end{eqnarray}
where $L$ is the core size, $v_{\rm A}$ is the Alfv\'en velocity, $n$ is the ICM density, and $B$ is the magnetic-field strength. The cooling time of the CR protons by the $pp$-interaction is larger than $T_{\rm A}$ and thus the cooling can be ignored. On the other hand, the electrons created via the $pp$-interaction lose their energy by synchrotron radio emissions. Their cooling time is given by
\begin{equation}
\label{eq:Ts}
T_s\sim 2.4\times 10^7{\rm\: yr}\left(\frac{\gamma}{10^4}\right)^{-1}
\left(\frac{B}{10\mu\rm\: G}\right)^{-2}\:,
\end{equation}
where $\gamma$ is the Lorentz factor, and it is much smaller than $T_{\rm A}$. 

\begin{figure}[tbp]
\begin{center}
\FigureFile(80mm,80mm){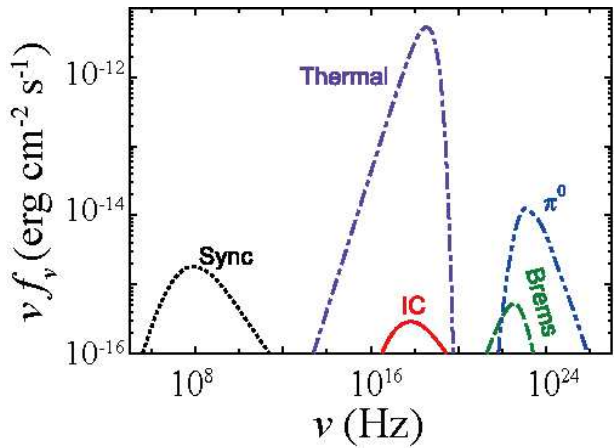}
\end{center}
\caption{
Predicted Broad-band spectra of RX~J1347--1145 based on the secondary CR scenario by \citet{2013MNRAS.428..599F}. Synchrotron radiation (dotted line), inverse Compton scattering off CMB (solid line), and non-thermal bremsstrahlung (dashed line) are of the secondary electrons. The $\pi^0$-decay gamma-rays are shown by the two-dot-dashed line. For comparison, the thermal bremsstrahlung from the ICM is shown by the dot-dashed line. Redshift has been corrected.
}\label{f11}
\end{figure}

These mean that the secondary electrons cool almost immediately after their birth ($\sim T_s$) while the CR protons injected by the central AGN can exist in the cluster core on a time scale of $\sim T_{\rm A}$. Thus, the synchrotron emissions from the secondary electrons should be generated near the site of the $pp$-interaction, which can be far away from the site of the proton injection (AGN). An prediction of broad band spectra of an hadronic model is shown in figure~\ref{f11}. In addition to the synchrotron emission, weak gamma-ray emission associated with $pp$-interaction should be generated, although it would be difficult to discriminate it from that from the central AGN in the near future.

In leptonic models, the short cooling time of the CR electrons (equation~\ref{eq:Ts}) means that the synchrotron emission is basically produced where the electrons are re-accelerated, because they do not have enough time to diffuse for a long distance. Thus, the synchrotron emission shows the position of turbulence. The turbulence that Hitomi has found in the core of the Perseus cluster may be strong enough to accelerate electrons to the energies required for the synchrotron emission \citep{2016Natur.535..117H}. However, ultimately, the spatial correlation between the synchrotron emission and turbulence must be confirmed to prove the leptonic models. The results of the Hitomi observations suggest that the turbulence in the Perseus cluster is not originated from the central AGN, because it is too weak to propagate from the central AGN. This may show that the turbulence is created via gas sloshing caused by minor cluster mergers \citep{2004ApJ...612L...9F, 2006ApJ...650..102A}.

%%%%%%%%%%%%%%%%%%%%%%%%%%%%%%%%%%%%%%%%%%%%
%%%%%%%%%%%%%%%%%%%%%%%%%%%%%%%%%%%%%%%%%%%%
\subsection{Radio Relic}
\label{mag.s8.ss3}

Radio relics are diffuse non-thermal synchrotron radio emitting regions, which are often found in the outskirts of merging clusters. They are typically arc-shaped and convex towards the outer regions of the cluster, whereas some of them show linear-shaped, or, knotty and irregular morphology \citep{2012A&ARv..20...54F}. Such variety of morphology likely indicates inhomogeneous distribution of CR electrons and/or magnetic fields, and could infer different formation processes. Their radio spectra show typically a power-law shape whose radio spectral index is $\alpha \sim 1$. However, some relics show significantly steeper spectra. In addition, a curved radio spectrum and spectral break are reported in recent detailed radio observations \citep{2013A&A...555..A110S, 2016A&A...455..2402S}.

\begin{figure}[tbp]
\begin{center}
\FigureFile(80mm,80mm){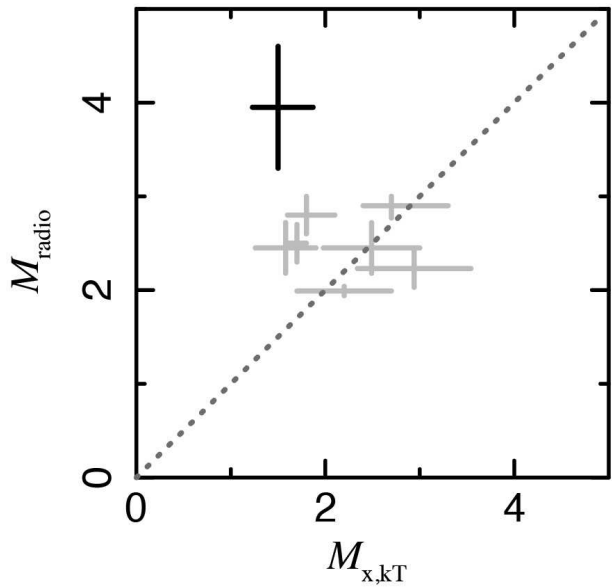}
\end{center}
\caption{
Mach numbers derived from the radio spectral index ($M_{\rm radio}$) plotted against those from the X-ray temperature measurements ($M_{X, kT}$) for 8 radio relics from \citet{2015PASJ...67..113I}. The gray dotted line represents $M_{\rm radio} = M_{X, kT}$. The result of ``toothbrush'' is shown by a black cross, which seems to be a rather extreme case. References for each relics are listed in \citet{2015PASJ...67..113I}.
}\label{f12}
\end{figure}

It is believed that CR electrons in radio relics are accelerated at shocks associated with cluster formation, which is consistent with the facts that shock structures are found in the ICM density and temperature distributions near the relics through X-ray observations \citep{2010ApJ...715..1143F, 2012PASJ...64..67A, 2013PASJ...65...16A} and that significant polarization degree is often observed in radio observations \citep{2010Sci...330..347V, 2012A&A...546A.124V, 2015PASJ...67..110O}. Diffusive shock acceleration (DSA) is the most promising particle acceleration process. Assuming a simple case of DSA, the Mach number of the shocks ($M_{\rm Radio}$) is estimated with the index ($\alpha_{\rm int}$) of integrated radio spectra as follows,
\begin{eqnarray}
M^2_{\rm Radio} = \frac{2 \alpha_{\rm int} + 2}{2 \alpha_{\rm int} - 2}.
\end{eqnarray}
On the other hand, X-ray observations of the ICM enable us to determine the Mach number ($M_{\rm X}$) through a temperature or density jump across the relics with the Rankine-Hugoniot conditions,
\begin{eqnarray}
\frac{T_2}{T_1} &=& \frac{5 M_{\rm X}^4 + 14 M_{\rm X}^2 - 3}{16 M_{\rm X}^2}, \\
\frac{\rho_2}{\rho_1} &=& \frac{4M_{\rm X}^2}{M_{\rm X}^2 + 3},
\end{eqnarray}
where $T_1$ and $T_2$ ($\rho_1$ and $\rho_2$) are the pre- and post-shock temperatures (densities), respectively, assuming that a specific heat ratio $\gamma$ is $5/3$. Both methods should lead to results consistent with each other if a simple DSA theory holds. \citet{2013PASJ...65...16A} is the first systematic study about this issue. The recent results from \citet{2015PASJ...67..113I} are shown in figure~\ref{f12}, where significant differences between $M_{\rm radio}$ and $M_{\rm X}$ are seen for some relics. This indicates some hints of particle acceleration process in the relics.

Non-thermal electrons attributed to radio relics also emit non-thermal hard X-rays through inverse Compton scattering of CMB photons. Comparison of synchrotron and inverse Compton fluxes enables us to estimate the magnetic-field strength. However, it is still difficult to detect such emission because thermal emission from the ICM is dominant in the X-ray band though relics are often in the outskirts where thermal emission is fainter. At present, only upper limits of the inverse Compton component and hence lower limits of the field strength are obtained \citep{2009ApJ...690..367A, 2010ApJ...725..1688A, 2009PASJ...61S..377K, 2009PASJ...61..339N, 2015PASJ...67..113I}.

In a theoretical model based on a simple DSA, index of integrated radio spectrum $\alpha_{\rm int}$ and that of just behind the shock $\alpha_{\rm inj}$, which reflect the CR electron energy spectrum just accelerated at the shocks, have a relation of $\alpha_{\rm int}=\alpha_{\rm inj} + 0.5$ because of synchrotron and inverse Compton cooling \citep{1999ApJ...520..529S}. However, recent detailed radio observations reveal that such a simple pictures cannot explain some relics at least. For example, spectral curvature is found in ``Sausage'' relic of CIZA J2242.8+5301 \citep{2013A&A...555..A110S}. In addition, ``toothbrush'' relic in 1RXS J0603.3+4214 shows spectral steepening in a higher frequency range,  which cannot be explained by the cooling \citep{2016A&A...455..2402S}. These facts as well as the Mach number discrepancy mentioned above mean that we need more elaborate theoretical modeling. For example, in a re-acceleration scenario \citep{2001MNRAS...320..365B}, where the electrons in the relic have already been accelerated once at shocks with a much higher Mach number such as virial shocks, the Mach number discrepancy could occur. Interplay between shock and turbulence acceleration is considered in \citet{2015ApJ...815..116F}.

%%%%%%%%%%%%%%%%%%%%%%%%%%%%%%%%%%%%%%%%%%%%
%%%%%%%%%%%%%%%%%%%%%%%%%%%%%%%%%%%%%%%%%%%%
\subsection{Cluster RM and Magnetic Turbulence}
\label{mag.s8.ss4}

Polarized emissions from radio sources inside or behind galaxy clusters mainly pass through three different components. Those are, the polarized radio source itself, the ICM, and the Milky Way (MW). Hence the total RM is the sum of RMs of them,
\begin{equation}
RM=RM_{\rm source}+RM_{\rm ICM}+RM_{\rm MW}.
\end{equation}

The first significant detection of the ICM RM was made by \citet{1982ApJ...252...81L}, using radio galaxies in dozens of clusters. They compared RMs of 12 radio galaxies seen in the inner part of the clusters with those of 46 radio galaxies seen in the outer part of the clusters, and found that the distribution of the RM values of the former population is broadened. In Abell 2319, \citet{1986A&A...156..386V} calculated the RMs of 10 radio sources inside and outside of the cluster core, and found that the RMs of the sources inside the cluster core show positive values, in contrast to those outside of the core. Both results indicate that the polarization is affected  by Faraday rotation in the ICM and clearly suggest the existence of IGMF.

\begin{figure}[tbp]
\begin{center}
\FigureFile(80mm,80mm){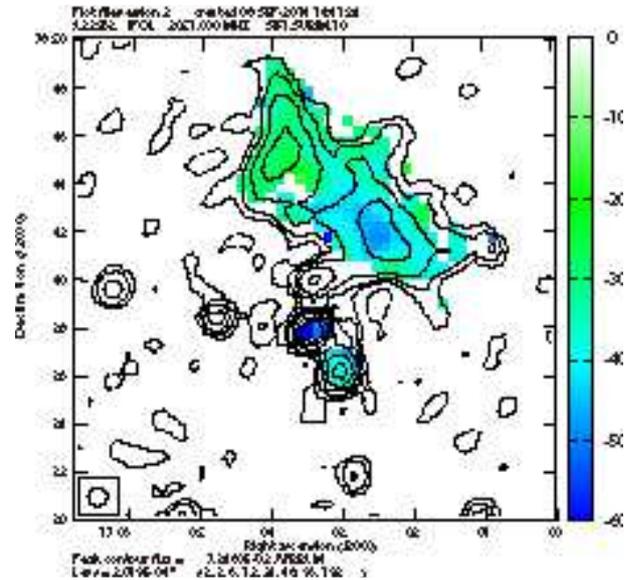}
\end{center}
\caption{
RM spatial distribution of Abell 2256 observed with the Karl G. Jansky Very Large Array (JVLA). Black contours show the total intensity at 2051 MHz. Northwest emission is a radio relic and two polarized radio sources are located near the center of the map.}
\label{f13}
\end{figure}

Thanks to high-sensitivity and high-resolution observation instruments, we can unveil spatial distribution of RM using individual polarized sources inside or behind clusters (see figure~\ref{f13}). High-resolution images discovered that RM spatial distribution is patchy and RM probability distribution is a Gaussian, indicating the existence of magnetic turbulence with scales of several kpc (e.g. \cite{2010A&A...513A..30B, 2010A&A...522A.105G, 2012A&A...540A..38V, 2015PASJ...67..110O}). Since the probability distribution indicates a non-zero mean RM, large scale magnetic fields would also be expected in addition to the small-scale magnetic fields due to turbulence.

Three-dimensional structure of the IGMF can be complicated due to the magnetic turbulence. In order to lead the profile of the magnetic turbulence, several authors analyzed the ICM RMs using a single scale cell model (e.g. \cite{1982ApJ...252...81L, 1991MNRAS.250..726T, 1995A&A...302..680F, 1996ASPCP...88...271, 2010A&A...522A.105G}). The model consists of a lot of cells with a uniform size, and each cell includes electrons with a uniform density and magnetic fields with a uniform strength with a single scale and a random direction. In this case, the RM probability distribution becomes a Gaussian with zero mean, and the variance of the RM in rad m$^{-2}$ is given by
\begin{equation}
\sigma_{\rm RM}^2=812^2 \Lambda_{\rm c} \int (n_{\rm e}B_{\parallel})^2 dl, \label{eq2}
\end{equation}
where $\Lambda_{\rm c}$ represents a single scale of the magnetic field in kpc, $n_e$ is the thermal electron density in cm$^{-3}$, and $B_\parallel$ is the magnetic field strengths along the LOS in $\mu$G, respectively. For the thermal electron density $n_{\rm e}$, we assume the $\beta$-model,
\begin{equation}
n_{\rm e}=n_0 \left( 1+\frac{r^2}{r_c^2} \right)^{-3\beta/2},
\end{equation}
where $n_0$ is the central electron density, $r$ is the distance from the X-ray center, and $r_c$ is the core radius of the ICM, respectively. Then, the equation (\ref{eq2}) is expressed as
\begin{equation}
\sigma_{\rm RM} = \frac{KBn_0 r_{\rm c}^{1/2} \Lambda_{\rm c}^{1/2}}{(1+r^2/r_{\rm c}^2)^{(6\beta-1)/4}} \sqrt{\frac{\Gamma(3\beta-0.5)}{\Gamma(3\beta)}},
\end{equation}
where $B=\sqrt{3B_\parallel}$ and $\Gamma$ represents the Gamma function. $K$ is the constant, which depends on the integration length over the thermal electron density distribution; $K=624$ if the radio source is located behind the cluster and $K=441$ if the radio source is located at a halfway of the cluster. Hence, we can estimate the magnetic-field strength by measuring the standard deviation of the RM if we assume the electron density of the ICM and $\Lambda_{\rm c}$. 

The above analysis leads reliable estimation if we assume the magnetic field correlation length $\Lambda_{\rm B}$ as $\Lambda_{\rm c}$ \citep{2004A&A...424..429M}. However, since the single scale cell model does not meet ${\rm div} B = 0$ and MHD simulations require wide-range fluctuation of the spatial scale of the magnetic turbulence \citep{2003A&A...401..835E, 2003A&A...412..373V}, a more realistic model is required. \citet{2004A&A...424..429M} developed a software FARADAY, which simulates a RM map from a three-dimensional electron density model and a multi-scale magnetic fields model. The model assumes random magnetic fields and a power-law power spectrum $|B_{k}|^{2} \propto k^{-n}$ with cutoffs at the minimum and maximum scales. It also assumes a correlation between the thermal electron density and the magnetic field strength. Comparison between a simulated RM map and a observed RM map can quantify the profile of the magnetic fields.  \citet{2008A&A...483..699G} analyzed magnetic fields in Abell 2382 using the FARADAY. They verified the model parameters in order to reproduce the observed RM map, and found that the magnetic field power spectrum index is the Kolmogorov one $n=11/3$, suggesting the presence of Kolmogorov-like magnetic turbulence.

The Kolmogorov-like magnetic turbulence is predicted in a MHD simulation performed by \citet{2008Sci...320..909R}. They proposed a scenario that seed magnetic fields are amplified by turbulent-flow motions in the IGM induced by the cascade of vorticity generated at cosmological shock waves during the large-scale structure formation. In this simulation, they found that the turbulence energy $\epsilon_{\rm turb}$ is converted into the magnetic energy $\epsilon_{\rm B}$ as
\begin{equation}
\epsilon_{\rm B} = \phi \left(\frac{t}{t_{\rm eddy}} \right)\epsilon_{\rm turb},
\end{equation}
where $\phi$ is the conversion factor and $t_{\rm eddy}$ is the eddy turnover time, respectively. The power spectra of the amplified magnetic fields roughly indicated the Kolmogorov like power spectra. Whatever the origin is, the magnetic fields should be affected by the turbulence in the ICM.

%%%%%%%%%%%%%%%%%%%%%%%%%%%%%%%%%%%%%%%%%%%%
%%%%%%%%%%%%%%%%%%%%%%%%%%%%%%%%%%%%%%%%%%%%
%%%%%%%%%%%%%%%%%%%%%%%%%%%%%%%%%%%%%%%%%%%%
\section{The Cosmic Web}
\label{mag.s9}

Magnetic fields appear whenever currents can be found. It means that magnetic fields potentially exist in the whole range of the Universe from scales of particle physics to cosmology. We expect that cosmological magnetic fields affect a variety of cosmological phenomena, big bang nucleosynthesis (BBN), the cosmic microwave background (CMB), the matter power spectrum (MPS), and the large-scale structure formation. In this section, we introduce several theories on the origin of cosmological magnetic fields, and their effects on the early Universe and the structure formation. We also summarize the current state of the art in the observations for cosmological magnetic fields.

%%%%%%%%%%%%%%%%%%%%%%%%%%%%%%%%%%%%%%%%%%%%
%%%%%%%%%%%%%%%%%%%%%%%%%%%%%%%%%%%%%%%%%%%%
\subsection{Primordial Magnetic Fields in Early Universe}
\label{mag.s9.ss1}

Can the early Universe create cosmological magnetic fields? Since we have not had any effective methods for observing primordial magnetic fields (PMFs) in the early Universe, it has been very hard to answer this question. However, many theorists have aggressively challenged to answer and break new ground in the modern cosmology. In this subsection, we introduce some plausible origins of PMFs in the early Universe (see \cite{2012SSRv..166....1R,2012SSRv..166...37W} for details).

It is difficult for the inflation models to provide electromagnetic quantum fluctuation, because the electromagnetic field is invariant in the conformal transformation. Therefore, generation and evolution of PMFs in the inflation era has been argued with an additional scalar field; the dilaton \citep{2004PhRvD..69d3507B} and the Higgs \citep{2004PhRvD..70d3004P}, and the field of gravity \citep{1988PhRvD..37.2743T}. These theories can explain the generations of coherent magnetic fields on cosmological scales. These fields, however, have too small amplitudes of $10^{-15}$ -- $10^{-25}$~G at the present.

When a magnetic field generated during the epoch of the inflation has a stronger energy density than the background inflaton energy density, this magnetic field destroys the cosmological homogeneity and isotropy. This is called the back-reaction problem of the PMF generation. The several Japanese researchers have contributed to building the generation theory of the PMF in consideration of the back-reaction problem \citep{2009JCAP...12..009K, 2012PhRvD..86b3512S,2014JCAP...03..013F}. \citet{2014JCAP...06..053F} indicated that the simplest gauge invariant models $f^2(\phi)F_{\mu\nu}F^{\mu\nu}$ suppress the generation of a PMF to $B\lesssim 10^{-30}$~G at Mpc. Such suppression can be avoided in PMFs generated by comic reheating \citep{2014JCAP...05..040K} and/or phase transitions \citep{1983PhRvL..51.1488H,1991PhLB..265..258V}.

The quantum chromodynamics and electroweak phase transition models \citep{2013PhRvD..87h3007K} can generate PMFs with an amplitude of $10^{-9}$~G on a coherence length of 50 kpc, and an amplitude of $10^{-10}$~G on 0.3 kpc, respectively, all at present ($z$=0). PMFs generated in the epoch of the inflation tend to affect the cosmological physical processes in larger scales, while those generated by phase transitions tend to affect those in smaller scales.

Finally, in the cosmological recombination era ($z\sim 1100$), there is also a significant generation process of magnetic fields \citep{2005PhRvL..95l1301T,2006Sci...311..827I}. The external force from CMB photons through Thomson scattering is exerted more preferentially on electrons than protons, which would separate electrons and protons in the primordial plasma. In cosmological timescale, such external force is balanced with the force from the electric fields induced by the charge separation, and these electric field generate magnetic fields through Maxwell equations.

%%%%%%%%%%%%%%%%%%%%%%%%%%%%%%%%%%%%%%%%%%%%
%%%%%%%%%%%%%%%%%%%%%%%%%%%%%%%%%%%%%%%%%%%%
\subsection{Impact of Primordial Fields on the Present Universe}
\label{mag.s9.ss2}

%%%%%%%%%%%%%%%%%%%%%%%%%%%%%%%%%%%%%%%%%%%%
\subsubsection{Big Bang Nucleosynthesis}
\label{mag.s9.ss2.sss1}

The abundances of the light elements in the Universe are well reproduced by the standard big bang nucleosynthesis (BBN) theory, except the $^7$Li abundance which is inconsistent with the observed one on the surface of metal-poor halo stars. Many researchers have been exploring the solutions for this ``$^7$Li problem". Previous studies have proposed three effective solutions to solve the $^7$Li problem; one of them focus on the PMF.

The first is to consider the photon cooling. In this model, the predicted baryon-to-photon ratio ($\eta_\mathrm{BBN}$) is smaller by a factor of $(2/3)^{3/4}$ before the end of the BBN epoch. In this case, $\eta_\mathrm{BBN} = (4.57\pm 0.10 )\times 10^{-10}$, and the D, $^3$He, and $^6$Li abundances are raised, while $^4$He and $^7$Li abundances are suppressed. The $^7$Li abundance predicted by the photon cooling model is consistent with the observed results, except that the D abundance is overproduced \citep{2012PhRvD..85f3520E}.

The second is to consider the decay of a long-lived $X$ particle. Non-thermal photons are produced by the radiative decay of a long-lived massive $X$ particle after the end of the BBN \citep{1979MNRAS.188P..15L, 1985NuPhB.259..175E, 1995PThPh..93..879K, 1995ApJ...452..506K}. The nuclei of $^7$Be and D produced in BBN are disintegrated by the nonthermal photons \citep{1979MNRAS.188P..15L, 1985NuPhB.259..175E, 2003PhRvD..67j3521C, 2005PhRvD..71h3502, 2006PhRvD..74j3509J, 2006PhRvD..74b3526K}. Considering the photon cooling and the decay particle model simultaneously (a previous hybrid model), we may solve the problem of deuterium overproduction.

The third is to consider the PMF. The cosmic expansion becomes faster if we consider the ensemble energy density of the PMF, $\rho_B$. Also, the weak reactions can freeze out earlier. Consequently, the neutron abundance increases. Since the faster cosmic expansion, leads to shorter time interval from the freeze-out to $^4$He production, more neutrons can survive from the $\beta$-decay before the epoch of $^4$He production. Therefore, the energy density of the PMF increases the $^4$He abundance significantly. On the other hand, the D and $^3$He abundances rise moderately, the $^6$Li abundance rises slightly, and the $^7$Li abundance is suppressed \citep{2012arXiv1204.6164K, 2012PhRvD..86l3006Y, 2014PhRvD..90b3001Y}.

\citet{2014PhRvD..90b3001Y} extended the standard BBN to a new hybrid BBN by taking into account the above three possible effects simultaneously. They used the maximum likelihood analysis to constrain the energy density of the PMF by the observed abundances of light elements up to Li. They found that the BBN model with a PMF gives a better likelihood than without a PMF, and the best-fit PMF energy density is given by
\begin{equation}
\rho_B (a) = \frac{\rho_B(a_0)}{a^4} = 6.82\times10^{-52} a^{-4} \mathrm{GeV}^4,
\end{equation}
where $a$ is the scale factor and $a_0 =1$ is the present value. This best-fit value corresponds to $B(a)=1.89a^{-2}\mu$G, where $\rho_B=B^2/8\pi =1.9084 \times 10^{-40}\times (B/1~\mathrm{G}$) GeV$^4$. An upper bound on the PMF energy density is also obtained, $\rho_B (a) < 1.45\times 10^{-51} a^{-4} \mathrm{GeV}^4$ (95\%~C.L.). This upper bound corresponds to $B(a) < 3.05a^{-2}\mu$G (95~\%~C.L.).

%%%%%%%%%%%%%%%%%%%%%%%%%%%%%%%%%%%%%%%%%%%%
\subsubsection{CMB and MSP}
\label{mag.s9.ss2.sss2}

The CMB temperature fluctuations and polarization anisotropies give the information of the cosmological parameters, e.g. the baryon and the dark matter abundances, the age of the Universe, the epoch of the reionization, the neutrino mass \citep{2000ApJ...538..473L, 2002PhRvD..66j3511L, 2013ApJS..208...19H, 2016A&A...594A..13P}. The matter power spectrum (MPS) shows the spatial distribution of the matter density fluctuation in the Universe. The time evolutions of MPS can be reproduced by theoretical estimations \citep{2000ApJ...538..473L} and observing the distributions of galaxies and clusters of galaxies with respect to each redshift $z$ \citep{2005MNRAS.362..505C, 2006PhRvD..74l3507T}. Since we can also extract the important information of the evolutions of the large-scale structures and the cosmological parameters by the MPS \citep{2002PhRvD..66j3511L, 2005MNRAS.362..505C, 2006PhRvD..74l3507T}, we can obtain the constrained cosmological parameters with a better likelihood using the CMB and MPS observation data simultaneously.

There are two main PMF effects in the cosmological linear perturbation theory. The first is the perturbative PMF effects \citep{2005ApJ...625L...1Y, 2006ApJ...646..719Y, 2010PhRvD..81b3008Y, 2012PhR...517..141Y}. Those effects make the other fluctuations of the CMB and the MPS. The strengths of the perturbative PMF at each wavenumber $k$ are given by the power law spectrum $\langle B(k)B^\ast(k)\rangle \propto k^{n_\mathrm{B}}$, where $ n_\mathrm{B}$ is the power spectrum index of the PMF. We assume that the local energy densities of the perturbative PMF comparable to or are less than the energy densities of the perturbative CMB photons at each $k$. The perturbative PMFs produce other perturbations of the ionized baryons by the Lorentz force before the recombination era. Since those baryons affect the CMB photons through Thomson scattering, and they also affect the dark matter through the gravity, the perturbative PMF can also produce the density fluctuations of the CMB photons and the dark matter indirectly \citep{2005ApJ...625L...1Y, 2006ApJ...646..719Y, 2010PhRvD..81b3008Y}. Furthermore, the perturbative PMF makes the polarization isotropies of the CMB \citep{2006ApJ...646..719Y, 2008PhRvD..77d3005Y}. The perturbative PMF effects tend to contribute to the CMB and the MPS in smaller-scale regions, e. g. wavenumber $k > 0.1h$~Mpc$^{-1}$, where $h$ is the Hubble parameter \citep{2006PhRvD..74l3518Y, 2008PhRvD..77d3005Y}.

\begin{figure}[tbp]
\begin{center}
\FigureFile(80mm,80mm){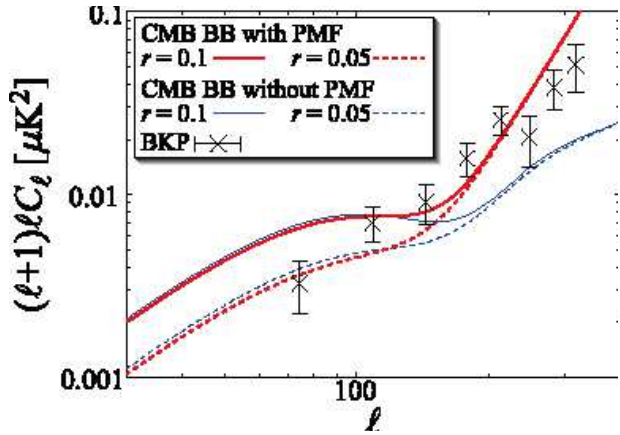}
\end{center}
\caption{
The total effects of the PMF on the CMB BB mode. In this figure, the power spectral index, $n_\mathrm{B}$ and the amplitude of The PMF, $B_\lambda$ are $(n_\mathrm{B}, B_\lambda) = (0.0, 3.0~\mathrm{nG})$. The bold and bold dotted curves  are the theoretical results with PMF on $r=0.1$ and $0.05$, respectively. The thin and thin dotted curves are the theoretical results without the PMF on $r=0.1$ and $0.05$, respectively. The points with the error bars are analysis results from \citep{2015PhRvL.114j1301B}.
}
\label{f14}
\end{figure}

The second is the background PMF effects. The PMF given by the power law spectrum has the background (ensemble) energy density $\rho_B$ as follows \citep{2014PhRvD..89j3528Y, 2016PhRvD..93d3004Y},
\begin{eqnarray}
\rho_B
\sim
\frac{1}{8\pi a^4}
\frac{
B^2_\lambda
}
{
  \Gamma
  \left(
     \frac{n_\mathrm{B}+5}{2}
  \right)
}
(\lambda k_\mathrm{max})^{n_\mathrm{B}+3},
\label{eq:BG_PL_PMF_energy_densityII}
\end{eqnarray}
where $B_\lambda=|\mathbf{B}_\lambda|$ is the comoving field strength by smoothing over Gaussian sphere of radius $\lambda=1$~Mpc ($k_\lambda = 2\pi/\lambda$), $\Gamma(x)$ is the gamma function, $k_\mathrm{max}$ is the cut off scale and it is defined by the PMF damping \citep{1998PhRvD..57.3264J, 1998PhRvD..58h3502S, 2002PhRvD..65l3004M}. The background PMF effects due to $\rho_B$ change the features of the CMB and the MPS through the Meszaros \citep{1974A&A....37..225M} and the potential damping \citep{1995ApJ...455....7M} effects. The CMB temperature fluctuations are suppressed around the 1st peak and their peaks are shifted to larger scales (smaller multipoles) by the background PMF \citep{2014PhRvD..89j3528Y}. The background PMF effects also suppress the amplitudes of the MPS in $k>0.02$~ Mpc$^{-1}$ \citep{2016PhRvD..93d3004Y}.

The CMB polarization anisotropies have the odd parity (curl) component. This is called the ``B" mode. The B mode of the CMB polarizations provide important information of the background gravitational wave, the inflation theory, and the reionizations. The B mode is affected by the weak lensing effects, which can be derived by the MPS. Since the PMF affects the MPS as mentioned above \citep{2016PhRvD..93d3004Y}, the PMF affects the B mode as figure~\ref{f14}.

%%%%%%%%%%%%%%%%%%%%%%%%%%%%%%%%%%%%%%%%%%%%
%%%%%%%%%%%%%%%%%%%%%%%%%%%%%%%%%%%%%%%%%%%%
\subsection{Structure Formation and Magnetic Fields}
\label{mag.s9.ss3}

%%%%%%%%%%%%%%%%%%%%%%%%%%%%%%%%%%%%%%%%%%%%
\subsubsection{Epoch of Reionization}
\label{mag.s9.ss3.sss1}

After recombination, the gas in the Universe had become almost neutral with the degree of ionization $x_e \sim 10^{-4}$. As the structure of the Universe evolved through gravitational instability, the gas was ionized by first luminous sources, and the electromagnetic action has become effective again.  Possible mechanisms for generating magnetic fields during this epoch have been intensively discussed in the literature. These are mainly based on two effects.

One is the anisotropic radiation pressure preferentially exerted on free electrons, which would separate positive and negative charges in the IGM. Anisotropic UV radiation fields are naturally realized in this epoch because neutral hydrogen clouds can absorb UV photons effectively, and therefore the magnetic fields are mostly generated behind the clouds \citep{2005A&A...443..367L,2010ApJ...716.1566A}. In cosmological timescale, the force of pressure gradient caused by UV photons from first-generation stars as well as CRs from supernovae explosions are balanced with the force of electric fields induced by the charge separation, and so the rotation component of those fields  induces magnetic fields \citep{2003PhRvD..67d3505L, 2004MNRAS.350..761C, 2005A&A...443..367L, 2010ApJ...716.1566A, 2011ApJ...729...73M, 2014ApJ...782..108S, 2015MNRAS.453..345D}. The field amplitude through these mechanisms in the IGM ranges from $\sim 10^{-19}$ G to $\sim 10^{-16}$ G, depending on the model parameters. Stronger fields can be locally expected near luminous sources. Based on radiation hydrodynamics simulations of proto-first-star formation, \citet{2015MNRAS.453..345D} found that $\sim 10^{-9}$ G fields are generated on the surface of the accretion disk around a proto-first-star. These fields would be blown out from the disk and diffuse into the low density regions, which may affect the next generation star formation activities \citep{2015MNRAS.453..345D}.

The other is based on the Biermann battery effect \citep{2000ApJ...539..505G,2011ApJ...741...93D, 2014ApJ...782..108S}.  The effect is expressed as 
$\dot{\bf{B}}\sim (ck_{\rm B}/en_{\rm e}) \nabla T_{\rm e} \times \nabla n_{\rm e}$, and therefore, it takes place under non-adiabatic conditions where directions of temperature and density gradients are different from each other. Typical sites are ionization fronts at the vicinities of protogalaxies and those propagating through the high-density filaments of the cosmological large-scale structure. \citet{2000ApJ...539..505G} estimated the amplitude of magnetic fields through the Biermann battery effect at the epoch of reionization by post-processing their reionization simulation data. They found that magnetic fields can be as large as $10^{-19}$ G by $z\sim 5$, and those fields are highly ordered on megaparsec scales. The field amplitude should be considered as a lower limit because of a finite spatial resolution of the simulation. These fields may be responsible for the magnetic fields in the IGM whose existence has been suggested from the recent gamma-ray observations \citep{2010Sci...328...73N}.

%%%%%%%%%%%%%%%%%%%%%%%%%%%%%%%%%%%%%%%%%%%%
\subsubsection{Shocks and Magnetic Turbulence}
\label{mag.s9.ss3.sss2}

Cosmological shock waves in the structure formation can generate seed magnetic fields by the Biermann battery \citep{1998A&A...335...19R}, the Wibel instability \citep{2003ApJ...599..964O}, and plasma instability \citep{2005MNRAS.364..247F}. \citet{2012Natur.481..480G} carried out plasma experiments for shock waves in galaxy clusters and claimed that the Biermann buttery effect can be an influential candidate of the seed magnetic fields. Seed magnetic fields of any origins could be further amplified through compression and eddy cascading (turbulence dynamo) in the cosmic web (e.g., \cite{2008Sci...320..909R, 2008A&A...482L..13D}). Therefore, the cosmic web is thought to be filled with the intergalactic magnetic field (IGMF). 

In galaxy clusters, it is well-studied that the IGMF attracts CRs and causes synchrotron radio emission (see \S\ref{mag.s8}). The IGMF, particularly magnetic turbulence, is important to understand thermal balance in cool-core clusters (e.g., \cite{2012ApJ...746...53F, 2013MNRAS.428..599F}) and structures of merging clusters (e.g., \cite{2005AdSpR..36..636A}, \cite{2008ApJ...687..951T}).

Meanwhile, the IGMF in filaments of galaxies is not well-known. A conservative range of the IGMF strength in filaments is $O(1$ -- $100)$~nG \citep{2008Sci...320..909R}. The amplitude, growth timescale, and characteristic scale of the IGMF was studied by means of MHD turbulence simulations \citep{2009ApJ...705L..90C}. It was suggested that turbulence is in the linear-growth and the saturation stages in filaments and clusters, respectively, based on the expected eddy turnover time. The integral scale, which is one of possible quantifications of the coherence length of magnetic fields, is 1/15 and 1/5 of the energy injection scales in filaments and galaxy clusters, respectively. Supposing the electron density of $10^{-5}$ cm$^{-3}$, the rms magnetic field strength of 300~nG, the integral scale of 300~kpc, and the depth of a filament to be 5~Mpc, the standard deviation of RM through a filament is estimated to be $\sim 1.5$~rad~m$^{-2}$.

Very recently, Hitomo observed the core region of Perseus galaxy cluster \citep{2016Natur.535..117H} and discovered the turbulence velocity of $\sim 150$~${\rm km/s}$ at $\sim 10$~kpc scales (the inertial range of turbulence expected in galaxy clusters). Such a velocity seems to be smaller than that expected (e.g. Heinz, et al. 2010, ApJ, 708, 462), implying a relatively low Raynolds number ($Re \sim 100$--$200$), i.e. high viscosity, of the IGM (\cite{2005MNRAS.357..242R}; see also \cite{2007MNRAS.378..662R}). Examining such key turbulence parameters (see Section 1.4) is crucially important to understand the significance of turbulence dynamo in the large-scale structure formation.

%%%%%%%%%%%%%%%%%%%%%%%%%%%%%%%%%%%%%%%%%%%%
\subsubsection{Galaxies}
\label{mag.s9.ss3.sss3}

Physical processes in the early Universe are not the only possibilities to generate the IGMF, but large-scale outflows from magnetized galaxies at late times can be an alternative (\cite{1987QJRAS..28..197R}; \cite{2013A&ARv..21...62D}). Magnetic fields in galaxies can be transported to outside of them in some ways. For instance, \citet{1997ApJ...477..560E} had analytic discussion whether radio galaxies in clusters can inject magnetic fields to the ICM, and found that fields can be transported to the ICM by jets within a radius of 1 Mpc. \citet{2001ApJ...556..619F} studied the possibility that the quasar outflow pollutes the IGM with magnetic field. They found that magnetized quasar bubble can further expand after the quasar activity has ceased due to overpressure, and that about 5 \% - 20 \% of the IGM volume is filled with magnetic field by $z\sim 3$, which strength is the level of 10\% of the thermal energy density of the IGM with the temperature of $10^4$ K. 

\citet{2001ApJ...560..178K} analytically demonstrated how magnetic fields ejected by AGN contribute to the IGM for the two cases: cluster embedded AGN and that outside of clusters. They suggested that 10 - 100 former type AGN can transport sufficient magnetic energy to the level observationally found ($\sim 10^{61}$ erg) beyond the inner cores of clusters, and that the expansion of lobes from the latter type AGN can magnetize a significant fraction of the IGM. \citet{2009ApJ...698L..14X} and \citet{2012ApJ...759...40X} performed cosmological MHD simulations and showed that the magnetic fields ejected by AGN can be transported throughout clusters. The radial profiles of the synthetic $|{\rm RM}|$ and $\sigma_{\rm RM}$ made from the simulation agree with observations.

Magnetic fields spread into the IGM by supernovae-driven galactic winds might be another possibility \citep{1999ApJ...511...56K}. \citet{2000ApJ...541...88V} analytically estimated that typical field strength in clusters can be $\geq10^{-7}$G by magnetized galactic winds. \citet{2009MNRAS.392.1008D} extended a semi-analytic study by \citet{2006MNRAS.370..319B} and performed cosmological MHD simulations to study the contribution of galactic outflows to the ICM magnetic fields. They showed that the galactic outflows can explain the properties of the $\mu$G scale fields observed in galaxy clusters.

The magnetization of void by galaxies in the vold has also been studied. \citet{2006MNRAS.370..319B} evaluated the effect of galactic winds to the IGM in voids and found that most regions affected by the winds have magnetic fields between $10^{-12}$ to $10^{-8}$ G. \citet{2013MNRAS.429L..60B} analytically discussed the transport of magnetic energy by CRs from galaxies and contributions of strong AGN bordering the voids. They argued that the lower limit on the void magnetic field claimed from observational results (e.g., \cite{2010Sci...328...73N}; \cite{2012ApJ...744L...7T}), $\gtrsim10^{-15}$ G, can be recovered by the galaxies and AGN.

%%%%%%%%%%%%%%%%%%%%%%%%%%%%%%%%%%%%%%%%%%%%
%%%%%%%%%%%%%%%%%%%%%%%%%%%%%%%%%%%%%%%%%%%%
\subsection{Can We Observe the Cosmic Web?}
\label{mag.s9.ss4}

%%%%%%%%%%%%%%%%%%%%%%%%%%%%%%%%%%%%%%%%%%%%
\subsubsection{RM Grids}
\label{mag.s9.ss4.sss1}

The IGMF strength of $B \sim 1-100$ nG is expected from cosmological simulations (e.g. \cite{2008Sci...320..909R}). Such a small field gives the rms value of RM of only $\sim 1$~rad~m$^{-2}$ for a single filament \citep{2010ApJ...723..476A}. The smallness is the main reason of the lack of the IGMF detection. However, \citet{2011ApJ...738..134A} demonstrated that the RM due to the IGMF for several filaments up to $z \sim$ a few becomes several rad~m$^{-2}$, which is comparable to RMs associated with the Milky Way (\S\ref{mag.s5}), DINGs (\S\ref{mag.s6}), and AGN (\S\ref{mag.s7}). 

The RM grid approach can be a powerful tool to prove such a small fields statistically \citep{2015A&A...575A.118O, 2015aska.confE..92J}. \citet{2014ApJ...790..123A} developed a method to estimate the statistics of RMs due to the IGMF from observed RMs. The simulation predicts that the IGMF in filaments would be distinguished from the GMF by $S_{\rm 2,RM}$ (see \S\ref{mag.s5.ss2.sss2}) which can be available with using RM grids in the SKA era. Similarly, the Bayesian approach allows us to detect $\sim$ nG magnetic fields using the SKA \citep{2016A&A...591A..13V}. 

Linearly-polarized fast radio bursts (FRBs) would be a new probe of the IGMF in filaments. \citet{2016ApJ...824..105A} pointed out that the classical estimator of the mean magnetic-field strength, $B_{||}^\dagger \sim RM/DM$, is incorrect in the cosmological context. They suggested to use $B_{||}^\ddagger \sim \langle 1+z \rangle B_{||}^\dagger/f_{\rm DM}$, where $\langle 1+z \rangle$ is the redshift of intervening gas weighted by the gas density in filaments, $f_{\rm DM}$ is the fraction of total DM due to the gas in filaments, and the redshift of the FRB is not required to be known. Recently, \citet{2016Sci...354.1249R} reported from FRB 150807 analysis that the net magnetization of the cosmic web along the LOS is constrained to be $<$ 21 $\mu$G under inference of the negligible magnetization of plasma around the source.

%%%%%%%%%%%%%%%%%%%%%%%%%%%%%%%%%%%%%%%%%%%%
\subsubsection{Faraday Tomography}
\label{mag.s9.ss4.sss2}

\begin{figure}[tbp]
\begin{center}
\FigureFile(80mm,80mm){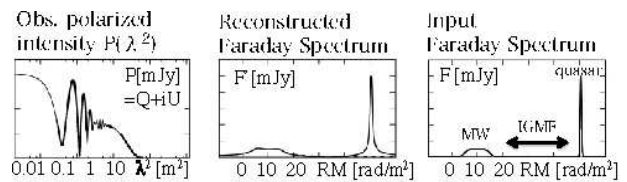}
\end{center}
\caption{
Exploring the IGMF by means of Faraday Tomography. (left) the mock polarized intensity spectrum, (middle) the reconstructed Faraday spectrum with the SKA2, (right) the input Faraday spectrum. see \citet{2014PASJ...66...65A} for details.
}
\label{f15}
\end{figure}

There are few attempts to explore the possibility to prove the IGMF in the cosmic web with direct usage of the Faraday tomography. \citet{2014PASJ...66...65A} first investigated the potential of the RM synthesis to find the IGMF. They developed a strategy where we observe two polarized sources along a LOS and tried to find the IGMF as a ``gap" between the two sources in Faraday depth space (figure~\ref{f15}). They found that the gap is detectable with the SKA if the RM due to the IGMF is $\sim$ 10~rad~m$^{-2}$. 

\citet{2014PASJ...66....5I} followed the strategy of \citet{2014PASJ...66...65A} and studied the potential of the QU-fitting technique to find the IGMF. They found that the IGMF with RM of $\sim$ 3~rad~m$^{-2}$ can be detected with the combination of the radio telescopes, ASKAP, LOFAR and GMRT. The strategy they adopted may be a rare situation, but we may find LOSs with such situation in the SKA era when we would obtain a large amount of polarized sources like a few hundreds of them per square degree \citep{2015aska.confE..92J}. Since robust detection of real components is essential to find the gap, it is also important to develop reliable Faraday tomography codes (e.g. \cite{2016PASJ...68...44M}).

%%%%%%%%%%%%%%%%%%%%%%%%%%%%%%%%%%%%%%%%%%%%
\subsubsection{Cross Correlation}
\label{mag.s9.ss4.sss3}

There are few works studying the cross correlation (CC) between synchrotron emission and large scale structure (LSS). For instance, \citet{2010MNRAS.402....2B} performed a CC between synchrotron emission observed with 1.4 GHz Bonn survey with galaxies as a tracer of the LSS for two redshifts slices ($0.03 < z < 0.04$ and $0.06 < z < 0.07$) using a $34^\circ\times34^\circ$ area of Two-Micron All-Sky-Survey (2MASS). Also, very recently, \citet{2017MNRAS.467.4914V} explored the CC using MWA for synchrotron emission and 2MASS and Widefield InfraRed Explorer for tracers of the LSS using a $21.76^\circ\times21.76^\circ$ area. Both works are turned out to be null correlation. 

The assumption of perfect correspondence between the distribution of galaxies and IGM synchrotron radiation is, however, oversimplification as mentioned in the two papers. Indeed, it has been reported that the synchrotron radiation associates with shock in LSS, not with matter structure, basically because electrons emitting synchrotron are accelerated at shocks via DSA process (e.g. \cite{2015ApJ...812...49H}). On the other hand, since it is known that RM associates with matter structure in LSS (\cite{2010ApJ...723..476A,2011ApJ...738..134A}), a CC between RM and LSS can be promising, as long as the removal of the Galactic foreground can be achieved \citep{2010MNRAS.408..684S}.

%%%%%%%%%%%%%%%%%%%%%%%%%%%%%%%%%%%%%%%%%%%%
%%%%%%%%%%%%%%%%%%%%%%%%%%%%%%%%%%%%%%%%%%%%
%%%%%%%%%%%%%%%%%%%%%%%%%%%%%%%%%%%%%%%%%%%%
\section{Summary}
\label{mag.s10}

The origin and evolution of cosmic magnetism is one of the fundamental questions in modern astrophysics. MHD simulations of global and turbulent magnetic fields have advanced our understanding of physics underling cosmic magnetism. Centimeter and meter radio observations of synchrotron emission and Faraday rotation measure are one of only the few established methods which allow us to measure magnetic fields in both galactic and extragalactic objects. In this review paper, we have reviewed progress of cosmic magnetism study mainly driven by centimeter and meter radio observations. We have highlighted magnetic fields in the ISM, the Milky Way Galaxy, external galaxies, AGN and jets, clusters of galaxies, and the cosmic web.

There are three key points in future studies of cosmic magnetism by centimeter and meter radio observations. The first one is to understand total radio intensity more precisely. The intensity gives basic information of magnetic-field strength. Moreover, it's spectrum provides information of CR energy distribution and will give a clue of understanding particle acceleration mechanisms. The second one is to understand depolarization. Depolarization is tightly related to magnetic-field structure, so that it will give a hint of the nature of both the regular and random magnetic fields. The third one is to understand Faraday tomography. Faraday tomography has the potential to deproject the LOS structure of magnetized plasma, so that it will give new discoveries of unknown such as the IGMF. Successful observations to confront the above three subjects would be ultra-wideband polarimetric observation. The UHF (300~MHz -- 3~GHz) band would be promising, based on the RM range of interests and corresponding energy range of CR electrons.

\begin{figure}[tbp]
\begin{center}
\FigureFile(75mm,75mm){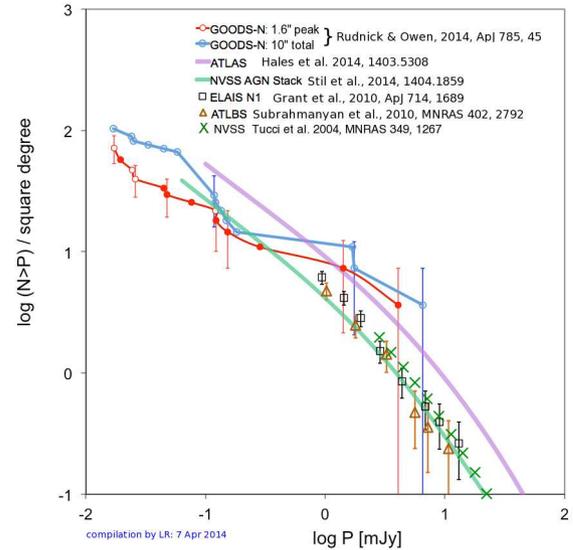}
\end{center}
\caption{
Population of polarized sources observed at 1.4 GHz (\cite{2014ApJ...785...45R, 2015aska.confE.113T}). The red and blue, purple, and green lines are based on the deep observations of the $\sim 0.2$ deg$^2$ VLA GOODS-N field \citep{2014ApJ...785...45R}, the $\sim 6$ deg$^2$ ATCA ATLAS field \citep{2014MNRAS.440.3113H}, and the all-sky VLA NVSS survey \citep{2014ApJ...787...99S}, respectively.
}
\label{f16}
\end{figure}

One of the advantages of centimeter and meter radio observations for the study of cosmic magnetism is a survey speed. All-sky polarization surveys can provide ``RM grid'' maps, which are useful for studying spatial structures of magnetic fields. Increasing the number of sources, i.e. denser RM grids, can reveal smaller spatial structures. Moreover, it can provide more observation points in redshift space, allowing us to study the population of AGN and cosmic evolution of magnetic fields. Table~\ref{t01} lists the expected density of RM grids to be obtained with the SKA1. These estimations are basically derived from figure~\ref{f16}, which summarizes plots of observed source populations (mostly FR galaxies and quasars) at 1.4 GHz. Because the measured populations have a large scatter, there is still a factor of four uncertainty in the estimation of RM grids \citep{2014skao.rept.....G}; the population below 10 $\mu$Jy is not well-known. Faint components such as star-forming galaxies, ULIRGs, merging galaxies, and quiet spiral galaxies, would be more important in the low luminosity regime \citep{2015aska.confE.113T}. Deep pilot observations (figure~\ref{f16}) are useful to foresee the density of RM grids in the SKA era. 

\begin{table}
\tbl{The expected density of RM grids at 1.4 GHz with the SKA.}{%
\begin{tabular}{rrl}
\hline
Pol. Intensity & Counts$^\dagger$ & Reference \\
\hline
4 $\mu$Jy & $\sim$230--450/deg$^2$ & \citet{2015aska.confE..92J}\\
0.75 $\mu$Jy & $\sim$5000/deg$^2$ & \citet{2015aska.confE.113T}\\
\hline
\end{tabular}}\label{t01}
\begin{tabnote}
$^\dagger$The number of extragalactic polarized sources.
\end{tabnote}
\end{table}

In conclusion, magnetic field is fundamental physics and ubiquitous in the Universe. It has been revealed that magnetic field play key roles in the formation and evolution of the Universe. The importance of understanding cosmic magnetism will increase more and more. Future centimeter and meter radio observations will allow us to adopt new breakthrough techniques, and will be essential in elucidating the origin and nature of cosmic magnetism in coming decades.

\vskip 12pt

The authors would like to thank Japan SKA Consortium Cosmic Magnetism Science Working Group members for their contributions onto magnetism sciences and their useful comments, suggestions, and encouragements. This article is in part motivated by the archival white book {\it "Resolving 4-D Nature of Magnetism with Depolarization and Faraday Tomography" (arXiv:1603.01974)} written by the members. The authors are also grateful very much to International SKA Cosmic Magnetism Science Working Group members for providing us opportunities of open discussion and cooperation. This work was supported in part by JSPS KAKENHI Grants: 26400218 (MT), 15H03639 (TA), 15K05080 (YF), 15H05896 (KT), 15K17614 (TA), 16H05999 (KT), 16K20927 (DGY), 17H01110 (TA, KI, KT), and by the National Research Foundation of Korea through grant: 2007-0093860 (SI).


\begin{thebibliography}{1}
\expandafter\ifx\csname natexlab\endcsname\relax\def\natexlab#1{#1}\fi

%%%%%%%%%%%%%%%%%%%%%%%%%%%%%%%%%%%%%%%%%%%%
%%%% A %%%%
%%%%%%%%%%%%%%%%%%%%%%%%%%%%%%%%%%%%%%%%%%%%
\bibitem[Abramowicz et al.(1995)]{1995ApJ...438L..37A}
	Abramowicz, M.~A., Chen, X., Kato, S., Lasota, J.-P., \& Regev, O.\ 
	1995, ApJL, 438, L37
\bibitem[Ajello et al.(2009)]{2009ApJ...690..367A} 
	Ajello, M., et al.
%	Ajello, M., Rebusco, P., Cappelluti, N., Reimer, O., 
%	B\"{o}hringer, H., Greiner, J., Gehrels, N., 
%	Tueller, J., \& Moretti, A.\ 
	2009, ApJ, 690, 367
\bibitem[Ajello et al.(2010)]{2010ApJ...725..1688A} 
	Ajello, M., Rebusco, P., Cappelluti, N., Reimer, O., 
	B\"{o}hringer, H., La Parola, V., \& Cusumano, G.\ 
	2010, ApJ, 725, 1688
\bibitem[Akahori \& Ryu(2010)]{2010ApJ...723..476A}
	Akahori, T., \& Ryu, D.\ 
	2010, ApJ, 723, 476
\bibitem[Akahori \& Ryu(2011)]{2011ApJ...738..134A}
	Akahori, T., \& Ryu, D.\ 
	2011, ApJ, 738, 134
\bibitem[Akahori et al.(2013)]{2013ApJ...767..150A}
	Akahori, T., Ryu, D., Kim J., \& Gaensler B. M.\ 
	2013, ApJ, 767, 150
\bibitem[Akahori et al.(2014a)]{2014PASJ...66...65A}
	Akahori, T., Kumazaki, K., Takahashi, K., \& Ryu, D.\ 
	2014a, PASJ, 66, 65
\bibitem[Akahori et al.(2014b)]{2014ApJ...790..123A}
	Akahori, T., Gaensler, B. M., \& Ryu, D.\ 
	2014b, ApJ, 790, 123
\bibitem[Akahori et al.(2016)]{2016ApJ...824..105A}
	Akahori, T., Ryu, D., \& Gaensler, B. M.\ 
	2016, ApJ, 824, 105
\bibitem[Akahori et al.(2017)]{2016arXiv161106647A}
	Akahori, T., et al.\
	submitted (arXiv:1611.06647)
\bibitem[Akahori et al.(2017b)]{Akahori17}
	Akahori, T., et al.\ 
	2017, in preparation
\bibitem[Akamatsu et al.(2012)]{2012PASJ...64..67A} 
	Akamatsu, H., Takizawa, M., Nakazawa, K., Fukazawa, Y., 
	Ishisaki, Y., \& Ohashi, T.\ 
	2012, PASJ, 64, 67
\bibitem[Akamatsu \& Kawahara(2013)]{2013PASJ...65...16A} 
	Akamatsu, H., \& Kawahara, H.\ 
	2013, PASJ, 65, 16
\bibitem[Altenhoff et al.(1978)]{1978A&AS...35...23A}
	Altenhoff, W. J., Downes, D., Pauls, D., \& Schraml, J.\
	1978, A\&AS, 35, 23
\bibitem[Alves et al.(2008)]{2008A&A...486L..13A}
	Alves, F. O., Franco, G. A. P., \& Girart, J. M.\ 
	2008, A\&A, 486, L13
\bibitem[Anantharamaiah et al.(1991)]{1991MNRAS.249..262A}
	Anantharamaiah, K. R., Pedlar, A., Ekers, R. D., \& Goss, W. M. \
	1991, MNRAS, 249, 262
\bibitem[Ando et al.(2010)]{2010ApJ...716.1566A}
	Ando, M., Doi, K., \& Susa, H.\ 
	2010, ApJ, 716, 1566
\bibitem[Antonucci \& Miller(1985)]{1985ApJ...297..621A}
	Antonucci, R.~R.~J., \& Miller, J.~S.\
	1985, ApJ, 297, 621
\bibitem[Armstrong et al.(1995)]{1995ApJ...443..209A}
	Armstrong, J.~W., Rickett, B.~J. \& Spangler, S. R.\ 
	1995, S.~R., ApJ, 443, 209
\bibitem[Arshakian et al.(2009)]{2009A&A...494...21A}
	Arshakian, T.~G., Beck, R., Krause, M., \& Sokoloff, D.\
	2009, A\&A, 494, 21
\bibitem[Arshakian \& Beck(2011)]{2011MNRAS.418.2336A}
	Arshakian, T. G. \& Beck, R.\ 
	2011, MNRAS, 418, 2336
\bibitem[Asada et al.(2014)]{2014ApJ...781L...2A}
	Asada, K., Nakamura, M., Doi, A., Nagai, H., \& Inoue, M.\ 
	2014, ApJL, 781, L2
\bibitem[Asada et al.(2002)]{2002PASJ...54L..39A}
	Asada, K., Inoue, M., Uchida, Y., et al.\ 
	2002, PASJ, 54, L39
\bibitem[Asai et al.(2005)]{2005AdSpR..36..636A}
	Asai, N., Fukuda, N., \& Matsumoto, R.\
	2005, Advances in Space Research, 36, 636
\bibitem[Asano \& Takahara(2009)]{2009ApJ...690L..81A}
	Asano, K., \& Takahara, F.\ 
	2009, ApJL, 690, L81
\bibitem[Ascasibar \& Markevitch(2006)]{2006ApJ...650..102A} 
	Ascasibar, Y., \& Markevitch, M.\ 
	2006, ApJ, 650, 102 

%%%%%%%%%%%%%%%%%%%%%%%%%%%%%%%%%%%%%%%%%%%%
%%%% B %%%%
%%%%%%%%%%%%%%%%%%%%%%%%%%%%%%%%%%%%%%%%%%%%
%%%\bibitem[Baade \& Minkowski(1954)]{1954ApJ...119..206B}
%%%	Baade, W., \& Minkowski, R.\
%%%	1954, ApJ, 119, 206
\bibitem[Balbus \& Hawley(1991)]{1991ApJ...376..214B}
	Balbus, S.~A., \& Hawley, J.~F.\
	1991, ApJ, 376, 214
\bibitem[Banba et al.(2003)]{2003ApJ...589..827B}
	Bamba, A., Yamazaki, R., Ueno, M., \& Koyama, K.\
	2003, ApJ, 589, 827
\bibitem[Bamba \& Yokoyama(2004)]{2004PhRvD..69d3507B}
	Bamba, K., \& Yokoyama, J.\ 
	2004, PRD, 69, 043507
\bibitem[Bandiera \& Petruk(2016)]{2016MNRAS.459..178B}
	Bandiera, R., \& Petruk, O.\ 
	2016, MNRAS, 459, 178
\bibitem[Barniol Duran \& Whitehead(2016)]{2016MNRAS.462L..31B}
	Barniol Duran, R., Whitehead, J. F., \& Giannios, D.\ 
	2016, MNRAS, 462, 31
\bibitem[Basu \& Mouschovias(1994)]{1994ApJ...432..720B}
	Basu, S., \& Mouschovias, T, Ch.\
	1994, ApJ, 432, 720
\bibitem[Basu \& Ciolek(2004)]{2004ApJ...607L..39B}
	Basu, S., \& Ciolek, G. E.\ 
	2004, ApJ, 607, L39
\bibitem[Basu et al.(2012)]{2012ApJ...756..141B}
	Basu, A., Roy, S., \& Mitra, D.\ 
	2012, ApJ, 756, 141
\bibitem[Beck et al.(1989)]{1989A&A...222...58B}
	Beck, R., et al.\
%	Beck, R., Loiseau, N., Hummel, E., et al.\
	1989, A\&A, 222, 58 
\bibitem[Beck(2001)]{2001SSRev...99..243B}
	Beck, R.\
	2001, SSRev, 99, 243
\bibitem[Beck et al.(2002)]{2002A&A...391...83B}
	Beck, R., Shoutenkov, V., Ehle, M., Harnett, J.~I., Haynes, R.~F., 
	Shukurov, A., Sokoloff, D.~D., \& Thierbach, M.\
	2002, A\&A, 391, 83
\bibitem[Beck et al.(2005)]{2005A&A...444..739B}
	Beck, R., Fletcher, A., Shukurov, A., Snodin, A., Sokoloff, D.~D., 
	Ehle, M., Moss, D., \& Shoutenkov, V.\
	2005, A\&A, 444, 739
\bibitem[Beck(2007)]{2007A&A...470..539B}
	Beck, R.\ 
	2007, A\&A, 470, 539
\bibitem[Beck et al.(2012)]{2012A&A...543A.113B}
	Beck, R., Frick, P., Stepanov, R., \& Sokoloff, D.\ 
	2012, A\&A, 543, 113
\bibitem[Beck et al.(2013)]{2013MNRAS.429L..60B}
	Beck, A. M., Hanasz, M., Lesch, H., 
	Remus, R. -S., \& Stasyszyn, F. A.\
	2013, MNRAS, 429, 60
\bibitem[Beck et al.(2013)]{2013MNRAS.435.3575B}
	Beck, A. M., Dolag, K., Lesch, H., \& Kronberg, P. P.\
	2013, MNRAS, 435, 3575
\bibitem[Beck(2015)]{2015A&A...578A..93B}
	Beck, R.\ 
	2015, A\&A, 578, 93 
\bibitem[Beck et al.(2016)]{2016JCAP...5..56B}
	Beck, M.~C., Beck, A.~M., Beck, R., Dolag, K., 
	Strong, A.~W., \& Nielaba, P.\ 
	2016, JCAP, 5, 56
\bibitem[Bell(2004)]{2004MNRAS.353..550B}
	Bell, A. R.\
	2004, MNRAS, 353, 550
\bibitem[Bell et al.(2011)]{2011A&A...535A..85B}
	Bell, M. R., Junklewitz, H., \& En{\ss}lin, T. A.\ 
	2011, A\&A, 535, A85
\bibitem[Bennett(1962)]{1962MmRAS..68..163B}
	Bennett, A.~S.\
	1962, MmRAS, 68, 163
\bibitem[Berkhuijsen et al.(2006)]{2006AN....327...82B}
	Berkhuijsen, E. M., Mitra, D., \& Mueller, P.\
	2006, AN, 327, 82
\bibitem[Bernet et al.(2008)]{2008Natur.454..302B}
	Bernet, M.~L., Miniati, F., Lilly, S.~J., 
	Kronberg, P.~P., \& Dessauges-Zavadsky, M.\
	2008, Nature, 454, 302
\bibitem[Bernet, Miniati, \& Lilly(2012)]{2012ApJ...761..144B}
	Bernet, M. L., Miniati, F., \& Lilly, S. J.\ 
	2012, ApJ, 761, 144
\bibitem[Bernet, Miniati, \& Lilly(2013)]{2013ApJ...772L..28B}
	Bernet, M. L., Miniati, F., \& Lilly, S. J.\ 
	2013, ApJ, 772, L28
\bibitem[Bertone et al.(2006)]{2006MNRAS.370..319B}
	Bertone, S., Vogt, C. \& En{\ss}lin, T.\ 
	2006, MNRAS, 370, 319
\bibitem[Beresnyak et al.(2013)]{2013ApJ...771..131B}
	Beresnyak, A., Xu, H., Li, H., \& Schlickeiser, R.\ 
	2013, ApJ, 771, 131
\bibitem[BICEP2/Keck and Planck Collaborations.(2015)]{2015PhRvL.114j1301B}
	BICEP2/Keck and Planck Collaborations\ 
	2015, PRL, 114, 101301
\bibitem[Biermann(1950)]{1950ZNatA...5...65B}
	Biermann, L.\ 
	1950, Zeitschrift Naturforschung Teil A, 5, 65
\bibitem[Blasi \& Colafrancesco(1999)]{1999APh...12..169B}
	Blasi, P., \& Colafrancesco, S.\ 
	1999, APh, 12, 169
\bibitem[Blasi(2013)]{2013A&ARv..21...70B}
	Blasi, P.\
	2013, A\&ARv, 21, 70
\bibitem[Bogdanov et al.(2011)]{2011ApJ...742...97B}
	Bogdanov, S. V., Archibald, A. M., Hessels, J. W. T., Kaspi, V. M., 
	Lorimer, D., McLaughlin, M. A.,Ransom, S. M., \& Stairs, I. H.\ 
	2011, ApJ, 742, 97
\bibitem[B\"{o}hringer \& Morfill(1988)]{1988ApJ...330..609B}
	B\"{o}hringer, H., \& Morfill, G.~E.\ 
	1988, ApJ, 330, 609
\bibitem[Bolton \& Stanley(1948)]{1948AuSRA...1...58B}
	Bolton, J.~G., \& Stanley, G.~J.\ 
	1948, Australian Journal of Scientific Research A Physical Sciences, 1, 58
\bibitem[Bonafede et al.(2010)]{2010A&A...513A..30B} 
	Bonafede, A., Feretti, L., Murgia, M., et al.\ 
	2010, A\&A, 513, 30
\bibitem[Bower et al.(2005)]{2005ApJ...618L..L29B}
	Bower, G. C., Falcke, H., Wright, M. C., \& Backer, D. C.\ 
	2005, ApJL, 618, L29
\bibitem[Brandenburg et al.(1993)]{1993A&A...271...36B}
	Brandenburg, A., et al.\ 
%	Brandenburg, A., Donner, K.~J., Moss, D., et al.\ 
	1993, A\&A, 271, 36  
\bibitem[Brandenburg \& Nordlund(2011)]{2011RPPh...74d6901B}
	Brandenburg, A., \& Nordlund, A.\ 
	2011, RPPh, 74, 046901
\bibitem[Brandenburg \& Stepanov(2014)]{2014ApJ...786...91B}
	Brandenburg, A. \& Stepanov, R.\
	2014, ApJ, 786, 91
\bibitem[Brandenburg(2015)]{2015ASSL..407..529B}
	Brandenburg, A.\ 
	2015, Magnetic Fields in Diffuse Media, 407, 529
\bibitem[Blandford \& Eichler(1987)]{1987PhR...154....1B}
	Blandford, R., \& Eichler, D. 1987, Phys. Rep., 154, 1
\bibitem[Brown \& Taylor(2001)]{2001ApJ...563..L31B}
	Brown, J.~C., \& Taylor, A. R.\ 
	2001, ApJL, 563, L31
\bibitem[Brown et al.(2007)]{2007ApJ...663..258B}
	Brown, J.~C., Haverkorn, M., Gaensler, B.~M., Taylor, A.~R., 
	Bizunok, N.~S., McCLure-Griffiths, N.~M., Dickey, J.~M, \& 
	Green, A.~J.\ 
	2007, ApJ, 663, 258
\bibitem[Brown et al.(2010)]{2010MNRAS.402....2B}
	Brown, S., Farnsworth, D., \& Rudnick, L., 2010\
	2010, MNRAS, 402, 2
\bibitem[Burkhart et al.(2012)]{2012ApJ...749..145B}
	Burkhart, B., Lazarian, A., \& Gaensler, B. M.\ 
	2012, ApJ, 749, 145
\bibitem[Burn(1966)]{1966MNRAS.133...67B}
	Burn, B. J.\ 
	1966, MNRAS, 133, 67
\bibitem[Brunetti et al.(2001)]{2001MNRAS...320..365B}
	Brunetti, G., Setti, G., Feretti, L., \& Giovannini, G.\ 
	2001, MNRAS, 320, 365
\bibitem[Brunetti et al.(2004)]{2004MNRAS...350..1174B}
	Brunetti, G., Blasi, P., Cassano, R., \& Gabici, S.\ 
	2004, MNRAS, 350, 1174
\bibitem[Brunetti et al.(2007)]{2007ApJ...670..L5B}
	Brunetti, G., Venturi, T., Dallacasa, D., Cassano, R., 
	Dolag, K, Giacintucci, S., \& Setti, G.\ 
	2007, ApJL, 670, L5
\bibitem[Brunetti \& Lazarian(2007)]{2007MNRAS...378..245B}
	Brunetti, G., \& Lazarian, A.\ 
	2007, MNRAS, 378, 245
\bibitem[Brunetti et al.(2009)]{2009A&A...507..661B}
	Brunetti, G., Cassano, R., Dolag, K., \& Setti, G.\ 
	2009, A\&A, 507, 661
\bibitem[Brunetti \& Lazarian(2011)]{2011MNRAS...410..127B}
	Brunetti, G., \& Lazarian, A.\ 
	2011, MNRAS, 410, 127
\bibitem[Brunetti \& Jones(2014)]{2014IJMP...23..1430007B}
	Brunetti, G., \& Jones, T.~W.\ 
	2014, International Journal of Modern Physics D, 23, 1430007

%%%%%%%%%%%%%%%%%%%%%%%%%%%%%%%%%%%%%%%%%%%%
%%%% C %%%%
%%%%%%%%%%%%%%%%%%%%%%%%%%%%%%%%%%%%%%%%%%%%
\bibitem[Cassano et al.(2006)]{2006MNRAS...369..1577C}
	Cassano, R., Brunetti, G., \& Setti, G.\ 
	2006, MNRAS, 369, 1577
\bibitem[Cassano(2010)]{2010A&A...517..10C}
	Cassano, R.\
	2010, A\&A, 517, 10
\bibitem[Cassano et al.(2011)]{2011JAA...32..519C}
	Cassano, R., Brunetti, G., Venturi, T.\ 
	2011, Journ. of Astroph. and Astr., 32, 519
\bibitem[Cayatte et al.(1990)]{1990AJ....100..604C}
	Cayatte, V., van Gorkom, J.~H., Balkowski, C., \& Kotanyi, C.\ 
	1990, AJ, 100, 604 
\bibitem[Chandrasekhar \& Fermi (1953)]{1953ApJ...118..113C}
	Chandrasekhar, S., \& Fermi, E.\ 
	1953, ApJ, 118, 113
\bibitem[Chapman et al. (2011)]{2011ApJ...741...21C}
	Chapman, N. L., Goldsmith, P. F., Pineda, J. L., Clemens, 
	D. P., Li, D., \& Kr\v{c}o, M.\ 
	2011, ApJ, 741, 21
\bibitem[Cho \& Ryu(2009)]{2009ApJ...705L..90C}
	Cho, J., \& Ryu, D.\ 
	2009, ApJL, 705, L90
\bibitem[Chomiuk et al.(2012)]{2012ApJ...761..173C}
	Chomiuk, L., et al.\ 
%	Chomiuk, L., Krauss, M. I., Rupen, M. P., Nelson, T., Roy, N.,
%	Sokoloski, J. L., Mukai, K., Munari, U., Mioduszewski, A., 
%	Weston, J., O'Brien, T. J., Eyres, S. P. S., \& Bode, M. F.
	2012, ApJ, 761, 173
\bibitem[Chuzhoy(2004)]{2004MNRAS.350..761C}
	Chuzhoy, L.\ 
	2004, MNRAS, 350, 761
\bibitem[Chy{\.z}y et al.(2000)]{2000A&A...355..128C}
	Chy{\.z}y, K.~T., Beck, R., Kohle, S., Klein, U., \& Urbanik, M.\ 
	2000, A\&A, 355, 128
\bibitem[Chy{\.z}y et al.(2003)]{2003A&A...405..513C}
	Chy{\.z}y, K.~T., Knapik, J., Bomans, D.~J., et al.\ 
	2003, A\&A, 405, 513
\bibitem[Clarke, Burns \& Norman(1992)]{1992ApJ...395..444C}
	Clarke, D.~A., Burns, J.~O., \& Norman, M.~L.\
	1992, ApJ, 395, 444
\bibitem[Clegg et al.(1992)]{1992ApJ...386..143C}
	Clegg, A.~W., Cordes, J.~M., Simonetti, J.~H., \& Kulkarni, S.~R.\ 
	1992, ApJ, 386, 143
\bibitem[Cole et al.(2005)]{2005MNRAS.362..505C}
	S. Cole, et al.\
	2005, MNRAS, 362, 505
\bibitem[Condon et al.(1991)]{1991ApJ...376...95C}
	Condon, J.~J., Anderson, M.~L., \& Helou, G.\
	1991, ApJ, 376, 95 
\bibitem[Conway \& Kronberg(1969)]{1969MNRAS.142...11C}
	Conway, R.~G., \& Kronberg, P.~P.\
	1969, MNRAS, 142, 11 
\bibitem[Cordes \& Lazio(2002)]{2002astro.ph..7156C}
	Cordes, J. M., \& Lazio, T. J. W.\ 
	2002, preprint (astro-ph/0207156)
\bibitem[Cordes \& Lazio(2003)]{2003astro.ph..1598C}
	Cordes, J. M., \& Lazio, T. J. W.\ 
	2003, preprint (astro-ph/0301598)
\bibitem[Crutcher(1999)]{1999ApJ...520..706C}
	Crutcher, R. M.\ 
	1999, ApJ, 520, 706
\bibitem[Crutcher(2012)]{2012ARA&A..50...29C}
	Crutcher, R. M.\ 
	2012, ARAA, 50, 29
\bibitem[Cyburt et al.(2003)]{2003PhRvD..67j3521C}
	 Cyburt, R. H., Ellis,J. R.,  Fields, B. D. \& Olive, K. A.\ 
	 2003, PRD, 67, 103521

%%%%%%%%%%%%%%%%%%%%%%%%%%%%%%%%%%%%%%%%%%%%
%%%% D %%%%
%%%%%%%%%%%%%%%%%%%%%%%%%%%%%%%%%%%%%%%%%%%%
\bibitem[Das et al.(2010)]{2010A&A...509A..23D}
	Das, K., Roy, A.~L., Keller, R., \& Tuccari, G.\ 
	2010, A\&A, 509, 23 
\bibitem[De Jager \& Djannati-Atai(2009)]{2009ASSL..357..451D}
	De Jager, O. C., \& Djannati-Atai, A.\
	 2009, ASSL, 357, 451
\bibitem[Dennison(1980)]{1980ApJ...239L..93D}
	Dennison, B.\ 
	1980, ApJ, 239, 93L
\bibitem[Diamond et al.(2010)]{2010rdla.book.....D}
	Diamond, P.~H., Garbet, X., Ghendrih, P., \& Sarazin, Y.\ 
	2010, Relaxation Dynamics in Laboratory and Astrophysical Plasmas: 
	Biennial Reviews of the Theory of Magnetized Plasmas, Ch. 4 
	(Singapore: World Scientific Publishing Co., Pte.~Ltd.)
\bibitem[Doi \& Susa(2011)]{2011ApJ...741...93D}
	Doi, K., \& Susa, H.\ 
	2011, ApJ, 741, 93 
\bibitem[Donnert et al.(2009)]{2009MNRAS.392.1008D}
	Donnert J., Dolag K., Lesch H., \& M\"uller E.\ 
	2009, MNRAS, 392, 1008
\bibitem[Dolag et al.(2008)]{2008SSRv..134..311D}
	Dolag K., Bykov A. M., \& Diaferio A.\ 
	2008, Sp. Sci. Rev., 134, 311
\bibitem[Dolag \& Stasyszyn(2009)]{2009MNRAS.398.1678D}
	Dolag K., \& Stasyszyn F.\ 
	2009, MNRAS, 398, 1678
\bibitem[Drzazga et al.(2016)]{2016A&A...589A..12D}
	Drzazga, R.~T., Chy{\.z}y, K.~T., Heald, G.~H., 
	Elstner, D., \& Gallagher, J.~S.\ 
	2016, A\&A, 589, 12
\bibitem[Dubois \& Teyssier(2008)]{2008A&A...482L..13D}
	Dubois Y., \& Teyssier R.\ 
	2008, A\&A, 482, L13
\bibitem[Durrer \& Neronov(2013)]{2013A&ARv..21...62D}
	Durrer, R., \& Neronov, A.\ 
	2013, A\&A Review, 21, 62 
\bibitem[Durrive \& Langer(2015)]{2015MNRAS.453..345D}
	Durrive, J.-B., \& Langer, M.\ 
	2015, MNRAS, 453, 345 

%%%%%%%%%%%%%%%%%%%%%%%%%%%%%%%%%%%%%%%%%%%%
%%%% E %%%%
%%%%%%%%%%%%%%%%%%%%%%%%%%%%%%%%%%%%%%%%%%%%
\bibitem[Eatough et al.(2013a)]{2013ATel.5040....1E}
	Eatough, R. P. et al.\ 
	2013a, The Astronomer’s Telegram, 5040, 1 
\bibitem[Eatough et al.(2013b)]{2013Natur.501..391E}
	Eatough, R. P. et al.\ 
	2013b, Nature, 501, 391
\bibitem[Edge et al.(1959)]{1959MmRAS..68...37E}
	Edge, D.~O., Shakeshaft, J.~R., McAdam, W.~B., 
	Baldwin, J.~E., \& Archer, S.\ 
	1959, MmRAS, 68, 37
%%\bibitem[Eichler(1979)]{1979ApJ...229..419E}
%%	Eichler, D.\
%%	1979, ApJ, 229, 419
\bibitem[Ellis et~al.(1985)]{1985NuPhB.259..175E}
	 Ellis, J. R., Nanopoulos, D. V. \& Sarkar, S.\ 
	 1985, Nucl. Phys. B, 259, 175
\bibitem[En{\ss}lin et al.(1997)]{1997ApJ...477..560E} 
	En{\ss}lin, T.~A., Biermann, P. L., Kronberg, P. P., \& Wu, X.-P.\ 
	1997, ApJ, 477, 560
\bibitem[En{\ss}lin \& Vogt(2003)]{2003A&A...401..835E} 
	En{\ss}lin, T.~A., \& Vogt, C.\ 
	2003, A\&A, 401, 835
\bibitem[En{\ss}lin et al.(2011)]{2011A&A...527A..99E}
	En{\ss}lin, T., Pfrommer, C., Miniati, F., \& Subramanian, K.\ 
	2011, A\&A, 527, 99
\bibitem[Erken et al.(2012)]{2012PhRvD..85f3520E}
	O. Erken, P. Sikivie, H. Tam, and Q. Yang\
	2012, Phys. Rev. D, 85, 063520

%%%%%%%%%%%%%%%%%%%%%%%%%%%%%%%%%%%%%%%%%%%%
%%%% F %%%%
%%%%%%%%%%%%%%%%%%%%%%%%%%%%%%%%%%%%%%%%%%%%
\bibitem[Fanaroff \& Riley(1974)]{1974MNRAS.167P..31F}
	Fanaroff, B.~L., \& Riley, J.~M.\ 
	1974, MNRAS, 167, 31P
\bibitem[Farnes et al.(2014a)]{2014ApJS..212...15F}
	Farnes, J. S., Gaensler, B. M., \& Carretti, E.\ 
	2014a, ApJS, 212, 15
\bibitem[Farnes et al.(2014b)]{2014ApJ...795...63F}
	Farnes, J. S., O'Sullivan, S. P., Corrigan, M. E., \& 
	Gaensler, B. M.\ 
	2014b, ApJ, 795, 63
%%%\bibitem[Fath(1909)]{1909LicOB...5...71F}
%%%	Fath, E.~A.\
%%%	1909, Lick Observatory Bulletin, 5, 71
\bibitem[Fatuzzo \& Adams(2002)]{2002ApJ...570..210F}
	Fatuzzo, M., \& Adams, F. C. \
	2002, ApJ, 570, 210
\bibitem[Felten(1996)]{1996ASPCP...88...271} 
	Felten, J. E.\ 
	1996, Clusters, Lensing, and the Future of the Universe, 88, 271
\bibitem[Ferreira et al.(2014)]{2014JCAP...06..053F}
	Ferreira, R.~J.~Z., Jain, R.~K., \& Sloth, M.~S.\ 
	2014, JCAP, 6, 53
\bibitem[Feretti et al.(1995)]{1995A&A...302..680F}
	Feretti, L., Dallacasa, D., Giovannini, G., \& Tagliani, A.\ 
	1995, A\&A, 302, 680
\bibitem[Feretti et al.(2012)]{2012A&ARv..20...54F}
	Feretti, L., Giovannini, G., Govoni, F., \& Murgia, M.\ 
	2012, A\&ARv, 20, 54
\bibitem[Field, Goldsmith, \& Habing(1969)]{1969ApJ...155L.149F}
	Field G.~B., Goldsmith D.~W., \& Habing H.~J.\
	1969, ApJ, 155, L149 
\bibitem[Finoguenov et al.(2010)]{2010ApJ...715..1143F} 
	Finoguenov, A., Sarazin, C. L., Nakazawa, K., 
	Wik, D. R., \& Clarke, T. E.\ 
	2010, ApJ, 715, 1143
\bibitem[Fletcher et al.(2011)]{2011MNRAS.412.2396F}
	Fletcher, A., Beck, R., Shukurov, A., 
	Berkhuijsen, E.~M., \& Horellou, C.\ 
	2011, MNRAS, 412, 2396 
\bibitem[Franco \& Alves (2015)]{2015ApJ...807....5F}
	Franco, G. A. P., \& Alves, F. O.\ 
	2015, ApJ, 807, 5
\bibitem[Frick et al.(2011)]{2011MNRAS.414.2540F}
	Frick, P., Sokoloff, D., Stepanov, R., \& Beck, R.\ 
	2011, MNRAS, 414, 2540
\bibitem[Fujishita et al.(2009)]{2009PASJ...61.1039F}
	Fujishita, M., Torii, K., Kudo, N., et al.\ 
	2009, PASJ, 61, 1039
\bibitem[Fujita et al.(2003)]{2003ApJ...584..190F}
	Fujita, Y., Takizawa, M., \& Sarazin, C.~L.\ 
	2003, ApJ, 584, 190
\bibitem[Fujita et al.(2004)]{2004ApJ...612L...9F}
	Fujita, Y., Matsumoto, T., \& Wada, K.\ 
	2004, ApJL, 612, L9 
\bibitem[Fujita \& Kato(2005)]{2005MNRAS.364..247F}
	Fujita, Y., \& Kato, T.~N.\ 
	2005, MNRAS, 364, 247 
\bibitem[Fujita et al.(2007)]{2007ApJ...663L..61F} 
	Fujita, Y., Kohri, K., Yamazaki, R., \& Kino, M.\ 
	2007, ApJL, 663, L61
\bibitem[Fujita \& Ohira(2011)]{2011ApJ...738..182F}
	Fujita, Y., \& Ohira, Y.\ 
	2011, ApJ, 738, 182
\bibitem[Fujita \& Ohira(2012)]{2012ApJ...746...53F}
	Fujita, Y., \& Ohira, Y.\ 
	2012, ApJ, 746, 53
\bibitem[Fujita \& Ohira(2013)]{2013MNRAS.428..599F}
	Fujita, Y., \& Ohira, Y.\ 
	2013, MNRAS, 428, 599
\bibitem[Fujita et al.(2013)]{2013MNRAS.432.1434F}
	Fujita, Y., Kimura, S., \& Ohira, Y.\
	2013, MNRAS, 432, 1434
\bibitem[Fujita et al.(2015)]{2015ApJ...815..116F}
	Fujita, Y., Takizawa, M., Yamazaki, R., Akamatsu, H., \& Ohno, H.\ 
	2015, ApJ, 815, 116
\bibitem[Fujita \& Yokoyama(2014)]{2014JCAP...03..013F}
	Fujita, T., \& Yokoyama, S.\ 
	2014, JCAP, 3, 13
\bibitem[Fukui et al.(2006)]{2006Sci...314..106F}
	Fukui, Y., Yamamoto, H., Fujishita, M., et al.\ 
	2006, Science, 314, 106
\bibitem[Fukumura et al.(2016)]{2016AN....337..454F}
	Fukumura, K., Tombesi, F., Kazanas, D., et al.\ 
	2016, Astronomische Nachrichten, 337, 454
\bibitem[Furlanetto \& Loeb(2001)]{2001ApJ...556..619F}
	Furlanetto, S. R. \& Loeb, A.\ 
	2013, ApJ, 556, 619

%%%%%%%%%%%%%%%%%%%%%%%%%%%%%%%%%%%%%%%%%%%%
%%%% G %%%%
%%%%%%%%%%%%%%%%%%%%%%%%%%%%%%%%%%%%%%%%%%%%
\bibitem[Gabuzda et al.(2015)]{2015MNRAS.450.2441G}
	Gabuzda, D.~C., Knuettel, S., \& Reardon, B.\ 
	2015, MNRAS, 450, 2441
\bibitem[Gaensler et al.(2005)]{2005Sci...307.1610G}
	Gaensler, B.~M., et al.\ 
%	Gaensler, B.~M., Haverkorn, M., Staveley-Smith, L., et al.\ 
	2005, Science, 307, 1610 
\bibitem[Gaensler et al.(2008)]{2008PASA...25..184G}
	Gaensler, B. M., Madsen, G. J., Chatterjee, S., \& Mao, S. A.\ 
	2008, PASA, 25, 184
\bibitem[Gaensler et al.(2011)]{2011Nature...478..214G}
	Gaensler, B. M., et al.\
%	Gaensler, B. M., Haverkorn, M., Burkhart, B., Newton-McGee, K. J., 
%	Ekers, R. D., Lazarian, A., McClure-Griffiths, N. M., Robishaw, T., 
%	Dickey, J. M., \& Green, A. J.\ 
	2011, Nature, 478, 214
\bibitem[Gaensler et al.(2015)]{2015aska.confE.103G}
	Gaensler, B. M., et al.\ 
%	Gaensler, B., Agudo, I., Akahori, T., et al.\ 
	2015, in Proc. of Advancing Astrophysics with the Square Kilometre 
	Array (Trieste: SISSA), PoS (AASKA14)103, id.103
\bibitem[Gammie et al.(2003)]{2003ApJ...589..444G}
	Gammie, C.~F., McKinney, J.~C., \& T{\'o}th, G.\ 
	2003, ApJ, 589, 444
\bibitem[Gao et al.(2010)]{2010A&A...515A..64G}
	Gao, X. Y., et al.\ 
	2010, A\&A, 515, 64
\bibitem[Gardner \& Whiteoak(1966)]{1966ARA&A...4..245G}
	Gardner, F. F.,  \& Whiteoak, J. B.\ 
	1966, ARA\&A, 4, 245
\bibitem[Gelfand et al.(2015)]{2015aska.confE..46G}
	Gelfand, J., et al.\ 
%	Gelfand, J., Breton, R., Ng, C. U. et al.\ 
	2015, in Proc. of Advancing Astrophysics with the Square Kilometre 
	Array (Trieste: SISSA), PoS (AASKA14)046, id.46
\bibitem[Gendelev \& Krumholz(2012)]{2012ApJ...745..158G}
	Gendelev, L. \& Krumholz, M.R.\ 
	2012, ApJ, 745, 158
\bibitem[Ghez et al.(2008)]{2008ApJ...689.1044G}
	Ghez, A. M., et al.\
	2008, ApJ, 689, 1044
\bibitem[Giacinti et al.(2010)]{2010JCAP...08..036G}
	Giacinti, G., Kachelrie{\ss}, M., Semikoz, D. V., \& Sigl, G.\ 
	2010, JCAP, 8, 36
\bibitem[Giacintucci et al.(2014)]{2014ApJ...781....9G}
	Giacintucci, S., et al.\ 
%	Giacintucci, S., Markevitch, M., Venturi, T., et al.\ 
	2014, ApJ, 781, 9
\bibitem[Gillessen et al.(2009)]{2009ApJ...692..1075G}
	Gillessen, S., et al.\ 
	2009, ApJ, 692, 1075
\bibitem[Ginzburg \& Syrovatskii(1965)]{1965ARA&A...3..297G}
	Ginzburg, V.~L., \& Syrovatskii, S.~I.\ 
	1965, ARA\&A, 3, 297
\bibitem[Ginzburg \& Syrovatskii(1969)]{1969ARA&A...7..375G}
	Ginzburg, V.~L., \& Syrovatskii, S.~I.\ 
	1969, ARA\&A, 7, 375
\bibitem[Girart et al. (2006)]{2006Sci...313..812G}
	Girart, J. M., Rao, R., \& Marrone, D. P.\
	2006, Science, 313, 812
\bibitem[Gitti et al.(2002)]{2002A&A...386..456G}
	Gitti, M., Brunetti, G., \& Setti, G.\ 
	2002, A\&A, 386, 456 
\bibitem[Gnedin et al.(2000)]{2000ApJ...539..505G}
	Gnedin N. Y., Ferrara A., \& Zweibel E. G.\ 
	2000, ApJ, 539, 505
\bibitem[Goldreich \& Sridher(1995)]{1995ApJ...438..763G}
	Goldreich, P., \& Sridhar, S.\ 
	1995, ApJ, 438, 763
\bibitem[Goldreich \& Sridhar(1997)]{1997ApJ...485..680G}
	Goldreich, P. \& Sridhar, S.\ 
	1997, ApJ, 485, 680
\bibitem[Goodman et al. (1990)]{1990ApJ...359..363G}
	Goodman, A. A., Bastien, P., Menard, F., \& Myers, P. C.\
	1990, ApJ, 359, 363
%%%\bibitem[Govoni \& Feretti(2004)]{2004IJMPD..13.1549G} 
%%%	Govoni, F., \& Feretti, L.\ 
%%%	2004, International Journal of Modern Physics D, 13, 1549
\bibitem[Govoni et al.(2010)]{2010A&A...522A.105G} 
	Govoni, F., et al.\ 
%	Govoni, F., Dolag, K., Murgia, M., et al.\ 
	2010, A\&A, 522, 105 
\bibitem[Govoni et al.(2014)]{2014skao.rept.....G}
	Govoni, F., et al.\ 
%	Govoni, F., Johnston-Hollitt, M., Agudo, I., et al.\ 
	2014, SKA Science Working Group Assessment Workshop Summary 
	(http://adsabs.harvard.edu/abs/2014skao.rept.....G)
\bibitem[Graeve \& Beck(1988)]{1988A&A...192...66G}
	Graeve, R., \& Beck, R.\ 
	1988, A\&A, 192, 66
\bibitem[Gray et al.(1999)]{1999ApJ...514..221G}
	Gray, A. D., Landecker, T. L., Dewdney, P. E., Taylor, A. R., 
	Willis, A. G., \& Normandeau, M.\ 
	1999, ApJ, 514, 221
\bibitem[Green(2009)]{2014BASI...42...47G}
	Green, D. A.\ 
	2009, Bull. Astron. Soc. India, 42, 47
\bibitem[Greenstein(1963)]{1963Natur.197.1041G}
	Greenstein, J.~L.\
	1963, Nature, 197, 1041
\bibitem[Gregori et al.(2012)]{2012Natur.481..480G}
	Gregori, G., et al.\ 
%	Gregori, G., Ravasio, A., Murphy, C.~D., et al.\ 
	2012, Nature, 481, 480
\bibitem[Gressel et al.(2013)]{2013A&A...560A..93G}
	Gressel, O., Elstner, D., \& Ziegler, U.\ 
	2013, A\&A, 560, 93
\bibitem[Guidetti et al.(2008)]{2008A&A...483..699G} 
	Guidetti, D., Murgia, M., Govoni, F., et al.\ 
	2008, A\&A, 483, 699
\bibitem[Guo \& Oh(2008)]{2008MNRAS.384..251G}
	Guo, F., \& Oh, S.~P.\ 
	2008, MNRAS, 384, 251

%%%%%%%%%%%%%%%%%%%%%%%%%%%%%%%%%%%%%%%%%%%%
%%%% H %%%%
%%%%%%%%%%%%%%%%%%%%%%%%%%%%%%%%%%%%%%%%%%%%
\bibitem[Hada et al.(2016)]{2016ApJ...817..131H}
	Hada, K., et al.\ 
%	Hada, K., Kino, M., Doi, A., et al.\ 
	2016, ApJ, 817, 131
\bibitem[Hales et al.(2014)]{2014MNRAS.440.3113H}
	Hales, C. A., Norris, R. P., Gaensler, B. M., \& Middelberg, E.\ 
	2014, MNRAS, 440, 3113
\bibitem[Hammond et al.(2012)]{1209.1438v3}
	Hammond, A. M., Robishaw, T., \& Gaensler, B. M.\ 
	2012 (arXiv:1209.1438)
\bibitem[Han \& Wielebinski(2002)]{2002ChJAA...2..293H}
	Han, J.~L., \& Wielebinski, R.\
	2002, ChJAA, 2, 293
\bibitem[Han et al.(2004)]{2004ApJ...610..820H}
	Han, J.~L., Ferriere, K., and Manchester, R.~N.\ 
	2004, ApJ, 610, 820
\bibitem[Han et al.(2006)]{2006ApJ...642..868H}
	Han, J.~L., Manchester, R.~N., Lyne, A.~G., 
	Qiao, G.~J., \& van Straten, W.\ 
	2006, ApJ, 642, 868
\bibitem[Han et al.(2015)]{2015AASKA14...041..41H}
	Han, J.~L., et al.\
%	Han, J.~L., Van Straten, W., Lazio, T. J. W., et al.\
	2015, in Proc. of Advancing Astrophysics with the Square Kilometre 
	Array (Trieste: SISSA), PoS (AASKA14)041, id.41
\bibitem[Hanasz et al.(2009)]{2009ApJ...706L.155H}
	Hanasz, M., W{\'o}lta{\'n}ski, D., \& Kowalik, K.\
	2009, ApJL, 706, L155
\bibitem[Harvey-Smith, Madsen, \& Gaensler(2011)]{2011ApJ...736..83H}
	Harvey-Smith, L., Madsen, G. J., \& Gaensler,B. M.\
	2011, ApJ, 736, 83
\bibitem[Haugen et al.(2003)]{2003ApJ...597L.141H}
	Haugen, N. E. L., Brandenburg, A., \& Dobler, W.\ 
	2003, ApJ, 597, L141
\bibitem[Haverkorn, Katgert, \& de Bruyn(2004)]{2004A&A...427..169H}
	Haverkorn, M., Katgert, P., \& de Bruyn, A.~G.\ 
	2004, A\&A, 427, 169
\bibitem[Haverkorn et al.(2008)]{2008ApJ...680..362H}
	Haverkorn, M., Brown, J.~C., Gaensler, B.~M., \& 
	McClure-Griffiths, N.~M.\ 
	2008, ApJ, 680, 362
\bibitem[Haverkorn(2015)]{2015ASSL...407..483H}
	Haverkorn, M.\
	2015, ASSL, 407, 483
\bibitem[Haverkorn et al.(2015)]{2015aska.confE.096G}
	Haverkorn, M., et al.\ 
%	Haverkorn, M., Akahori, T., Ettorre, C., et al.\ 
	2015, in Proc. of Advancing Astrophysics with the Square Kilometre 
	Array (Trieste: SISSA), PoS (AASKA14)103, id.96
\bibitem[Hawley et al.(1995)]{1995ApJ...440..742H}
	Hawley, J.~F., Gammie, C.~F., \& Balbus, S.~A.\
	1995, ApJ, 440, 742
\bibitem[Hazard et al.(1963)]{1963Natur.197.1037H}
	Hazard, C., Mackey, M.~B., \& Shimmins, A.~J.\
	1963, Nature, 197, 1037
\bibitem[Heald et al.(2009)]{2009A&A...503..409H}
	Heald, G., Braun, R., \& Edmonds, R.\ 
	2009, A\&A, 503, 409
\bibitem[Heesen et al.(2009)]{2009A&A...506.1123H}
	Heesen, V., Krause, M., Beck, R., \& Dettmar, R.-J.\
	2009, A\&A, 506, 1123 
\bibitem[Heiles \& Chu(1980)]{1980ApJL...235L..105H}
	Heiles, C. \& Chu, Y.\ 
	1980, ApJL, 235, L105
\bibitem[Heiles, Chu, \& Troland(1981)]{1981ApJL...247L..77H}
	Heiles, C., Chu, Y., \& Troland, T. H.\ 
	1981, ApJL, 247, L77
\bibitem[Heiles(1984)]{1984ApJS...55..585H}
	Heiles, C.\
	1984, ApJS, 55, 585
\bibitem[Herron et al.(2016)]{Herron2016}
	Herron, C.~A., Burkhart, B., Lazarian, A., Gaensler, B.~M., \& 
	McClure-Griffiths, N.~M.\ 
	2016, ApJ, 822, 13
\bibitem[Hezareh et al.(2010)]{2010ApJ...720..603H}
	Hezareh, T., Houde, M., McCoey, C., Li, Hua-bai\
	2010, ApJ, 720, 603
\bibitem[Hill et al.(2008)]{2008ApJ...686..363H}
	Hill, A. S., Benjamin, R. A., Kowel, G., Reynolds, R. J., 
	Haffner, L. M., \& Lazarian, A.\ 
	2008, ApJ, 686, 363
\bibitem[Hinshaw et al. (2013)]{2013ApJS..208...19H}
	Hinshaw, G., et al.\
	2013, ApJS. 208, 19
\bibitem[Hitomi Collaboration et al.(2016)]{2016Natur.535..117H}
	Hitomi Collaboration; Aharonian, F., et al.\ 
%	Hitomi Collaboration, Aharonian, F., Akamatsu, H., et al.\ 
	2016, Nature, 535, 117
\bibitem[Hogan(1983)]{1983PhRvL..51.1488H}
	Hogan, C.~J.\ 
	1983, PRL, 51, 1488 
\bibitem[Hong et al.(2015)]{2015ApJ...812...49H}
	Hong, S. E., Kang, H., \& Ryu, D.\
	2015, ApJ, 812, 49
\bibitem[Horiuchi et al.(2011)]{2011ApJ...738..154H}
	Horiuchi, S., Beacom, J. F., Kochanek, C. S., Prieto, J. L., 
	Stanek, K. Z., \&  Thompson, T. A.\
	2011, ApJ, 738, 154
\bibitem[Houde et al.(2013)]{2013ApJ...766...49H}
	Houde, M., Fletcher, A., Beck, R., Hildebrand, R. H., 
	Vaillancourt, J. E., \& Stil, J. M. \
	2013, ApJ, 766, 49

%%%%%%%%%%%%%%%%%%%%%%%%%%%%%%%%%%%%%%%%%%%%
%%%% I %%%%
%%%%%%%%%%%%%%%%%%%%%%%%%%%%%%%%%%%%%%%%%%%%
\bibitem[Iacobelli et al.(2014)]{2014A&A...566..5I}
	Iacobelli, M., et al.\
%	Iacobelli, M., Burkhart, B., Haverkorn, M., Lazarian, A., Carretti, 
%	E., Staveley-Smith, L., Gaensler, B.~M., Bernardi, G., 
%	Kesteven, M.~J.,\& Poppi, S.\ 
	2014, A\&A, 566, 5
\bibitem[Ichiki et al.(2006)]{2006Sci...311..827I}
	Ichiki, K., Takahashi, K., Ohno, H., Hanayama, H., \& Sugiyama, N.\ 
	2006, Science, 311, 827
\bibitem[Ideguchi et al.(2014a)]{2014PASJ...66....5I}
	Ideguchi, S., Takahashi, K., Akahori T., Kumazaki, K., \& Ryu, D.\ 
	2014a, PASJ, 66, 5
\bibitem[Ideguchi et al.(2014b)]{2014ApJ...792...51I}
	Ideguchi, S., Tashiro, Y., Akahori, T., Takahashi, K. \& Ryu, D.\ 
	2014b, ApJ, 792, 51
\bibitem[Ideguchi et al.(2016)]{Ideguchi16}
	Ideguchi, S., Tashiro, Y., Akahori, T., Takahashi, K., \& Ryu, D.\ 
	2016, ApJ, submitted
\bibitem[Inoue et al.(1984)]{1984PASJ...36..633I}
	Inoue, M., Takahashi, T., Tabara, H., Kato, T., \& Tsuboi, M.\
	1984, PASJ, 36, 633
\bibitem[Inoue et al.(2012)]{2012ApJ...744...71I}
	Inoue, T., Yamazaki, R., Inutsuka, S., \& Fukui, Y. 2012, ApJ, 744, 71
\bibitem[Inoue \& Inutsuka(2016)]{2016ApJ...833...10I}
	Inoue, T., \& Inutsuka, S.-I.\
	2016,ApJ, 833, 10 
\bibitem[Iroshnikov(1964)]{1964SvA.....7..566I}
	Iroshnikov, P. S.\ 
	1964, Sov. Astron., 7, 566
\bibitem[Itahana et al.(2015)]{2015PASJ...67..113I}
	Itahana, M., Takizawa, M., Akamatsu, H., Ohashi, T., Ishisaki, Y., 
	Kawahara, H., \& Van Weeren, R. J.\ 
	2015, PASJ, 67, 113
\bibitem[Iwamoto \& Takahara(2004)]{2004ApJ...601...78I}
	Iwamoto, S., \& Takahara, F.\ 
	2004, ApJ, 601, 78

%%%%%%%%%%%%%%%%%%%%%%%%%%%%%%%%%%%%%%%%%%%%
%%%% J %%%%
%%%%%%%%%%%%%%%%%%%%%%%%%%%%%%%%%%%%%%%%%%%%
\bibitem[Jacob \& Pfrommer(2017)]{2017MNRAS.467.1449J}
	Jacob, S., \& Pfrommer, C.\
	2017, MNRAS, 467, 1449
\bibitem[Jaffe et al.(2010)]{2010MNRAS...401..1013J}
	Jaffe, T.~R., Leahy, J.~P., Banday, A.~J., Leach, S.~M., 
	Lowe, S.~R., \& Wilkinson, A.\ 
	2010, MNRAS, 401, 1013
\bibitem[Jansson \& Farrar(2012)]{2012ApJ...757..14J}
	Jansson, R., \& Farrar, G,~R.\ 
	2012, ApJ, 757, 14
\bibitem[Jedamzik et al.(1998)]{1998PhRvD..57.3264J}
	Jedamzik, K., Katalinic, V., \& Olinto, A.\ 
	1998, PRD, 57, 3264
\bibitem[Jedamzik(2006)]{2006PhRvD..74j3509J}
	Jedamzik, K.\ 
	2006, PRD, 74, 103509
\bibitem[Johnson et al.(2015)]{2015Sci...350..1242J}
	Johnson, M., D., et al.\ 
	2015, Science, 350, 1242
\bibitem[Johnston-Hollitt et al.(2015)]{2015aska.confE..92J}
	Johnston-Hollitt, M., et al.\ 
%	Johnston-Hollitt, M., Govoni, F., Beck, R., et al.\ 
	2015, in Proc. of Advancing Astrophysics with the Square Kilometre 
	Array (Trieste: SISSA), PoS (AASKA14)092, id.92
\bibitem[Jokipii \& Lerche(1969)]{1969ApJ...157..1137J}
	Jokipii, J.~R., \& Lerche, I.\ 
	1969, ApJ, 157, 1137

%%%%%%%%%%%%%%%%%%%%%%%%%%%%%%%%%%%%%%%%%%%%
%%%% K %%%%
%%%%%%%%%%%%%%%%%%%%%%%%%%%%%%%%%%%%%%%%%%%%
\bibitem[Kahn \& Breitschwedt(1989)]{1989MNRAS.242..209K}
	Kahn, F. D. \& Breitschwedt, D.\ 
	1989, MNRAS, 242, 209
\bibitem[Kahniashvili et al.(2013)]{2013PhRvD..87h3007K}
	Kahniashvili, T., Tevzadze, A. G., Brandenburg, A., \& 
	Neronov, A.\ 
	2013, PRD, 87, 083007
\bibitem[Kamaya et al.(1996)]{1996ApJ...458L..25K}
	Kamaya, H., Mineshige, S., Shibata, K., \& Matsumoto, R.\
	1996, ApJL, 458, L25
\bibitem[Kanno et al.(2009)]{2009JCAP...12..009K}
	Kanno, S., Soda, J., \& Watanabe, M.-a.\ 
	2009, JCAP, 12, 9
\bibitem[K\"{a}pyl\"{a} et al.(2008)]{2008A&A...491..353K}
	K\"{a}pyl\"{a}, P. J., Korpi, M. J. \& Brandenburg, A.\ 
	2008, A\&A, 491, 353
\bibitem[Kawano et al.(2009)]{2009PASJ...61S..377K} 
	Kawano, N., et al.\ 
%	Kawano, N., Fukazawa, Y., Nishino, S., et al.\ 
	2009, PASJ, 61S, 377
%%%\bibitem[Kawasaki et al.(1994)]{Kawasaki:1994af}
%%%	Kawasaki, M., et al. 1995, Prog. Theor. Phys., 93, 879
\bibitem[Kawasaki \& Moroi(1995a)]{1995PThPh..93..879K}
	Kawasaki, M., \& Moroi, T.\ 
	1995, Prog. Theor. Phys., 93, 879
\bibitem[Kawasaki \& Moroi(1995b)]{1995ApJ...452..506K}
	Kawasaki, M., \& Moroi, T.\ 
	1995, ApJ. 452, 506
\bibitem[Kawasaki et al.(2005)]{2005PhRvD..71h3502}
	Kawasaki, M., Kohri, K. \& Moroi, T.\ 
	2005, PRD, 71, 083502
\bibitem[Kawasaki \& Kusakabe(2012)]{2012arXiv1204.6164K}
	Kawasaki, M., \& Kusakabe, M.\ 
	2012, PRD, 86, 063003
\bibitem[Kawashima et al.(2017)]{2017arXiv170207903K}
	Kawashima, T., Matsumoto, Y., \& Matsumoto, R.\ 
	2017, arXiv:1702.07903
\bibitem[Kennea et al.(2013)]{2013ApJ...770L..24K}
	Kennea, J. A. et al.\ 
	2013, ApJL, 770, L24
\bibitem[Keshet \& Loeb(2010)]{2010ApJ...722..737K}
	Keshet, U., \& Loeb, A.\ 
	2010, ApJ, 722, 737 
\bibitem[Kim et al.(1990)]{1990ApJ...355..29K}
	Kim, K.-T., Kronberg, P.~P., Dewdney, P.~E., \& Landecker, T.~L.\ 
	1990, ApJ, 355, 29
\bibitem[Kim et al.(2012)]{2012ApJ...751..124K}
	Kim, W.-T., \& Stone, J.~M.\
	2012, ApJ, 751, 124
\bibitem[Kimura et al.(2015)]{2015ApJ...806..159K}
	Kimura, S.~S., Murase, K., \& Toma, K.\ 
	2015, ApJ, 806, 159
\bibitem[Kobayashi et al.(2007)]{2007PhRvD..75j3501K}
	Kobayashi, T., Maartens, R., Shiromizu, T., \& Takahashi, K.\
	2007, PRD, 75, 103501
\bibitem[Kobayashi(2014)]{2014JCAP...05..040K}
	Kobayashi, T.\ 
	2014, JCAP, 5, 40
\bibitem[Koide et al.(2002)]{2002Sci...295.1688K}
	Koide, S., Shibata, K., Kudoh, T., \& Meier, D.~L.\ 
	2002, Science, 295, 1688
\bibitem[Kolmogorov(1941)]{1941DoSSR..30..301K}
	Kolmogorov, A.\ 
	1941, DoSSR, 30, 301
\bibitem[Kothes et al.(2006)]{2006A&A...457.1081K}
	Kothes, R., Fedotov, K., Foster, T. J., \& Uyanıker, B.\
	2006, A\&A, 457, 1081
\bibitem[Koyama et al.(1995)]{1995Natur.378..255K}
	Koyama, K., Petre, R., Gotthelf, E. V., Hwang, U., 
	Matsuura, M., Ozaki, M., \&  Holt, S. S.\ 
	1995, Nature, 378, 255
\bibitem[Kraichnan(1965)]{1965PhFl....8.1385K}
	Kraichnan, R. H.\ 
	1965, Phys. Fluids, 8, 1385
\bibitem[Kronberg et al.(1999)]{1999ApJ...511...56K}
	Kronberg, P. P.,Lesch, H., \& Hopp, U.\
	1999, ApJ, 511, 56
\bibitem[Kronberg et al.(2001)]{2001ApJ...560..178K}
	Kronberg, P. P., Dufton, Q. W., Li, H., \& Colgate, S. A.\
	2001, ApJ, 560, 178
\bibitem[Kronberg et al.(2008)]{2008ApJ...676...70K}
	Kronberg, P. P., Bernet, M. L., Miniati, F., Lilly, S. J., 
	Short, M. B., \& Higdon, D. M.\ 
	2008, ApJ, 676, 70
\bibitem[Kronberg \& Newton-McGee(2011)]{2011PASA...28..171K}
	Kronberg, P. P., \& Newton-McGee, K. J.\
	2011, PASA, 28, 171
\bibitem[Krumholz et al.(2007)]{2007ApJ...671..518K}
	Krumholz, M. R., Stone, J. M., \& Gardiner, T. A.\ 
	2007, ApJ, 671, 518
\bibitem[Kudoh et al.(1998)]{1998ApJ...508..186K}
	Kudoh, T., Matsumoto, R., \& Shibata, K. \
	1998, ApJ, 508, 186
\bibitem[Kudoh et al.(2007)]{2007MNRAS.380..499K}
	Kudoh, T., Basu, S., Ogata, Y., \& Yabe, T. \ 
	2007, MNRAS, 380, 499
\bibitem[Kudoh \& Basu(2011)]{2011ApJ...728..123K}
	Kudoh, T., \& Basu, S. \ 
	2011, ApJ, 728, 123
\bibitem[Kudo et al.(2011)]{2011PASJ...63..171K}
	Kudo, N., et al.\ 
%	Kudo, N., Torii, K., Machida, M., et al.\ 
	2011, PASJ, 63, 171
\bibitem[Kulesza-{\.Z}ydzik et al.(2009)]{2009A&A...498L..21K}
	Kulesza-{\.Z}ydzik, B., Kulpa-Dybe{\l}, K., Otmianowska-Mazur, K., 
	Kowal, G., \& Soida, M.\
	2009, A\&A, 498, 21
\bibitem[Kulesza-{\.Z}ydzik et al.(2010)]{2010A&A...522A..61K}
	Kulesza-{\.Z}ydzik, B., Kulpa-Dybe{\l}, K., Otmianowska-Mazur, K., 
	Soida, M. \& Urbanik, M.\
	2010, A\&A, 522, 61
\bibitem[Kumazaki et al.(2014)]{2014PASJ...66...61K}
	Kumazaki, K., Akahori, T., Ideguchi, S., Kurayama, T., \& 
	Takahashi, K.\ 
	2014, PASJ, 66, 61
\bibitem[Kulpa-Dybe{\l} et al.(2011)]{2011ApJ...733L..18K}
	Kulpa-Dybe{\l}, K., Otmianowska-Mazur, K., Kulesza-{\.Z}ydzik, B. 
	Hanasz, M., Kowal, G., W{\'o}lta{\'n}ski, D., \& Kowalik, K.\
	2011, ApJ, 733, 18
\bibitem[Kulpa-Dybe{\l} et al.(2015)]{2015A&A...575A..93K}
	Kulpa-Dybe{\l}, K., Nowak, N., Otmianowska-Mazur, K.,
	Hanasz, M., Siejkowski, H., \& Kulesza-{\.Z}ydzik, B.\
	2015, A\&A, 575, 93
\bibitem[Kusakabe et~al.(2006)]{2006PhRvD..74b3526K}
	Kusakabe, M., Kajino, T., \& Mathews, G. J.\ 
	2006, PRD, 74, 023526

%%%%%%%%%%%%%%%%%%%%%%%%%%%%%%%%%%%%%%%%%%%%
%%%% L %%%%
%%%%%%%%%%%%%%%%%%%%%%%%%%%%%%%%%%%%%%%%%%%%
\bibitem[Laing \& Bridle(1987)]{1987MNRAS.228..557L}
	Laing, R.~A., \& Bridle, A.~H.\ 
	1987, MNRAS, 228, 557
\bibitem[Langer et al.(2003)]{2003PhRvD..67d3505L}
	Langer, M., Puget, J.-L., \& Aghanim, N.\ 
	2003, Phys. Rev. D, 67, 043505  
\bibitem[Langer et al.(2005)]{2005A&A...443..367L}
	Langer M., Aghanim N., Puget J. L.\ 
	2005, A\&A, 443, 367
\bibitem[LaRosa et al.(2000)]{2000AJ....119..207L}
	LaRosa, T. N., Kassim, N. E., Lazio, T. J. W., \& Hyman, S. D.\
	2000, ApJ, 119, 207
\bibitem[Lawler \& Dennison(1982)]{1982ApJ...252...81L}
	Lawler, J.~M., \& Dennison, B.\ 
	1982, ApJ, 252, 81 
\bibitem[Leahy, Pooley \& Jagers(1986)]{1986A&A...156..234L}
	Leahy, J.~P., Pooley, G.~G., \& Jagers, W.~J.\ 
	1986, A\&A, 156, 234
\bibitem[Lewis et al.(2000)]{2000ApJ...538..473L}
	Lewis, A., Challinor, A., Lasenby, A.\ 
	2000, ApJL, 538, 473
\bibitem[Lewis and Bridle(2002)]{2002PhRvD..66j3511L}
	Lewis, A., Bridle, S.\ 
	2002, PRD, 66, 103511
\bibitem[Li \& Nakamura(2004)]{2004ApJ...609L..83L}
	Li, Zhi-Yun, \& Nakamura, F.\ 
	2004, ApJ, 609, L83
\bibitem[Li \& Houde(2008)]{2008ApJ...677.1151L}
	Li, Hua-bai, \& Houde, M.\
	2008, ApJ, 677, 1151
\bibitem[Li et al.(2009)]{2009ApJ...704..891L}
	Li, Hua-bai, Dowell, C. D., Goodman, A., Hildebrand, R., \& 
	Novak, G.\ 
	2009, ApJ, 704, 891
\bibitem[Lindley(1979)]{1979MNRAS.188P..15L}
	Lindley, D.\ 
	1979, NMRAS, 188, 15

%%%%%%%%%%%%%%%%%%%%%%%%%%%%%%%%%%%%%%%%%%%%
%%%% M %%%%
%%%%%%%%%%%%%%%%%%%%%%%%%%%%%%%%%%%%%%%%%%%%
\bibitem[Ma \& Bertschinger(1995)]{1995ApJ...455....7M}
	Ma, C.-P., \& Bertschinger, E.\ 
	1995, ApJ, 455, 7
\bibitem[Machida et al.(2008)]{2008ApJ...676.1088M}
	Machida, M. N., Inutsuka, S., \& Matsumoto, T. \
	2008, ApJ, 676, 1088
\bibitem[Machida et al.(2009)]{2009PASJ...61..411M}
	Machida, M., et al.\ 
%	Machida, M., Matsumoto, R., Nozawak, S., et al.\ 
	2009, PASJ, 61, 411
\bibitem[Machida et al.(2013)]{2013ApJ...764...81M}
	Machida, M., Nakamura, K. E., Kudoh, T., Akahori, T., 
	Sofue, Y., \& Matsumoto, R.\ 
	2013, ApJ, 764, 81
\bibitem[Mack et al.(2002)]{2002PhRvD..65l3004M}
	Mack, A., Kahniashvili, T., \& Kosowsky, A.\
	2002, PRD, 65, 123004
\bibitem[Macquart et al.(2006)]{2006ApJ...646L.111M}
	Macquart, J.-P., Bower, G. C., Wright, M. C. H., 
	Backer, D. C., \& Falcke, H.\ 
	2006, ApJL, 646, L111
\bibitem[Manchester(1974)]{1974ApJ...188..637M}
	Manchester, R.~N.\ 
	1974, ApJ, 188, 637
\bibitem[Mao et al.(2008)]{2008ApJ...688.1029M}
	Mao, S.~A., et al.\ 
%	Mao, S.~A., Gaensler, B.~M., Stanimirovi{\'c}, S., et al.\ 
	2008, ApJ, 688, 1029
\bibitem[Mao et al.(2010)]{2010ApJ...714.1170M}
	Mao, S. A., et al.\ 
%	Mao, S. A., Gaensler, B. M., Haverkorn, M., et al.\ 
	2010, ApJ, 714, 1170
\bibitem[Mao et al.(2012)]{2012ApJ...755...21M}
	Mao, S. A., et al.\
%	Mao, S. A., McClure-Griffiths, N. M., Gaensler, B. M., Brown, J. C., 
%	van Eck, C. L., Haverkorn, M., Kronberg, P. P., Stil, J. M.,
%	Shukurov, A., \& Taylor, A. R.\
	2012, ApJ, 755, 21
\bibitem[Mao et al.(2012)]{2012ApJ...759...25M}
	Mao, S.~A., et al.\ 
%	Mao, S.~A., McClure-Griffiths, N.~M., Gaensler, B.~M., et al.\ 
	2012, ApJ, 759, 25 
\bibitem[Marin(2014)]{2014MNRAS.441..551M}
	Marin, F.\ 
	2014, MNRAS, 441, 551
\bibitem[Marinacci et al.(2015)]{2015MNRAS.453.3999M}
	Marinacci, F., Vogelsberger, M., Mocz, P., \& Pakmor, R.\
	2015, MNRAS, 453, 3999
\bibitem[Marrone et al.(2007)]{2007ApJ...654L..57M}
	Marrone, D. P., Moran, J. M., Zhao, J.-H., \& Rao, R.\ 
	2007, ApJL, 654, L57
\bibitem[Mart{\'{\i}}-Vidal et al.(2015)]{2015Sci...348..311M}
	Mart{\'{\i}}-Vidal, I., Muller, S., Vlemmings, W., 
	Horellou, C., \& Aalto, S.\ 
	2015, Science, 348, 311 
\bibitem[Matsumoto et al.(1988)]{1988PASJ...40..171M}
	Matsumoto, R., Horiuchi, T., Shibata, K., \& Hanawa, T.\
	1988, PASJ, 40, 171
\bibitem[Matsumoto et al.(2017)]{2017ma}
	Matsumoto, Y., et al. submitted
\bibitem[McClure-Griffiths et al.(2005)]{2005ApJS..158..178M}
	McClure-Griffiths, N.~M., et al.\ 
%	McClure-Griffiths, N.~M., Dickey, J.~M., Gaensler, B.~M., et al.\ 
	2005, ApJS, 158, 178 
\bibitem[McClure-Griffiths et al.(2009)]{2009ApJS..181..398M}
	McClure-Griffiths, N.~M., et al.\
%	McClure-Griffiths, N.~M., Pisano, D.~J., Calabretta, M.~R., et al.\
	2009, ApJS, 181, 398 
\bibitem[Men et al.(2008)]{2008A&A...486..819M}
	Men, H., Ferri\`{e}re, K., \& Han, J.~L.\ 
	2008, A\&A, 486, 819
\bibitem[Mestel \& Spitzer(1956)]{1956MNRAS.116..503M}
	Mestel, L., \& Spitzer, L., Jr. \ 
	1956, MNRAS, 116, 503
\bibitem[Meszaros(1974)]{1974A&A....37..225M}
	Meszaros, P.\ 
	1974, A\&A, 37, 225
\bibitem[Mezger et al.(1996)]{1996A&ARv...7..289M}
	Mezger, P. G., Duschl, W. J., \& Zylka, R.\
	1996, The Astron Astrophys Rev, 7, 289
\bibitem[Migliori et al.(2007)]{2007ApJ...668..203M}
	Migliori, G., Grandi, P., Palumbo, G.~G.~C., 
	Brunetti, G., \& Stanghellini, C.\
	2007, ApJ, 668, 203
\bibitem[Miniati et al.(2001)]{2001ApJ...562..233M}
	Miniati, F., Jones, T.~W., Kang, H., \& Ryu, D.\ 
	2001, ApJ, 562, 233
\bibitem[Minter \& Spangler(1996)]{1996ApJ...458..194M}
	Minter, A.~H., \& Spangler, S.~R.\ 
	1996, ApJ, 458, 194
\bibitem[Miniati \& Bell(2011)]{2011ApJ...729...73M}
	Miniati F., Bell A. R.\ 
	2011, ApJ, 729, 73
\bibitem[Mitra et al.(2003)]{2003A&A...398..993M}
	Mitra, D., Wielebinski, R., Kramer, M., \& Jessner, A.\ 
	2003, A\&A, 398, 993
\bibitem[Miyashita et al.(2016)]{2016PASJ...68...44M}
	Miyashita, Y., Ideguchi, S. \& Takahashi, K.\ 
	2016, PASJ, 68, 44
\bibitem[Mizuno et al.(2014)]{2014JAI.....350010M}
	Mizuno, I., Kameno, S., Kano, A., et al.\ 
	2014, Journal of Astronomical Instrumentation , 3, 1450010 
\bibitem[Mori et al.(2013)]{2013ApJ...770L..23M}
	Mori, K., et al.\ 
	2013, ApJL, 770, L23
\bibitem[Morris et al.(1992)]{1992ApJ...399L..63M}
	Morris, M., Davidson, J. A., Werner, M., Dotson, J., Figer, D. F.,\
	Hildebrand, R., Novak, G., \& Platt, S.\
	1992, ApJ, 399, L63
\bibitem[Morris \& Serabyn(1996)]{1996ARA&A..34..645M} 
	Morris, M. \& Serabyn, E.\
	1996, ARA\&A, 34, 645
\bibitem[Morris et al.(2006)]{2006Natur.440..308M}
	Morris, M., Uchida, K., \& Do, T.\
	2006, Nature, 440, 308
\bibitem[Moss \& Shukurov(1996)]{1996MNRAS.279..229M}
	Moss, D., \& Shukurov, A.\ 
	1993, MNRAS, 279, 229
\bibitem[Mouschovias \& Paleologou(1980)]{1980ApJ...237..877M}
	Mouschovias, T. Ch., \& Paleologou, E. V. \ 
	1980, 237, 877
\bibitem[Mouschovias(1999)]{1999ASIC..540..305M}
	Mouschovias, T. Ch. \& Ciolek, G. E. \ 1999, in 
	``The Origin of Stars and Planetary Systems."
	Edited by Charles J. Lada and Nikolaos D. Kylafis. 
	Kluwer Academic Publishers, 305
\bibitem[Murgia et al.(2004)]{2004A&A...424..429M} 
	Murgia, M., et al.\ 
%	Murgia, M., Govoni, F., Feretti, L., et al.\ 
	2004, A\&A, 424, 429
\bibitem[Murphy(2009)]{2009ApJ...706..482M}
	Murphy, E.~J.\ 
	2009, ApJ, 706, 482

%%%%%%%%%%%%%%%%%%%%%%%%%%%%%%%%%%%%%%%%%%%%
%%%% N %%%%
%%%%%%%%%%%%%%%%%%%%%%%%%%%%%%%%%%%%%%%%%%%%
\bibitem[Nagai et al.(2014)]{2014ApJ...785...53N}
	Nagai, H., et al.\ 
%	Nagai, H., Haga, T., Giovannini, G., et al.\ 
	2014, ApJ, 785, 53 
\bibitem[Nakamura \& Li(2007)]{2007ApJ...662..395N}
	Nakamura, F., \& Li, Z.-Y.\
	2007, ApJ, 662, 395
\bibitem[Nakamura et al.(2017)]{Nakamura2017}
	Nakamura, F., et al.\ 
	2017, in preparation
\bibitem[Nakanishi et al.(2006)]{2006ApJ...651..804N}
	Nakanishi, H., Kuno, N., Sofue, Y., et al.\ 
	2006, ApJ, 651, 804
\bibitem[Nakanishi et al.(2017)]{Nakanishi17} 
	Nakanishi, H., et al. 
	2017, in preparation
\bibitem[Nakano \& Nakamura(1978)]{1978PASJ...30..671N}
	Nakano, T., \& Nakamura, T.\
	1978, PASJ, 30, 681
\bibitem[Nakano(1989)]{1989MNRAS.241..495N}
	Nakano, T. \
	1989, MNRAS, 241, 495
\bibitem[Nakazawa et al.(2009)]{2009PASJ...61..339N} 
	Nakazawa, K. et al.\ 
	2009, PASJ, 61, 339
\bibitem[Neronov \& Vovk(2010)]{2010Sci...328...73N}
	Neronov, A. \& Vovk, I.\
	2010, Science, 328, 73 
\bibitem[Netzer(2015)]{2015ARA&A..53..365N}
	Netzer, H.\
	2015, ARA\&A, 53, 365 
\bibitem[Niklas \& Beck(1997)]{1997A&A...320...54N}
	Niklas, S., \& Beck, R.\
	1997, A\&A, 320, 54 
\bibitem[Nishikori et al.(2006)]{2006ApJ...641..862N}
	Nishikori, H., Machida, M., \& Matsumoto, R.\
	2006, ApJ, 641, 862 
\bibitem[Nishiyama et al.(2009)]{2009ApJ...690.1648N}
	Nishiyama, S., et al.\
	2009, ApJ, 690, 1648
\bibitem[Nishiyama et al.(2010)]{2010ApJ...722L..23N}
	Nishiyama, S., et al.\
%%%	Nishiyama, S., Hatano, H., Tamura, M., et al.\
	2010, ApJL, 722, L23
\bibitem[Nomura et al.(2016)]{2016PASJ...68...16N}
	Nomura, M., Ohsuga, K., Takahashi, H.~R., 
	Wada, K., \& Yoshida, T.\ 
	2016, PASJ, 68, 16
\bibitem[Nota \& Katgert(2010)]{2010A&A...513..65N}
	Nota, T., \& Katgert, P.\ 
	2010, A\&A, 513, 65
\bibitem[Noutsos et al.(2008)]{2008MNRAS...386..1881N}
	Noutsos, A., Johnston, S., Kramer, M., \& Karasterigou, A.\ 
	2008, MNRAS, 386, 1881

%%%%%%%%%%%%%%%%%%%%%%%%%%%%%%%%%%%%%%%%%%%%
%%%% O %%%%
%%%%%%%%%%%%%%%%%%%%%%%%%%%%%%%%%%%%%%%%%%%%
\bibitem[O'Dea \& Owen(1987)]{1987ApJ...316...95O}
	O'Dea, C.~P., \& Owen, F.~N.\
	1987, ApJ, 316, 95
\bibitem[Ohno \& Shibata(1993)]{1993MNRAS...262..953O}
	Ohno, H., \& Shibata, S.\ 
	1993, MNRAS, 262, 953
\bibitem[Ohno et al.(2002)]{2002ApJ...577..658O}
	Ohno, H., Takizawa, M., \& Shibata, S.\ 
	2002, ApJ, 577, 658
\bibitem[Okabe \& Hattori(2003)]{2003ApJ...599..964O}
	Okabe N., \& Hattori M.\ 
	2003, ApJ, 599, 964
\bibitem[Okabe et al.(2011)]{2011ApJ...741..116O}
	Okabe, N., Bourdin, H., Mazzotta, P., \& Maurogordato, S.\ 
	2011, ApJ, 741, 116
\bibitem[Oke(1963)]{1963Natur.197.1040O}
	Oke, J.~B.\ 
	1963, Nature, 197, 1040
\bibitem[Oppermann et al.(2011)]{2011A&A...530A..89O}
	Oppermann, N., Junklewitz, H., Robbers, G., \& En\ss lin, T. A.\
	2011, A\&A, 530, 89
\bibitem[Oppermann et al.(2015)]{2015A&A...575A.118O}
	Oppermann, N., et al., \
	2015, A\&A, 575, 118
\bibitem[O'Sullivan et al.(2012)]{2012MNRAS.421.3300O}
	O'Sullivan, S. P., et al.\ 
%	O'Sullivan, S. P., Brown, S., Robishaw, et al.\ 
	2012, MNRAS, 421, 3300
\bibitem[Ota et al.(2012)]{2012RAA...12..973O}
	Ota, N.\ 
	2012, Research in Astronomy and Astrophysics, 12, 973
\bibitem[Ota et al.(2013)]{2013A&A...562..60O}
	Ota, N., Nagayoshi, K., Pratt, G.~W., Kitayama, T., 
	Oshima, T., \& Reiprich, T.~H.\ 
	2013, A\&A, 562, 60
\bibitem[Ozawa et al.(2015)]{2015PASJ...67..110O} 
	Ozawa, T., et al.
%	Ozawa, T.,  Nakanishi, H., Akahori, T., Anraku, K., Takizawa, M., 
%	Takahashi, I., Onodera, S., Tsuda, Y., \& Sofue, Y.\ 
	2015, PASJ, 67, 110

%%%%%%%%%%%%%%%%%%%%%%%%%%%%%%%%%%%%%%%%%%%%
%%%% P %%%%
%%%%%%%%%%%%%%%%%%%%%%%%%%%%%%%%%%%%%%%%%%%%
\bibitem[Pakmor et al.(2014)]{2014ApJ...783L..20P}
	Pakmor, R., Marinacci, F., \& Springel, V.\
	2014, ApJL, 783, L20 
\bibitem[Pandey et al.(2015)]{2015MNRAS.451.1692P}
	Pandey, K.~L., Choudhury, T.~R., Sethi, S.~K., \& Ferrara, A.\ 
	2015, MNRAS, 451, 1692
\bibitem[Parker(1971)]{1971ApJ...163..255P}
	Parker, E.~N.\ 
	1971, ApJ, 163, 255
\bibitem[Pavel(2011)]{2011ApJ...740..21P}
	Pavel, M.~D.\ 
	2011, ApJ, 740, 21
\bibitem[Pavel, Clemens, \& Pinnick(2012)]{2012ApJ...749..71P}
	Pavel, M.~D., Clemens, D.~P., \& Pinnick, A.~F.\ 
	2012, ApJ, 749, 71
\bibitem[Pellegrini et al.(2007)]{2007ApJ...658.1119P}
	Pellegrini, E., et al.\
	2007, ApJ, 658, 1119
\bibitem[Perley, Bridle \& Willis(1984)]{1984ApJS...54..291P}
	Perley, R.~A., Bridle, A.~H., \& Willis, A.~G.\
	1984, ApJS, 54, 291
\bibitem[Peterson \& Webber(2002)]{2002ApJ...575..217P}
	Peterson, J. D., \& Webber, W. R.\ 
	2002, ApJ, 575, 217
\bibitem[Petrosian(2001)]{2001ApJ...557..560P}
	Petrosian, V.\ 
	2001, ApJ, 557, 560
\bibitem[Pfrommer \& En{\ss}lin(2004)]{2004A&A...413...17P}
	Pfrommer, C., \& En{\ss}lin, T.~A.\ 
	2004, A\&A, 413, 17 
\bibitem[Piotrovich et al.(2015)]{2015MNRAS.454.1157P}
	Piotrovich, M.~Y., Gnedin, Y.~N., Silant'ev, N.~A., 
	Natsvlishvili, T.~M., \& Buliga, S.~D.\ 
	2015, MNRAS, 454, 1157
\bibitem[Planck Collaboration(2016)]{2016A&A...594A..13P}
	Planck Collaboration; Ade, P. A. R., et al.\
	2016, A\&A, 594, 13
\bibitem[Plante et al.(1995)]{1995ApJ...445L..113P}
	Plante, R. L., Lo, K. Y.,  \& Crutcher, R. M.\ 
	1995, ApJL, 445, L113
\bibitem[Pounds et al.(2003)]{2003MNRAS.345..705P}
	Pounds, K.~A., Reeves, J.~N., King, A.~R., et al.\ 
	2003, MNRAS, 345, 705
\bibitem[Prokopec \& Puchwein(2004)]{2004PhRvD..70d3004P}
	Prokopec, T., \& Puchwein, E.\ 
	2004, PRD, 70, 043004
\bibitem[Prouza \& ${\rm \check{S}m\acute{i}da}$(2003)]{2003A&A...410....1P}
	Prouza, M., \& ${\rm \check{S}m\acute{i}da}$, R.\ 
	2003, A\&A, 410, 1
\bibitem[Pshirkov et al.(2011)]{2011ApJ...738..192P}
	Pshirkov, M. S., Tinyakov, P. G., Kronberg, P. P., \& 
	Newton-McGee, K. J.\
	2011, ApJ, 738, 192
\bibitem[Pudritz \& Norman(1986)]{1986ApJ...301..571P}
	Pudritz, R. E., \& Norman, C. A.\
	1986, ApJ, 301, 571
\bibitem[Purcell et al.(2015)]{2015ApJ...804...22P}
	Purcell, C. R., et al.\ 
%	Purcell, C. R., Gaensler, B. M., Sun, X. H., et al.\ 
	2015, ApJ, 804, 22

%%%%%%%%%%%%%%%%%%%%%%%%%%%%%%%%%%%%%%%%%%%%
%%%% Q %%%%
%%%%%%%%%%%%%%%%%%%%%%%%%%%%%%%%%%%%%%%%%%%%

%%%%%%%%%%%%%%%%%%%%%%%%%%%%%%%%%%%%%%%%%%%%
%%%% R %%%%
%%%%%%%%%%%%%%%%%%%%%%%%%%%%%%%%%%%%%%%%%%%%
\bibitem[Rand \& Kulkarni(1989)]{1989ApJ...343..760R}
	Rand, R.~J., \& Kulkarni, S.~R.\ 
	1989, ApJ, 343, 760
\bibitem[Rao et al. (1998)]{1998ApJ...502L..75R}
	Rao, R., Crutcher, R. M., Plambeck, R. L., \& Wright, M. C. H.\ 
	1998, ApJ, 502, L75
\bibitem[Ravi et al.(2016)]{2016Sci...354.1249R}
	Ravi, V., et al.\
	2016, Science, 354, 1249
\bibitem[Reber(1944)]{1944ApJ...100..279R}
	Reber, G.\
	1944, ApJ, 100, 279
\bibitem[Rees(1987)]{1987QJRAS..28..197R}
	Rees, M. J.\
	1987, QJRAS, 28, 197
\bibitem[Reich(2007)]{reich2007.springer.63}
	Reich, W.\
	2007 in 'Mapping the Galaxy and Nearby Galaxies', ed. K. Wada and F. Combes, Springer 2007, pp63-70.
\bibitem[Rephaeli et al.(2008)]{2008SSR...134..71R}
	Rephaeli, Y., Nevalainen, J., Ohashi, T., \& Bykov, A.~M.\ 
	2008, Sp. Sci. Rev., 134, 71
\bibitem[Reynolds et al.(2005)]{2005MNRAS.357..242R}
	Reynolds, C. S., McKernan, B., Fabian, A. C., Stone, J. M., \& 
	Vernaleo, J. C.\ 
	2005, MNRAS, 357, 242
\bibitem[Reynolds et al.(2012)]{2012SSRv..166..231R}
	Reynolds, S. P., Gaensler, B. M., \& Bocchino, F.\ 
	2012, Sp. Sci. Rev., 166, 231
\bibitem[Ressler et al.(2014)]{2014ApJ...790...85R}
	Ressler, S. M., Katsuda, S., Reynolds, S. P., Long, K. S., 
	Petre, R., Williams, B. J., \& Winkler, P. F.\ 
	2014, ApJ, 790, 85
\bibitem[Rodriguez, Gomez, \& Tafoya(2012)]{2012MNRAS.420..279R}
	Rodriguez, L. F., Gomez, Y.,  \& Tafoya, D.\ 
	2012, MNRAS, 420, 279
%%%\bibitem[Rickett et al.(2009)]{2009MNRAS...395..1391R}
%%%	Rickett, R., Johnston, S., Tomlinson, T., \& Reynolds, J.\ 
%%%	2009, MNRAS, 395, 1391
\bibitem[Rybicki \& Lightman(1979)]{1979rpa..book.....R}
	Rybicki, G. B., \& Lightman, A. P.\ 
	1979, Radiation processes in astrophysics 
	(New York: Wiley-Interscience)
\bibitem[Rudnick \& Owen(2014)]{2014ApJ...785...45R}
	Rudnick, L. \& Owen, F. N.\ 
	2014, ApJ, 785, 45
\bibitem[Russell et al.(2013)]{2013MNRAS.432..530R}
	Russell, H.~R., et al.\
%	Russell, H.~R., McNamara, B.~R., Edge, A.~C., et al.\ 
	2013, MNRAS, 432, 530
\bibitem[Ruszkowski et al.(2007)]{2007MNRAS.378..662R}
	Ruszkowski, M., En{\ss}lin, T. A., Br\"{u}ggen, M., Heinz, S., \& 
	Pfrommer, C.\ 
	2007, MNRAS, 378, 662
\bibitem[Ryu, Kang, \& Biermann(1998)]{1998A&A...335...19R}
	Ryu, D., Kang, H. \& Biermann, P.~L.\ 
	1998, A\&A, 335, 19
\bibitem[Ryu et al.(2008)]{2008Sci...320..909R}
	Ryu, D., Kang, H., Cho, J., \& Das, S.\ 
	2008, Science, 320, 909
\bibitem[Ryu et al.(2012)]{2012SSRv..166....1R}
	Ryu D., Schleicher D. R. G., Treumann R. A., Tsagas C. G., \& 
	Widrow, L. M.\
	2012, Sp. Sci. Rev., 166, 1

%%%%%%%%%%%%%%%%%%%%%%%%%%%%%%%%%%%%%%%%%%%%
%%%% S %%%%
%%%%%%%%%%%%%%%%%%%%%%%%%%%%%%%%%%%%%%%%%%%%
\bibitem[Serabyn \& Gusten(1991)]{1991A&A...242..376S}
	Serabyn, E.  \& Gusten, R.\ 
	1991, A\&A, 242, 376
\bibitem[Sarazin(1999)]{1999ApJ...520..529S}
	Sarazin, C.~L.\ 
	1999, ApJ, 520, 529
\bibitem[Schekochihin et al. (2005)]{2005ApJ...625L.115S}
	Schekochihin, A. A., Haugen, N. E. L., Brandenburg, A., 
	Cowley, S. C., Maron, J. L., \& McWilliams, J. C.\ 
	2005, ApJ, 625, L115
%%%\bibitem[Seyfert(1943)]{1943ApJ....97...28S}
%%%	Seyfert, C.~K.\
%%%	1943, ApJ, 97, 28
\bibitem[Shiromoto et al.(2014)]{2014ApJ...782..108S}
	Shiromoto, Y., Susa, H., \& Hosokawa, T.\ 
	2014, ApJ, 782, 108 
\bibitem[Schleicher et al.(2010)]{2010A&A...522..115S}
	Schleicher D. R. G., Banerjee R., et al.\ 
	2010, A\&A, 522, 115
\bibitem[Schlickeiser(2002)]{2002CosmicRayAstrophysicsS}
	Schlickeiser, R.\ 
	2002, Cosmic Ray Astrophysics, Springer-Verlag, Berlin Heidelberg
\bibitem[Schmidt(1963)]{1963Natur.197.1040S}
	Schmidt, M.\
	1963, Nature, 197, 1040
\bibitem[Schnitzeler(2010)]{2010MNRAS.409L..99S}
	Schnitzeler, D. H. F. M.\ 
	2010, MNRAS, 409, 99
\bibitem[She \& Leveque(1994)]{1994PRL....72..336S}
	She, Z. \& Leveque, E.\ 
	1994, PRL, 72, 336
\bibitem[Shibata \& Uchida(1987)]{1987PASJ...39..559S}
	Shibata, K., \& Uchida, Y.\
	1987, PASJ, 39, 559
\bibitem[Shibata \& Sekiguchi(2005)]{2005PhRvD..72d4014S}
	Shibata, M., \& Sekiguchi, Y.-I.\
	2005, PRD, 72, 044014
\bibitem[Shulevski et al.(2015)]{2015A&A...579A..27S}
	Shulevski, A., et al.\ 
%	Shulevski, A., Morganti, R., Barthel, P.~D., et al.\ 
	2015, A\&A, 579, 27
\bibitem[Sikora et al.(2005)]{2005ApJ...625...72S} 
	Sikora, M., Begelman, M.~C., Madejski, G.~M., \& Lasota, J.-P.\ 
	2005, ApJ, 625, 72 
\bibitem[Simard-Normandin \& Kronberg(1980)]{1980ApJ...242..74S}
	Simard-Normandin, M., \& Kronberg, P.~P.\ 
	1980, ApJ, 242, 74
\bibitem[Simonetti et al.(1984)]{1984ApJ...284..126S}
	Simonetti, J.~H., Cordes, J.~M., \& Spangler, S.~R.\ 
	1984, ApJ, 284, 126
\bibitem[Silant'ev et al.(2013)]{2013AstBu..68...14S}
	Silant'ev, N.~A., Gnedin, Y.~N., Buliga, S.~D., 
	Piotrovich, M.~Y., \& Natsvlishvili, T.~M.\ 
	2013, Astrophysical Bulletin, 68, 14
\bibitem[Skilling(1975)]{1975MNRAS.173..255S}
	Skilling, J.\ 
	1975, MNRAS, 173, 255
\bibitem[Smith et al.(2002)]{2002MNRAS.335..773S}
	Smith, J.~E., et al.\ 
%	Smith, J.~E., Young, S., Robinson, A., et al.\ 
	2002, MNRAS, 335, 773
\bibitem[Sokoloff et al.(1998)]{1998MNRAS.299..189S}
	Sokoloff, D. D., Bykov, A. A., Shukurov, A., Berkhuijsen, E. M.,
	Beck, R., \& Poezd, A. D.\
	1998, MNRAS, 299, 189
\bibitem[Sofue et al.(1980)]{1980A&A....91..335S}
	Sofue, Y., Takano, T., \& Fujimoto, M.\ 
	1980, A\&A, 91, 335
\bibitem[Sofue \& Takano(1981)]{1981PASJ...33...47S}
	Sofue, Y., \& Takano, T.\ 
	1981, PASJ, 33, 47 
\bibitem[Sofue \& Fujimoto(1983)]{1983ApJ...265..722S}
	Sofue, Y., \& Fujimoto, M.\ 
	1983, ApJ, 265, 722
\bibitem[Sofue \& Handa(1984)]{1984Natur.310..568S}
	Sofue, Y. \& Handa, T.\
	1984, Nature, 310, 568 
\bibitem[Sofue et al.(1986)]{1986ARA&A..24..459S} 
	Sofue, Y., Fujimoto, M., \& Wielebinski, R.\ 
	1986, ARA\&A, 24, 459
\bibitem[Sofue et al.(1987)]{1987PASJ...39...95S}
	Sofue, Y., Reich, W., Inoue, M. \& Seiradakis, J.H.,\
	1987, PASJ, 39, 95
\bibitem[Sofue(1989)]{1989IAUS..136..213S}
	Sofue, Y.\
	1989, in The Center of the Galaxy, ed. M.Morris (Dordrecht: Kluwer), 213
\bibitem[Sofue et al.(1994)]{1994AJ....108.2102S}
	Sofue, Y., Wakamatsu, K.-I., \& Malin, D.~F.\
	1994, AJ, 108, 2102 
\bibitem[Sofue et al.(2010)]{2010PASJ...62.1191S}
	Sofue, Y., Machida, M.,  \& Kudoh, T.\ 
	2010, PASJ, 62, 1191 
\bibitem[Sofue(2017)]{sofue2017.springer.6}
	Sofue, Y.\ 
	2017 in 'Galactic Radio Astronomy', Springer, chapter 6.
\bibitem[Sofue \& Nakanishi(2017)]{2017MNRAS.464..783S}
	Sofue, Y., \& Nakanishi, H.\
	2017, MNRAS, 464, 783 
%%%\bibitem[Springel \& Hanquist(2003a)]{2003MNRAS.339..289S}
%%%	Springel, V., \& Hernquist, L.\ 
%%%	2003, MNRAS, 339, 289
%%%\bibitem[Springel \& Hanquist(2003b)]{2003MNRAS.339..312S}
%%%	Springel, V., \& Hernquist, L.\ 
%%%	2003, MNRAS, 339, 312
\bibitem[Stasyszyn et al.(2010)]{2010MNRAS.408..684S}
	Stasyszyn F., Nuza S. E., Dolag K., Beck R., Donnert J.\ 
	2010, MNRAS, 408, 684
\bibitem[Stepanov et al.(2008)]{2008A&A...480...45S}
	Stepanov, R., Arshakian, T.~G., Beck, R., Frick, P., \& Krause, M.\ 
	2008, A\&A, 480, 45
\bibitem[Stepanov et al.(2014)]{2014MNRAS.437.2201S}
	Stepanov, R., et al.\ 
%	Stepanov, R., Shukurov, A., Fletcher, A., et al.\ 
	2014, MNRAS, 437, 2201 
\bibitem[Stil et al.(2011)]{2011ApJ...726....4S}
	Stil, J. M., Taylor, A. R., \& Sunstrum, C.\ 
	2011, ApJ, 726, 4
\bibitem[Stil et al.(2014)]{2014ApJ...787...99S}
	Stil, J. M., Keller, B. W., Geoge, S. J., \& Taylor, A. R.\ 
	2014, ApJ, 787, 99
\bibitem[Stone \& Norman(1992)]{1992ApJS...80..791S}
	Stone, J.~M., \& Norman, M.~L.\
	1992, ApJS, 80, 791 
\bibitem[Stone et al.(2008)]{2008ApJS..178..137S}
	Stone, J.~M., Gardiner, T.~A., Teuben, P., 
	Hawley, J.~F., \& Simon, J.~B.\
	2008, ApJS, 178, 137
\bibitem[Stroe et al.(2013)]{2013A&A...555..A110S} 
	Stroe, A., van Weeren, R. J., Intema, H. T., 
	R\"{o}ttgering, H. J. A., Br\"{u}ggen, M., \& Hoeft, M.\ 
	2013, A\&A, 555, 110
\bibitem[Stroe et al.(2016)]{2016A&A...455..2402S} 
	Stroe, A., Shimwell, T., Rumsey, C., et al.\ 
	2016, MNRAS, 455, 2402
\bibitem[Strom \& Jaegers(1988)]{1988A&A...194...79S}
	Strom, R.~G., \& Jaegers, W.~J.\ 
	1988, A\&A, 194, 79
\bibitem[Subramanian \& Barrow(1998)]{1998PhRvD..58h3502S}
	Subramanian, K., \& Barrow, J. D., 
	1998, PRD, 58, 083502
\bibitem[Subramanian(2016)]{2016RPPh...79g6901S}
	Subramanian, K.\ 
	2016, Reports on Progress in Physics, 79, 076901
\bibitem[Sugawara et al.(2009)]{2009PASJ...61..1293S}
	Sugawara, C., Takizawa, M., \& Nakazawa, K.\ 
	2009, PASJ, 61, 1293
\bibitem[Sun et al.(2007)]{2007A&A...463..993S}
	Sun, X. H., Han, J. I., Reich, W., Reich, P., Shi, W. B., 
	Wielebinski, R., \& F\"urst, E.\
	2007, A\&A, 463, 993
\bibitem[Sun et al.(2008)]{2008A&A...477..573S}
	Sun, X. H., Reich, W., Waelkens, A., \& En{\ss}lin, T. A.\ 
	2008, A\&A, 477, 573
\bibitem[Sun \& Reich(2009)]{2009A&A...507..1087S}
	Sun, X. H., \& Reich, W.\ 
	2009, A\&A, 507, 1087
\bibitem[Sun \& Reich(2010)]{2010RAA....10.1287S}
	Sun, X. H., \& Reich, W.\ 
	2010, Research in Astronomy and Astrophysics, 10, 1287
\bibitem[Sun et al.(2015)]{2015ApJ...811...40S}
	Sun, X. H., et al.\ 
%	Sun, X. H., Landecker, T. L., Gaensler, B. M., et al.\ 
	2015, ApJ, 811, 40
\bibitem[Sun et al.(2015)]{2015AJ...149...60S}
	Sun, X. H., Rudnick, L., Akahori, T., et al.\ 
	2015, AJ, 149, 60
\bibitem[Suyama \& Yokoyama(2012)]{2012PhRvD..86b3512S}
	Suyama, T., \& Yokoyama, J.\ 
	2012, PRD, 86, 023512

%%%%%%%%%%%%%%%%%%%%%%%%%%%%%%%%%%%%%%%%%%%%
%%%% T %%%%
%%%%%%%%%%%%%%%%%%%%%%%%%%%%%%%%%%%%%%%%%%%%
\bibitem[Takahashi et al.(2005)]{2005PhRvL..95l1301T}
	Takahashi, K., Ichiki, K., Ohno, H., Hanayama, H.\ 
	2005, PRL 95, 121301
\bibitem[Takahashi et al.(2009)]{2009PASJ...61..957T}
	Takahashi, K., et al.\ 
%	Takahashi, K., Nozawa, S., Matsumoto, R., et al.\ 
	2009, PASJ, 61, 957
\bibitem[Takahashi et al.(2012)]{2012ApJ...744L...7T}
	Takahashi, K., Mori, M., Ichiki, K., \& Inoue, S.\
	2012, ApJ, 744, 7
\bibitem[Takahashi et al.(2016)]{2016ApJ...826...23T}
	Takahashi, H.~R., Ohsuga, K., Kawashima, T., \& Sekiguchi, Y.\ 
	2016, ApJ, 826, 23
\bibitem[Takami \& Sato(2010)]{2010ApJ...724.1456T}
	Takami, H., \& Sato, K.\ 
	2010, ApJ, 724, 1456
\bibitem[Takizawa \& Naito(2000)]{2000ApJ...535..586T}
	Takizawa, M., \& Naito, T.\ 
	2000, ApJ, 535, 586
\bibitem[Takizawa(2008)]{2008ApJ...687..951T}
	Takizawa, M.\ 
	2008, ApJ, 687, 951
\bibitem[Tanaka \&Takahara(2010)]{2010ApJ...715.1248T}
	Tanaka, S. J., \& Takahara, F.\
	2010, ApJ, 715, 1248
\bibitem[Taylor et al.(1990)]{1990ApJ...360...41T}
	Taylor, G.~B., Perley, R.~A., Inoue, M., Kato, T., 
	Tabara, H., \& Aizu, K.\ 
	1990, ApJ, 360, 41
\bibitem[Taylor et al.(2009)]{2009ApJ...702.1230T}
	Taylor, A. R., Stil, J. M., \& Sunstrum, C.\ 
	2009, ApJ, 702, 1230
\bibitem[Taylor et al.(2015)]{2015aska.confE.113T}
	Taylor, R., et al.\ 
%	Taylor, R., Agudo, I., Akahori, T., et al.\ 
	2015, in Proc. of Advancing Astrophysics with the Square Kilometre 
	Array (Trieste: SISSA), PoS (AASKA14)113, id.113
\bibitem[Tegmark et al.(2006)]{2006PhRvD..74l3507T}
	Tegmark, M., et al.\
	2006, PRD, 74, 123507
\bibitem[Thomson \& Nelson(1980)]{1980MNRAS...191..863T}
	Thomson, R.~C., \& Nelson, A.~H.\ 
	1980, MNRAS, 191, 863
\bibitem[Toma \& Takahara(2012)]{2012ApJ...754..148T}
	Toma, K., \& Takahara, F.\ 
	2012, ApJ, 754, 148
\bibitem[Tombesi et al.(2010)]{2010A&A...521A..57T}
	Tombesi, F., et al.\ 
%	Tombesi, F., Cappi, M., Reeves, J.~N., et al.\ 
	2010, A\&A, 521, 57
\bibitem[Tomisaka(2002)]{2002ApJ...575..306T}
	Tomisaka, K.\
	2002, ApJ, 575, 306
\bibitem[Torii et al.(2010)]{2010PASJ...62..675T}
	Torii, K., et al.\ 
%	Torii, K., Kudo, N., Fujishita, M., et al.\ 
	2010, PASJ, 62, 675
%%%\bibitem[Tornatore et al.(2007)]{2007MNRAS.382.1050T}
%%%	Tornatore, L.; Borgani, S.; Dolag, K.; Matteucci, F.\
%%%	2007, MNRAS, 382, 1050
\bibitem[Tosa \& Fujimoto(1978)]{1978PASJ...30..315T}
	Tosa, M., \& Fujimoto, M.\
	1978, PASJ, 30, 315 
\bibitem[Tribble(1991)]{1991MNRAS.250..726T}
	Tribble, P.~C.\ 
	1991, MNRAS, 250, 726
\bibitem[Troland \& Crutcher (2008)]{2008ApJ...680..457T}
	Troland, T. H., \& Crutcher, R. M.\ 
	2008, ApJ, 680, 457
\bibitem[Tsuboi et al.(1986)]{1986AJ.....92..818T}
	Tsuboi, M., Inoue, M., Handa, T., Tabara, H., Kato, T., 
	Sofue, Y., \& Kaifu, N.\
	1986, AJ, 92, 818
\bibitem[Tsuboi et al.(2015)]{2015ApJ...798L...6T}
	Tsuboi, M., et al.\ 
	2015, ApJL, 798, L6
\bibitem[Turner \& Widrow(1988)]{1988PhRvD..37.2743T}
	Turner, M.~S., \& Widrow, L.~M.\ 
	1988, PRD, 37, 2743
\bibitem[Turner(2004)]{2004ApJ...605L..45T}
	Turner, N.~J.\
	2004, ApJL, 605, L45

%%%%%%%%%%%%%%%%%%%%%%%%%%%%%%%%%%%%%%%%%%%%
%%%% U %%%%
%%%%%%%%%%%%%%%%%%%%%%%%%%%%%%%%%%%%%%%%%%%%

\bibitem[Uchida \& Shibata(1984)]{1984PASJ...36..105U}
	Uchida, Y., \& Shibata, K.\
	1984, PASJ, 36, 105
\bibitem[Uchida \& Shibata(1985)]{1985PASJ...37..515U}
	Uchida, Y., \& Shibata, K.\ 
	1985, PASJ, 37, 515
\bibitem[Uchida et al.(1985)]{1985Natur.317..699U}
	Uchida, Y., Shibata, K., \& Sofue Y.,\
	1985, Nature, 317, 699
\bibitem[Uchiyama et al.(2007)]{2007Natur.449..576U}
	Uchiyama, Y., Aharonian, F. A., Tanaka, T., 
	Takahashi, T., \& Maeda, Y.\
	2007, Nature, 449, 576
\bibitem[Ugai \& Tsuda(1977)]{1977JPlPh..17..337U}
	Ugai, M., \& Tsuda, T.\
	1977, Journal of Plasma Physics, 17, 337
\bibitem[Uyaniker et al.(2002)]{2002ApJ...565.1022U}
	Uyaniker, B., Kothes, R., \& Brunt, C. M.\ 
	2002, ApJ, 565, 1022

%%%%%%%%%%%%%%%%%%%%%%%%%%%%%%%%%%%%%%%%%%%%
%%%% V %%%%
%%%%%%%%%%%%%%%%%%%%%%%%%%%%%%%%%%%%%%%%%%%%
\bibitem[Vacca et al.(2012)]{2012A&A...540A..38V} 
	Vacca, V., et al.\ 
%	Vacca, V., Murgia, M., Govoni, F., et al.\ 
	2012, A\&A, 540, 38
\bibitem[Vacca et al.(2016)]{2016A&A...591A..13V}
	Vacca, V., et al. 
%	Vacca, V., Oppermann, N., En{\ss}lin, T., et al. 
	2016, A\&A, 591, 13
\bibitem[Vachaspati(1991)]{1991PhLB..265..258V}
	Vachaspati, T.\ 
	1991, Physics Letters B, 265, 258
\bibitem[Vall\'{e}e et al.(1986)]{1986A&A...156..386V}
	Vall\'{e}e, J.~P., MacLeod, M.~J., \& Broten, N.~W.\ 
	1986, A\&A, 156, 386 
\bibitem[Vall\'{e}e(2005)]{2005ApJ...619..297V}
	Vall\'{e}e, J.~P.\ 
	2005, ApJ, 619, 297
\bibitem[Vall\'{e}e(2008)]{2008ApJ...728..303V}
	Vall\'{e}e, J.~P.\ 
	2008, ApJ, 681, 303
\bibitem[Van Eck et al.(2011)]{2011ApJ...728...97V}
	Van Eck, C.~L., et al.\ 
%	Van Eck, C.~L., Brown, J.~C., Stil, J.~M., et al.\ 
	2011, ApJ, 728, 97
\bibitem[van Weeren et al.(2010)]{2010Sci...330..347V}
	van Weeren, R. J., R\"{o}ttgering, H. J. A., Br\"{u}ggen, \& 
	M., Hoeft, M.\ 
	2010, Science, 330, 347
\bibitem[van Weeren et al.(2012)]{2012A&A...546A.124V}
	van Weeren, R. J., R\"{o}ttgering, H. J. A., Intema, H. T., 
	Rudnick, L., Br\"{u}ggen, M., Hoeft, M. , \& Oonk, J. B. R.\ 
	2012, A\&A, 546, 124
\bibitem[Vazza et al.(2009)]{2009MNRAS...395..1333V}
	Vazza, F., Brunetti, G., \& Gheller, C.\ 
	2009, MNRAS, 395, 1333
\bibitem[Vazza et al.(2011)]{2011A&A...529..17V}
	Vazza, F., Brunetti, G., \& Gheller, C., Brunino, R., \& 
	Br\"{u}ggen, M.\ 
	2011, A\&A, 529, 17
\bibitem[Vazza et al.(2016)]{2016MNRAS...459..70V}
	Vazza, F., Br\"{u}ggen, M., Wittor, D., Gheller, C., 
	Eckert, D., \& Stubbe, M.\ 
	2016, MNRAS, 459, 70
\bibitem[Vernstrom et al.(2017)]{2017MNRAS.467.4914V}
	Vernstrom, T., Gaensler, B. M., Brown, S., 
	Lenc, E., \& Norris, R. P.\
	2017, MNRAS, 467, 4914
\bibitem[Voelk(1989)]{1989A&A...218...67V}
	Voelk, H.~J.\ 
	1989, A\&A, 218, 67 
\bibitem[Vogt \& En{\ss}lin(2003)]{2003A&A...412..373V} 
	Vogt, C., \& En{\ss}lin, T.~A.\ 
	2003, A\&A, 412, 373
\bibitem[V\"olk \& Atoyan et al.(2000)]{2000ApJ...541...88V}
	V\"olk, H. J. \& Atoyan, A. M.\ 
	2000, ApJ, 541, 88 
\bibitem[Vollmer et al.(2013)]{2013A&A...553A.116V}
	Vollmer, B., Soida, M., Beck, R., et al.\ 
	2013, A\&A, 553, 116 

%%%%%%%%%%%%%%%%%%%%%%%%%%%%%%%%%%%%%%%%%%%%
%%%% W %%%%
%%%%%%%%%%%%%%%%%%%%%%%%%%%%%%%%%%%%%%%%%%%%
\bibitem[Waelkens et al.(2009)]{2009A&A...495..697W}
	Waelkens, A., Jaffe, T., Reinecke, M., Kitaura, F. S., \& 
	En{\ss}lin, T. A.\ 
	2009, A\&A, 495, 697
\bibitem[West et al.(2016)]{2016A&A...587A.148W}
	West, J. L., Safi-Harb, S., Jaffe, T., Kothes, R., 
	Landecker, T. L., \& Foster, T.\ 
	2016, A\&A, 587, 148
\bibitem[Weymann et al.(1991)]{1991ApJ...373...23W}
	Weymann, R.~J., Morris, S.~L., Foltz, C.~B., \& Hewett, P.~C.\ 
	1991, ApJ, 373, 23
\bibitem[We{\.z}gowiec et al.(2012)]{2012A&A...545A..69W}
	We{\.z}gowiec, M., Urbanik, M., Beck, R., 
	Chy{\.z}y, K.~T., \& Soida, M.\
	2012, A\&A, 545, 69
\bibitem[Widrow et al.(2012)]{2012SSRv..166...37W}
	Widrow L. M., Ryu D., Schleicher D. R. G., Subramanian K., 
	Tsagas C. G., Treumann R. A.\ 
	2012, Sp. Sci. Rev., 166, 37
\bibitem[Wik et al.(2009)]{2009ApJ...696..1700W}
	Wik, D.~R., Sarazin, C.~L., Finoguenov, A., Matsushita, 
	K., Nakazawa, K., \& Clarke, T.~E.\ 
	2009, ApJ, 696, 1700
\bibitem[Wilson \& Weiler(1976)]{1976A&A....53...89W}
	Wilson, A. S. \& Weiler, K. W.\
	1976, A\&A, 53, 89

%%%%%%%%%%%%%%%%%%%%%%%%%%%%%%%%%%%%%%%%%%%%
%%%% X %%%%
%%%%%%%%%%%%%%%%%%%%%%%%%%%%%%%%%%%%%%%%%%%%
\bibitem[Xu et al.(2009)]{2009ApJ...698L..14X}
	Xu, H., Li, H., Collins, D. C., Li, S., Norman, M. L.\ 
	2009, ApJL, 698, L14
\bibitem[Xu et al.(2012)]{2012ApJ...759...40X}
	Xu, H. et al.\ 
	2012, ApJ, 759, 40
\bibitem[Xu \& Han(2014)]{2014MNRAS.442.3329X}
	Xu, J., \& Han, J. L.,\
	2014, MNRAS, 442, 3329

%%%%%%%%%%%%%%%%%%%%%%%%%%%%%%%%%%%%%%%%%%%%
%%%% Y %%%%
%%%%%%%%%%%%%%%%%%%%%%%%%%%%%%%%%%%%%%%%%%%%
\bibitem[Yamazaki et al.(2005)]{2005ApJ...625L...1Y}
	Yamazaki, D. G., Ichiki, K., \& Kajino, T.\ 
	2005, ApJ, 625, L1
\bibitem[Yamazaki et al.(2006)]{2006ApJ...646..719Y}
	Yamazaki, D. G., Ichiki, K., Kajino, T., \& Mathews, G. J.\ 
	2006, ApJ, 646, 719
\bibitem[Yamazaki et al.(2006)]{2006PhRvD..74l3518Y}
	Yamazaki, D. G., Ichiki, K., Umezu, K., \& Hanayama, H.\ 
	2006, PRD, 74, 123518
\bibitem[Yamazaki et al.(2008)]{2008PhRvD..77d3005Y}
	Yamazaki, D. G., Ichiki, K., Kajino, T., \& Mathews, G. J.\ 
	2008, PRD, 77, 043005
%%%\bibitem[Yamazaki, et al.(2008)]{2008PhRvD..78l3001Y}
%%%	Yamazaki, D. G., Ichiki, K., Kajino, T., \& Mathews, G. J.\ 
%%%	2008, PRD, 78, 123001
\bibitem[Yamazaki et al.(2010)]{2010PhRvD..81b3008Y}
	Yamazaki, D. G., Ichiki, K., Kajino, T., \& Mathews, G. J.\ 
	2010, PRD, 81, 023008
%%%\bibitem[Yamazaki et al.(2010)]{2010PhRvD..81j3519Y}
%%%	Yamazaki, D. G., Ichiki, K., Kajino, T., \& Mathews, G. J.\ 
%%%	2010, PRD, 81, 103519
%%%\bibitem[Yamazaki et al.(2010)]{2010AdAst2010E..80Y}
%%%	Yamazaki, D. G., Ichiki, K., Kajino, T., \& Mathews, G. J.\ 
%%%	2010, Advances in Astronomy, Vol. 2010, 586590
\bibitem[Yamazaki et al.(2011)]{2011PhRvD..84l3006Y}
	Yamazaki, D. G., Ichiki, K., \& Takahashi, K.\ 
	2011, PRD, 84, 123006
\bibitem[Yamazaki et al.(2012)]{2012PhR...517..141Y}
	Yamazaki, D. G., Kajino, T., Mathews, G. J., \& Ichiki, K.\ 
	2012, Phys. Rep., 517, 141
\bibitem[Yamazaki \& Kusakabe(2012)]{2012PhRvD..86l3006Y}
	Yamazaki, D. G., \& Kusakabe, M.\ 
	2012, PRD, 86, 123006
%%%\bibitem[Yamazaki et al.(2013)]{2013PhRvD..88j3011Y}
%%%	Yamazaki, D. G., Ichiki, K., \& Takahashi, K.\ 
%%%	2013, PRD, 88, 103011
\bibitem[Yamazaki(2014)]{2014PhRvD..89j3528Y}
	Yamazaki, D. G.\ 
	2014, PRD, 89, 083528
\bibitem[Yamazaki et al.(2014)]{2014PhRvD..90b3001Y}
	Yamazaki, D. G., Kusakabe, M., Kajino, T., Mathews, G. J., \& 
	Cheoun, M. K.\ 
	2014, PRD, 90, 023001
\bibitem[Yamazaki(2016)]{2016PhRvD..93d3004Y}
	Yamazaki, D. G.\ 
	2016, PRD, 93, 043004
\bibitem[Yusef-Zadeh et al.(1984)]{1984Natur.310..557Y}
	Yusef-Zadeh, F., Morris, M., \& Chance, D.\
	1984, Nature, 310, 557
\bibitem[Yusef-Zadeh(1986)]{1986PhDT.........3Y}
	Yusef-Zadeh, F.\ 
	1986, Ph.D. thesis, Columbia University, New York
\bibitem[Yusef-Zadeh \& Morris(1988)]{1988ApJ...329..729Y}
	Yusef-Zadeh, F. \& Morris, M.\
	1988, ApJ, 329, 729

%%%%%%%%%%%%%%%%%%%%%%%%%%%%%%%%%%%%%%%%%%%%
%%%% Z %%%%
%%%%%%%%%%%%%%%%%%%%%%%%%%%%%%%%%%%%%%%%%%%%
\bibitem[Zhang et al.(2016)]{Zhang2016}
	Zhang, J., Lazarian, A., Lee, H.,\& Cho, J.\ 
	2016, ApJ, 825, 154
\bibitem[Zhu \& Merard(2013)]{2013ApJ...770..130Z}
	Zhu, G., \& M\'enard, B.\ 
	2013, ApJ, 770, 130
\bibitem[ZuHone et al.(2013)]{2013ApJ...762...78Z}
	ZuHone, J.~A., Markevitch, M., Brunetti, G., \& Giacintucci, S.\ 
	2013, ApJ, 762, 78


\end{thebibliography}
\end{document}